\documentclass[preprint,authoryear,3p]{elsarticle}

\usepackage{amssymb,amsthm,amsmath}
\usepackage{algpseudocode,algorithm,algorithmicx}
\usepackage{graphicx}
\usepackage{subfig}
\usepackage{float}

\usepackage{multirow}
\usepackage{longtable,lscape}
\usepackage{cases}
\usepackage{url}
\usepackage{color}
\usepackage{xcolor}
\usepackage{comment}
\usepackage{enumitem}
\usepackage{multirow}
\usepackage{makecell}
\usepackage{longtable}

\usepackage{fancyhdr}
\usepackage{stix}

\newcommand{\tcr}[1]{{\textcolor{red}{#1}}}

\newcommand{\tco}[1]{{\textcolor{cyan}{#1}}}

\usepackage{slashbox}
\usepackage{diagbox}

\theoremstyle {plain}
\newtheorem{thm}{Theorem}

\newtheorem{prop}{Proposition}

\theoremstyle{definition}

\usepackage{fancyhdr}

\newtheorem{lemma}{Lemma}

\theoremstyle{remark}
\newtheorem*{rem}{Remark}
\newtheorem*{note}{Note}

\numberwithin{equation}{section}

\def\ps@pprintTitle{%
	\let\@oddhead\@empty
	\let\@evenhead\@empty
	\def\@oddfoot{\reset@font\hfil\thepage\hfil}
	\let\@evenfoot\@oddfoot
}
\makeatother

\makeatletter
\def\ps@pprintTitle{%
	\let\@oddhead\@empty
	\let\@evenhead\@empty
	\let\@oddfoot\@empty
	\let\@evenfoot\@oddfoot
}
\makeatother

\journal{ } 

\begin{document}

\begin{frontmatter}



\title{A Unified Network Equilibrium for \\
E-Hailing Platform Operation and Customer Mode Choice}

\author[CUaddr]{Xu Chen}
\author[CUaddr,dsi]{Xuan Di \corref{cor:Xuan}}\ead{sharon.di@columbia.edu}	

\cortext[cor:Xuan]{Corresponding author, Tel.: +1-212-853-0435.}
\address[CUaddr]{Department of Civil Engineering and Engineering Mechanics, Columbia University}
\address[dsi]{Data Science Institute, Columbia University}

\begin{abstract}
This paper aims to combine both economic and network user equilibrium for ride-sourcing and ride-pooling services, while endogenously optimizing the pooling sequence of two origin-destination (OD) pairs. 
With the growing popularity of ride-sourcing and ride-pooling services provided by transportation network companies (TNC), there lacks a theoretical network equilibrium model that accounts for the emerging ride-pooling service, due to the challenge in enumerating all possible combinations of OD pairs pooling and sequencing. 
This paper proposes a unified network equilibrium framework that integrates three modules, including travelers' modal choice between e-pooling and e-solo services, e-platforms' decision on vehicle dispatching and driver-passenger matching, and network congestion. 
To facilitate the representation of vehicle and passenger OD flows and pooling options, a layered OD graph is created encompassing ride-sourcing and ride-pooling services over origins and destination nodes. Numerical examples are performed on both small and large road networks to demonstrate the efficiency of our model. 
The proposed equilibrium framework can efficiently assist policymakers and urban planners to evaluate the impact of TNCs on traffic congestion, and also help TNCs with pricing and fleet sizing optimization.
\end{abstract}


\end{frontmatter}

\allowdisplaybreaks

\section{Introduction}

E-hail or ride-sourcing service provided by transportation network companies (TNC) has gained growing popularity in recent years \citep{national2016between}.  
However, the emergence of TNCs likely increases vehicle deadhead miles traveled \citep{erhardt2019transportation},
which thus catalyzes public agencies to consider enforcing policies and regulation to TNCs for congestion mitigation \citep{joshi2019hail}. 
To counteract the adversarial effect of offering single rides, 
TNCs begin offering shared rides, called ``for-profit pooling" or ``ridesplitting" (UberPool and LyftLine for instance), which allegedly reduces the total number of vehicles on public roads and thus relieve traffic congestion. 
However, pooling service only accounts for $20\%$ of total trips \citep{Uber2018}. 
Such a smaller than expected market could likely hinder the benefits of congestion mitigation claimed by TNC companies. 
Thus, we need to investigate the relationship among TNC service, customers' mode choices, and traffic efficiency.  
However, there lacks a theoretical network equilibrium model that accounts for the emerging pooling service so as to assess its long-term impact on network wide congestion and help promote such service. 
This paper aims to develop a generic equilibrium model that integrates economic equilibrium 
into the framework of traffic assignment problem (TAP) and lay a theoretical ground for regulation of TNCs.

The for-profit pooling service has been controversial since its inception. 
On one hand, it could reduce the number of vehicles needed to serve demands; on the other hand, inefficient matching between drivers and riders and improper operation of detouring could lengthen passengers' waiting time and worsen social welfare. 
It is thus crucial to quantify the impact of the for-profit pooling service on transportation system performance, 
so that pricing or regulation policies can be selected to not only profit TNC service providers, but also benefit the society as a whole. 
This paper is mainly focused on the long-term planning that inspects how interventions impact service operations in terms of supply and demand, as well as road congestion. 

\subsection{Literature Review} 

The ridesourcing services face a bilateral market where its supply and demand need to be matched under a certain market clearance mechanism.  
Depending on how such a supply-demand imbalance is modeled, the existing literature bifurcates into two steams: 
the first stream of research is focused on the economic analysis of the shared mobility market, 
while the second is on the spatial network equilibrium of various shared mobility modes. 

The first school of studies analyze the relation between supply and demand in order to understand how surge pricing \citep{castillo2017surge,ke2020pricing} or commission \citep{shou2020optimal,shou2020marl} can be used as a lever to balance supply and demand. 
There are a large body of studies on modeling supply-demand equilibria 
\citep{yang1998network,banerjee2016dynamic,xu2020supply,ke2020pricing}.
These studies, however, assume that the experienced travel time of vehicles remains constant and thus the spatial imbalance is omitted. 
With a growing number of ridesourcing service vehicles brought to public roads, their presence could influence traffic congestion and thus cannot be ignored in network equilibrium modeling. 
Plus, vehicles' routing behavior in congested networks would impact travel time and waiting time of passengers and thus should be accounted for. 

The above concerns lead to the second school of studies 
that integrate new mobility modes into the TAP framework,
in which not only the matching between drivers and passengers in a two-sided market is modeled, so is the movement of vehicles on congested networks. 
In other words, drivers and passengers not only face an imbalance in sizes, but also an imbalance in space.  
When drivers and passengers are distributed in a spatial network with imbalanced magnitudes, it is not only the economic imbalance of supply and demand that influences the market equilibrium.  
The spatial imbalance could also impact travel times and thus passengers' waiting time, resulting in different modal choice distributions, market economic equilibria, and network user equilibria. 

Along this line of research, 
a group of pioneers integrated modal choice and taxi movements in congested road networks \citep{wong2001modeling,wong_modeling_2008,yang2010equilibria,yang2014taxi}. 
Bearing the same characteristics of two-sided markets for both traditional taxi and TNC services, the network equilibrium models initially developed for traditional taxi markets have gained increasing attentions in ride-sourcing markets \citep{chen2020dynamic,ke2020rscongestion} on congested road networks. 
Recent studies \citep{ban2019general,di2019unified} incorporated e-hailing taxis and TNC services into TAP and developed a generalized Nash game leveraging the market economic clearance mechanism.
These studies have opened up the field of modeling the interactions among multiple traffic entities into one game-theoretic framework. However, the e-pooling service is not in the picture. 
In other words, these studies are primarily focused on one-to-one driver-passenger matching, without considering the pooling service provided by TNCs.

The existing literature on ridesouring services is primarily focused on order dispatch in on-demand platforms \citep{wang2019ridesourcing}. Some literature investigates the single-ride service, in other words, one passenger is matched to one driver \citep{wang2017RS, dickerson2018RS,ozkan2019RS,Bertsimas2019OnlineVRP,lyu2019online}. There are many studies focusing on multi-rider trips where passengers can split a ride (i.e., ride-splitting services) in ride-sourcing systems \citep{alonso2017RS,moody2019RS,simonetto2019RS,fielbaum2021split,jacob2021RS}.
We acknowledge that the dynamic nature of on-demand services renders dynamic operations, but it is still important to perform equilibrium analysis that assists city planners and policymakers in understanding and assessing the systematic impact of emerging shared mobility.

Table~\ref{tab:lit} summarizes the related work that employs the equilibria framework for various ridesharing services and indicates where this paper is positioned. 
The related work is categorized into three branches based on the equilibrium concepts: the economic equilibrium, the network user equilibrium, and the fusion of these two equilibria. The majority of literature regarding economic market equilibrium focuses on pricing of ride-sourcing services from supply and demand curves \citep{zha2016RS,zha2017surge,zha2018pricing,cachon2017surge,castillo2017surge,korolko2019RS,chen2020dynamic,ke2020pricing,zhou2021pool,yang2020RSprice,xu2020supply}. However, these studies normally do not account for traffic flows on real road networks. Recent years have seen a growing number of studies on network equilibria for ride-sharing markets \citep{xu2015complementarity,bahat2016RSequili,di2017ridesharing,di2018link,di2019unified,ban2019general,li2019rue,li2020path,chen2021RUE,li2021ue}, which account for travelers' mode and route choice as well as user equilibria. 
These studies, however, do not include ride-splitting and are primarily focused on self-organized carpooling or peer-to-peer organized ridesharing \citep{chen2021RUE}.  
There exist a few studies that model ride-splitting with network equilibria. 
\cite{noruzoliaee2022section} proposes a section-based formulation of user equilibrium for ride-splitting services, but it predetermines the sequence of travelers' origins and destinations. 
This paper aims to combine both economic and network user equilibrium for ride-sourcing and ride-splitting services, while endogenously optimizing the OD pooling sequence.


\begin{table}[H]
  	\centering
  	\fontsize{9}{0}\selectfont
  	\caption{Literature on ride-sourcing, ride-spilling, ride-sharing}
  	\label{tab:lit}
  	\begin{tabular}{p{2.5cm}||p{4cm}|p{4cm}|p{4cm}}
  		\hline
 		 Method & Ride-sourcing & Ride-splitting (or pooling) & Ride-sharing\\ \hline\hline 
  		 Economic equilibrium &  \cite{zha2016RS,zha2017surge,zha2018pricing,xu2020supply,ke2020pricing,ke2020rscongestion,zhou2021pool,cachon2017surge,castillo2017surge} & \cite{korolko2019RS,chen2020dynamic} & \cite{wang2018RS,yang2020RSprice} \\ \hline
 		 Network user equilibrium & \cite{ban2019general,di2019unified} & \cite{noruzoliaee2022section} & \cite{xu2015complementarity,bahat2016RSequili,di2017ridesharing,di2018link,li2019rue,li2020path,chen2021RUE,li2019rue,ma2021rue}  \\  \hline 
 		 
 		 Combined economic network equilibrium &  \multicolumn{2}{c|}{ \makecell{ This paper}} & Future work of this paper \\ \hline 
 	\end{tabular}
 \end{table}

\begin{note}
Subsequently, for the sake of simplicity, we call the single rides provided by TNC companies as e-solo and the shared rides as e-pooling. TNCs are simplified as e-platforms.
\end{note}




	
	
	

\subsection{Contributions of this paper}


The goal of this paper is to understand how two types of e-hailing vehicles interact with one another on a congested traffic network and how traffic congestion influences passengers' mode choices, as well as the platform's pickup plan. Accordingly, we need to consider a large number of TNC vehicles moving on a network simultaneously, whose actions would collectively influence the network traffic pattern, customers' mode choices, and the platform's matching decision, and vehicles' OD sequencing. 
Building on such a framework, we would like to 
evaluate the impact of e-platforms' operational strategies
(including vehicle dispatching, order matching, pricing, fleet sizing) on system performances (including deadhead miles traveled, total system travel time). 

This paper can be regarded as a significant extension of \cite{ban2019general,di2019unified} that only model the single ride service provided by TNCs. 
A major difference of this paper from our previous papers \citep{di2017ridesharing,di2018link,li2019rue,chen2021RUE} is that here we include the e-pooling service, which requires to explicitly pair ODs and match two orders and thus, exhibits distinction from ad hoc ridesharing. 

Before delving into our contributions, we would like to first explain the computational complexity involving ride-pooling vehicle flows. 
In an e-haling system, the sequence of pick-ups and drop-offs of two ODs is a permutation problem, which is, $P^2_2=4$ possible cases. 
When there are $N$ pairs, the total possible permutation is $N(N-1)\times P^2_2$ with a time complexity of $O(N^2)$. 
Direct brute force is practically infeasible, especially as $N$ grows, not mentioning when 3 or more rides can be shared in one trip. 
We believe that a quite different and innovative framework is warranted to accommodate the e-pooling service. 
The core contribution of this paper is the characterization of pooled trips without explicit enumeration of order pooling and sequencing.

In a nutshell, the contributions of the paper include: 
\begin{enumerate}
    \item Rather than formulating vehicle dispatching and vehicle-passenger matching as combinatorial problems, we establish an equivalence of these two problems to two sequential network flow problems over properly defined OD graphs, in order to avoid enumeration of matching and routing plans for TNC vehicles in pooling service;
    \item The topology and structure of a layered OD graph is designed over which vehicle dispatching and vehicle-passenger matching can be simplified as link-node based minimum-cost flow problems;
    \item Extensive numerical examples in small and large road networks are performed to illustrate the efficiency of our model, and offer insights into e-platform operation, planning and policymaking. 
\end{enumerate}

The remainder of the paper is organized as follows. 
Section~\ref{sec:prob} presents the problem statement and lays out three modules: e-platform optimization, customer mode choice, and network congestion. 
Section~\ref{sec:od} proposes the structure of the layered OD graph that will be used to efficiently solve combination and sequencing of vehicle and passenger flows.  
Section~\ref{sec:1_OD}-\ref{sec:3_mode} detail each module. Section~\ref{sec:sol} discusses solution properties. In Section~\ref{sec:exp}, numerical examples are presented to demonstrate the developed model on a simple 3-node network and on the Sioux Falls network.
Section~\ref{sec:conclu} concludes this study.

	
	
	

\section{Problem statement}\label{sec:prob}

\subsection{Game-theoretic scheme for the e-hailing system}
\label{subsec:game}

An e-hailing system consists of multiple agents, including customers, e-hailing platforms, and traffic entities on roads. 
Customers select e-solo or e-pooling to move from A to B. 
E-haling platforms match idle vehicles and customers, once orders are placed. 
For an e-solo order, one vehicle is matched to one passenger; while for an e-pooling order, one vehicle is matched to two passengers. 
Traffic entities are comprised of passengers and vehicles in e-solo and e-pooling modes. 
For vehicles, they can be vacant or occupied, of which the occupancy ratio of passengers can be one or two, depending on whether a vehicle serves an e-solo or e-pooling order, and if e-pooling, whether it carries one or already pools two passengers. 
Each agent has its objective (or disutility or payoff) function, strategy, and constraints. 
At equilibrium, 
 \begin{enumerate}
     \item an e-platform cannot improve its net profit by unilaterally switching its vehicle dispatching nor vehicle-passenger matching plans; 
     \item the e-hailing fleet size is no less than the order requests. In other words, supply and demand reaches an equilibrium;
     \item no traveler can reduce her generalized travel disutility by unilaterally switching her travel mode choice; 
     \item no vehicle can reduce its travel time by unilaterally switching its route choice.  
 \end{enumerate}
 
An equilibrium state is achieved during a long term planning horizon. That being said, we are not interested in a transient period with the dynamic operation of TNCs. Instead, we are focused on a stead state when customer order arrival rates, platform operational strategies, and traffic congestion remain stationary. 
Moreover, the supply would fulfill the demand at the equilibrium. In other words, the vehicle number is no less than the order number.

We make the following modeling assumptions: 
\begin{enumerate}[label=(A\arabic*)]
	\item One idle vehicle can pick up passengers from at maximum two different origin-destination (OD) pairs in a single trip. In other words, a maximum of two groups of riders are allowed to pool.
	\item 
	One idle driver can only carry one group of passengers from the same OD pair. 
\end{enumerate}

\begin{rem}
\begin{enumerate}[label=(A\arabic*)]
	\item This assumption is made for both practical and computational reasons. 
	First, according to empirical evidence \citep{Uber2018}, pooling service only accounts for $20\%$ of total trips, not mentioning pooling three or more rides. The challenge of matching rides persists in the real-world, given that the probability of more than two incoming order requests within the same time window with spatial proximity could be small. 
	Moreover, pooling more than two rides usually refers to as on-demand shuttle service, rather than ride-sourcing.
	Second, From the computational perspective, pooling more than two rides involves solving a more complex problem over a higher-dimensional OD graphs. Thus, it is left for future research.
	\item This assumption is proposed not to simplify the model, rather to make it less trivial to model. Allowing riders from the same OD to share rides is the central focus of a series of work in ridesharing user equilibrium (RUE) cited in Table~\ref{tab:lit}. Since RUE has been extensively studied, we impose such an assumption to avoid repeating similar modeling. 
\end{enumerate}
\end{rem}

\subsection{Three modules: Coupling among e-platform operation, customer choice, and network congestion}

\begin{figure}[H]
  	\centering
  	\includegraphics[scale=.4]{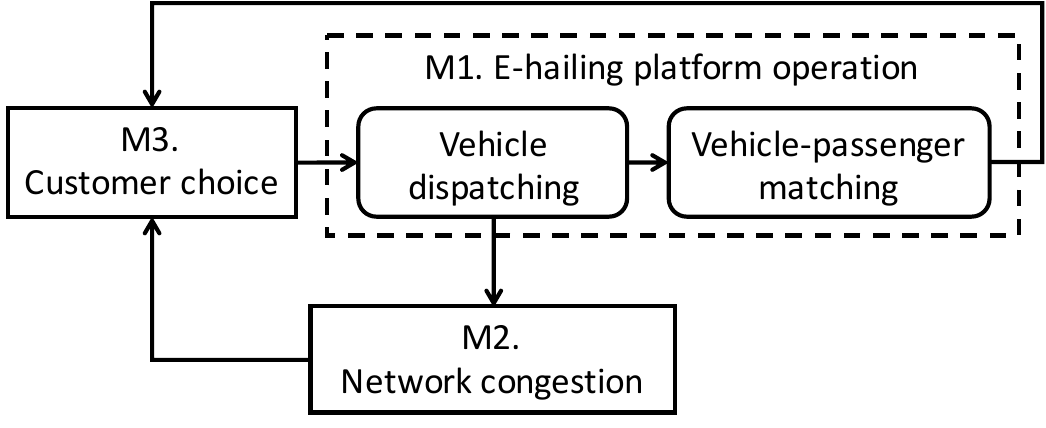}
  	\caption{Three modules: 
  	(Module 1) The e-hailing platform operation module contains two submodules: first dispatch idle vehicles and then match passengers and vehicles. 
  	  The vehicle dispatching and matching of passengers and vehicles would incur waiting cost and matching friction, which would contribute to passengers' travel disutility. (Module 2) The customer choice module outputs passengers' mode choice selection between e-solo and e-pooling, resulting in orders for the e-hailing platform to serve and thus incurring the dispatching of idle vehicles.
  	(Module 3) The dispatched vehicles driving on roads would contribute to traffic network congestion and such a congestion effect would further influence travel mode utilities and ultimately customers' choices.
  	}
  	\label{fig:3modules}					
\end{figure}

This paper develops a methodological framework encompassing three methodological modules, which are (1) e-hailing platform operation including vehicle dispatching and passenger-vehicle matching, (2) network congestion, and (3) customer choice.
Fig.~\ref{fig:3modules} demonstrates the linkage and logical order among these three modules. 

\subsection{Notations}

Before proceeding to each individual module, we will first introduce notations that will be used in this paper. 
The notations 
are summarized in Table~\ref{tab:notations}.

{\small
	\begin{longtable}{c|l|l|l}
		\hline
		Road network& set &$\mathcal{N}_{(net)}$ & node set on a road network \\
		&   &$\mathcal{L}_{(net)}$ & link set on a road network \\
		&  &$\mathcal{G}_{(net)}$ &  network $\mathcal{G}_{(net)}=\{\mathcal{N}_{(net)},\mathcal{L}_{(net)} \}$\\
		& & $\mathcal{W}$ & OD pair set $\mathcal{W}=\{1,...,w,... \}$\\
		& components &$x_{ij}^{d}$ & link flow from $i$ to $j$ on a road network destined at $d$ \\
		& &$x_{ij}$ & aggregate link flow from $i$ to $j$ on a road network \\
		& &$t_{ij}$ & travel time when traveling from $i$ to $j$\\
		& &$\tau_{i}^{d}$ & node potential of $i$ or shortest travel time from $i$ to $d$ \\
		& & $q_{w}$ & travel demand of OD pair $w$  \\
		& & $m$ & travel mode, $m \in \mathcal{M}, \mathcal{M}=\{e,p \}$ where $e$: e-solo, $p$: e-pooling \\
		& & $q_{w}^{[m]}$ & travel demand of OD pair $w$ regarding travel mode $m$ \\
		\hline
		OD graph & node set& $\mathcal{N}_O$ & origin node set\\
		& & $\mathcal{N}_D$ & destination node set \\
		& &$\mathcal{N}^{+}$ & virtual origin node set\\
		& & $\mathcal{N}^{-}$ & virtual destination node set \\
		& &$\mathcal{N}(m)$ & node set in the OD graph for mode $m$ - vehicle layer \\
		& & $\mathcal{N}$ & node set in the OD graph\\
		& edge set &$\mathcal{L}(m)$ & edge set in the OD graph for mode $m$ - vehicle layer \\
		& & $\mathcal{L}$ & edge set in the OD graph\\
		& graph & $\mathcal{G}$ & OD graph $\mathcal{G}=\{ \mathcal{N}, \mathcal{L} \}$ \\
		 &  & $\mathcal{G}^{[m]}$ & OD graph regarding mode $m$  $\mathcal{G}^{[m]}= \{ \mathcal{N}(m), \mathcal{L}(m) \}$\\ 
		&  vehicle flow & $z_{uv}^{[m]}$ & vehicle flow in mode $m$ from node $u$ to $v$  \\
		& & $z_{\bar{w}\bar{w}_{[m]}}$ & vehicle flow from virtual origins to origins  \\
		& & $	z_{\underline{w}_{[m]}\underline{w}}$ & vehicle flow from destinations to virtual destinations \\
		& & $z_{v}^{u}$ & rebalancing flow from virtual node $v$ to $u$  \\
		&  passenger flow & $y^{(w)[m]}_{uv}$ & passenger flow on edge $(u,v)$  belonging to OD pair $w$\\
		& cost & $C_{vu}^{[m]}$ & cost of mode $m$ on edge $(v,u)$ \\
		&   & $C_{v}^{u}$ & cost of the rebalancing edge $(v,u)$ \\
		&   & $R_{w}$ & base fare for orders of OD pair $w$  \\
		&   & $F_{w}$ & fixed fare for orders of OD pair $w$  \\
		&   & $t_{vu}$ & travel time from node $v$ to $u$  \\
		&   & $\hat{t}_{vu}$ & free-flow travel time from node $v$ to $u$  \\
		&   & $l_{vu}$ & travel distance from node $v$ to $u$  \\
		&   & $\alpha_1,\alpha_2$ & cost coefficients in the base fare $R_{w}$ \\
		&   & $r^{[m]}_{w}$ & trip fare of mode $m$\\
		&   & $\gamma_p$ & discount multiplier for e-pooling service \\
		&   & $c^{[m](se-veh)}_{\bar{w}}$ & search friction, i.e., time spent on seeking riders\\
		&   & $\beta_{(se-veh)}$ & cost coefficient of search friction\\
		& & $S$ & fleet size\\
		\hline
		\multirow{5}{*}{\makecell{Linkage \\between \\ Road network \\ \& OD graph}}
		&  indicator & $\delta^{(o)}_{u \bar{w}}$ &  whether $u$ is the origin of OD pair $w$ \\
		&  & $\delta^{(d)}_{u \underline{w}}$ &  whether $u$ is the destination of OD pair $w$ \\
		&  & $\delta^{(od)}_{(u,v) w}$ & whether $(u,v)$ is OD pair $w$ \\
		&  & $\delta^{(oo)}_{(u,v) wk}$ & whether $(u,v)$ is OO pair $(\bar{w},\bar{k})$ \\
		&  & $\delta^{(od)}_{(u,v) wk}$ & whether $(u,v)$ is OD pair $(\bar{w},\underline{k})$ \\
		&  & $\delta^{(dd)}_{(u,v) wk}$ & whether $(u,v)$ is DD pair $(\underline{w},\underline{k})$ \\
		&  & $\delta^{(do)}_{(u,v) wk}$ &  whether $(u,v)$ is DO pair $(\underline{w},\bar{k})$ \\
		\hline
		\multirow{2}{*}{\makecell{Customer \\choice}} & disutility  & $U^{[m]}_{w}$ & travel disutility of passengers  \\
		&  nodal cost & $c^{[m](nod)}_{\bar{w}}$ & nodal cost regarding travel mode $m$ \\
		&   & $c^{[m](wt)}_{\bar{w}}$, $\beta^{[m]}_{(wt)}$ & cost of waiting for vehicles to reach nodes, coefficient\\
		&   & $c^{[m](se-pas)}_{\bar{w}}$, $\beta_{(se-pas)}$ & passengers' waiting time caused by search friction, coefficient\\
		& edge cost  & $c^{[m](surp-pas)}_{w}$, $\beta^{[m]}_{(surp-pas)}$ &  supply-demand surplus for passenger flows, coefficient  \\
		& in-vehicle cost  & $c^{(in-veh)}_{w}$, $\beta_{(in-veh)}$ & travel time for passengers to stay within a vehicle, coefficient  \\
		& inconvenience cost   & $c^{[p](inc)}_{w}$, $\beta_{(inc)}$ & inconvenience cost for e-pooling passengers, coefficient\\
		\hline
		\caption{Notation Glossary}
		\label{tab:notations}
	\end{longtable}
}

\begin{rem}
We will explain our naming conventions here.
\begin{enumerate}
\item Letters $w,k$: represent OD pairs, and the bar or underline of $w,k$ to represent its origin or destination node. 
\item Letters $u,v$: denote generic nodes in an OD graph, can be origin or destination nodes. 
\item Sub- or superscripts $m$: represent travel modes associated with nodes.
\begin{itemize}
    \item Vehicle flow on edge $z_{uv}^{[m]} \triangleq z_{u_{[m]}v_{[m]}}$: if both nodes of an edge contain a subscript of mode, we move the mode to the superscript.
    \item Vehicle flow on virtual edge $z_{uu_{[m]}}$: if only one node contains a subscript of mode, we leave the mode as the subscript of the node index.
    \item Vehicle flow on rebalancing edge $z_{v}^{u} \triangleq z_{uv}$: we place the origin as the subscript and the destination as the superscript.
\end{itemize}
\end{enumerate}
\end{rem}



\subsection{Overview of the unified model}\label{sec:unified}

\begin{figure}
	\centering
	\includegraphics[scale=.4]{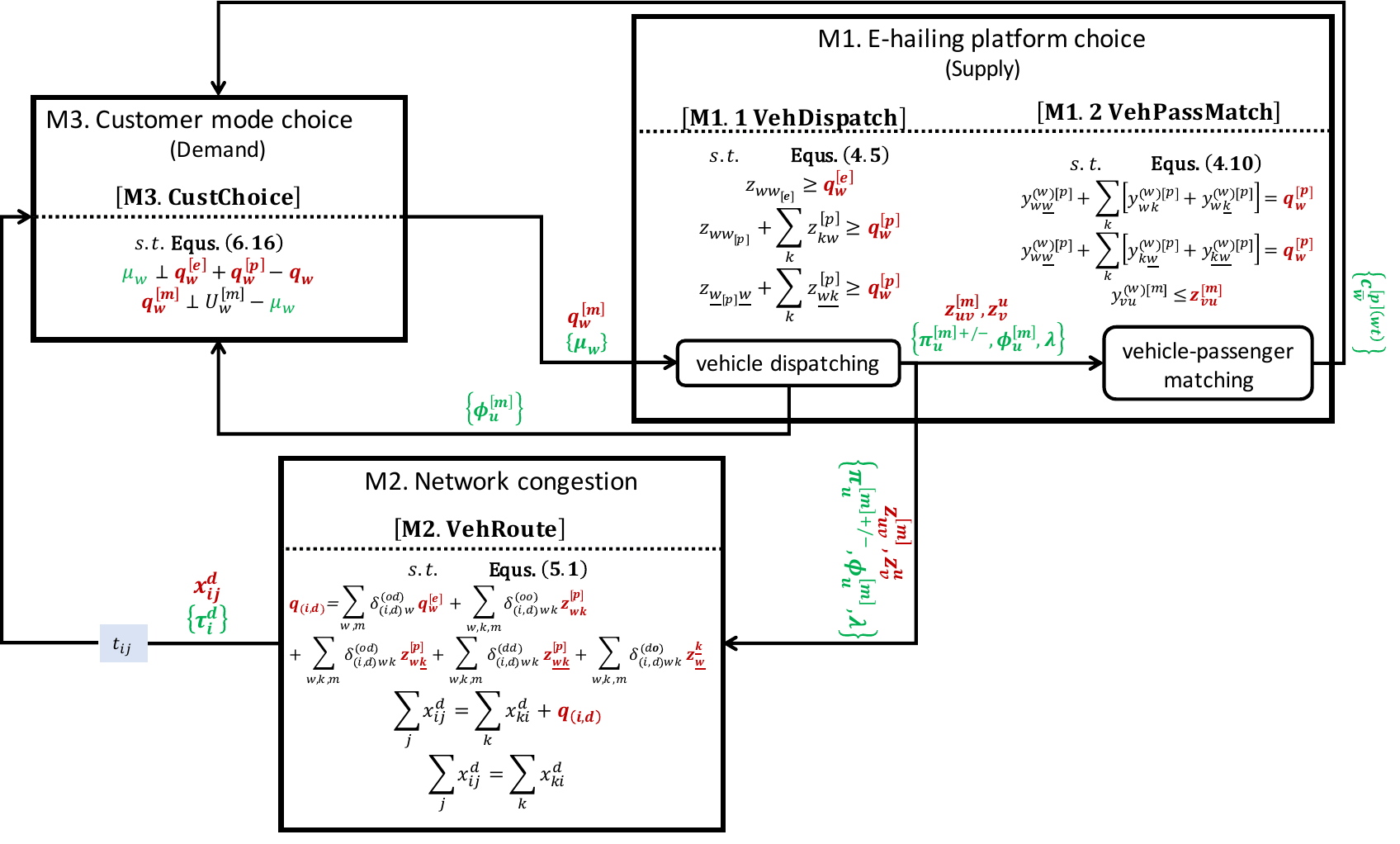}
	\caption{Total model summary [\textbf{All}.\mbox{NCP}\mbox{-ESys}]}
	\label{fig:model_sum}					
\end{figure}

Before delving into the mathematical details of each module, we first provide an overview of the unified model.
Fig.~\ref{fig:model_sum}, a mathematical expansion of Fig.~\ref{fig:3modules}, depicts all the modules and their relations, as well as variables exchange across modules. 
The layout of three modules enclosed by solid boxes and their connection by directed arrows follow the same pattern as in Fig.~\ref{fig:3modules}. 
The variables on each arrow represent variable exchange between two modules. The top variables in red are travel demand or traffic flow variables, and the bottom ones in green are auxiliary multiples.  
Within each module, we only highlight the constraints that contain variables linked to those from other modules, and leave the rest denoted by Equation indices for simplicity. 
Let us now introduce the flow within three modules. 
Given travel time and cost, (M3. Customer choice) outputs travel demands across available travel modes. 
Once receiving passengers orders, the e-hailing platform dispatches idle vehicles from destination nodes (based on (M1.1. Vehicle Dispatch)) to pick up orders and determines the sequence of e-pooling order pick-up. 
(M1.1) outputs vehicle flows, depicting which origin nodes the idle vehicles move to, what OD pairs are pooled, and in what sequence. 
The vehicle flows, along with passenger orders from (M3), are fed into (M1.2. Vehicle-Passenger Matching) to solve passenger flows by vehicle-passenger matching. 
(M1.1. Vehicle dispatch) generates traffic congestion on a road network. 
The vehicle flows from (M1.1) are sources of vehicle demands to the traffic assignment problem in (M2) and update flow-dependent travel time on each road segment, which are fed back to (M3. Customer Choice). 
Meanwhile, the way vehicles are dispatched in (M1.1. Vehicle dispatch) influences passengers' waiting time, which is also fed back to (M3. Customer Choice) for customers to account for additional waiting cost while making mode choices.

\section{Origin-destination (OD) graphs}\label{sec:od}

In this section, we will introduce an important tool developed in this paper, namely, origin-destination (OD) graphs, which is the building block for solving passenger and vehicle pooling flows. 
The description of traffic entities on roads including passenger flows and vehicle flows are detailed in \ref{append:entity}. We introduce notations in a road network: The road network is denoted by a directed graph $\mathcal{G}_{(net)}=\left\lbrace \mathcal{N}_{(net)},\mathcal{L}_{(net)} \right\rbrace$ where $\mathcal{N}_{(net)}$ is the node set and $\mathcal{L}_{(net)}$ is the link set. Origin and destination sets are denoted by ${\cal N}_O$ and ${\cal N}_D$, respectively. We have ${\cal N}_O, {\cal N}_D \subset \mathcal{N}_{(net)}$.

We now demonstrate how to construct layered OD graphs according to origins and destinations on a road network.

\subsection{Layered OD graphs}

\begin{figure}
	\centering
	\includegraphics[scale=.55]{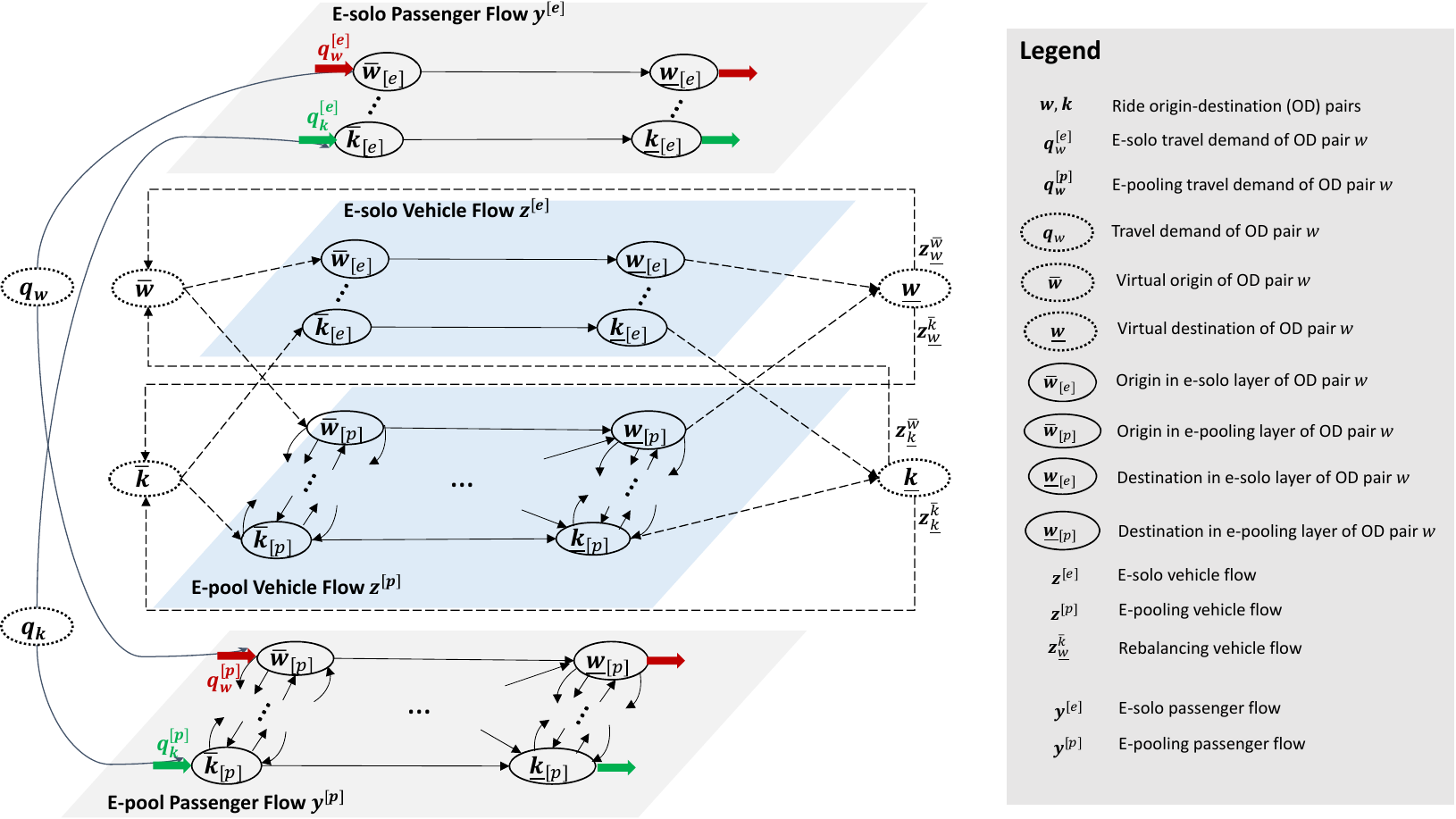}
	\caption{Layered OD graph for an e-hailing system}
	\label{fig:linknode_e}					
\end{figure}

To model the complex vehicle and passenger flows and simplify enumerating permutations for e-pooling vehicle flows, we develop a layered OD graph, demonstrated in Fig.~\ref{fig:linknode_e}. Before introducing vehicle and passenger flows in the OD graph, we first look into augmented OD pairs generated from the original node sets in the road network.

\subsubsection{Augmented OD pairs}

An OD graph, denoted as $\mathcal{G}=\left\lbrace \mathcal{N},\mathcal{L}\right\rbrace$, is constructed by origin and destination node sets ${\cal N}_O$ and ${\cal N}_D$. To distinguish nodes in an OD graph from those on the road network, in the OD graph, we use $\mathcal{N}^+(m)$ and $\mathcal{N}^-(m)$ to denote the origin and destination sets regarding travel mode $m$, respectively. With this mapping, we then have ${\cal N}^{+}(m)= {\cal N}_O, {\cal N}^{-}(m)= {\cal N}_D$. We further denote the virtual origin and destination node sets by $\mathcal{N}^{+}$ and $\mathcal{N}^{-}$, respectively. 
Summarizing the above notations, 
in a layered OD graph, we have $\mathcal{N}= \mathcal{N}^{+}(m) \bigcup \mathcal{N}^{-}(m) \bigcup \mathcal{N}^{+} \bigcup \mathcal{N}^{-}$ and $\mathcal{L}\subset\mathcal{N}\times\mathcal{N}$.

Let ${\cal W}$ denote a node pair set for e-hailing orders. 
To simplify notations, define the origin node of OD pair $w$ by adding a bar over the letter, denoted as $\bar{w}$, and the destination node of OD pair $w$ by adding an underline below the letter, denoted as $\underline{w}$. Accordingly, we can write OD pair $w$ as a pair of nodes composed of its origin $\bar{w}\in {\cal N}^+(m)$ and destination $\underline{w}\in {\cal N}^-(m)$, denoted as $(\bar{w},\underline{w}), \forall w\in {\cal W}$. The total number of orders or the aggregate OD demand between OD pair $w$ is $q_w>0$. We summarize all the possible OD pairs needed in an e-hailing system in Tab.~\ref{tab:od_aug}. 

\begin{table}[H]\centering
    \fontsize{9}{0}\selectfont
	\caption{Augmented OD pair set}\label{tab:od_aug}
	\begin{tabular}{l|l|l}
		\hline
		Mode  & Node pair & Description \\\hline\hline
		E-solo   & $(\bar{w}_{[e]}, \underline{w}_{[e]}), \forall w \in {\cal W}$ & OD pair in e-solo\\ \hline
		E-pooling  & $(\bar{w}_{[p]}, \bar{k}_{[p]}), \forall w,k \in {\cal W}, w \neq k$ & OO pair in e-pooling \\
		 & $(\bar{w}_{[p]}, \underline{k}_{[p]}), \forall w,k \in {\cal W}$ & OD pair in e-pooling \\
		 & $(\underline{w}_{[p]},\underline{k}_{[p]}), \forall w,k \in {\cal W}, w\neq k$ & DD pair in e-pooling\\ \hline
    Rebalancing  & $(\underline{w}, \bar{k}), \forall w,k \in {\cal W}$ & DO pair in rebalancing\\ \hline
	\end{tabular}
\end{table}

\begin{rem}
\begin{itemize}
    \item One edge in the OD graph $e=(u,v)\in\mathcal{L}, \forall u,v\in\mathcal{N}$ represents a path connecting two nodes (either origin or destination) on a road network, which links a sequence of nodes on a road network.
    \item We have ${\cal N}^+(m)=\{\bar{w}_{[m]}, \forall w \in {\cal W} \}$, ${\cal N}^-(m)=\{\underline{w}_{[m]}, \forall w \in {\cal W} \}, \forall m \in \{e,p\}$. ${\cal N}^+=\{\bar{w}, \forall w \in {\cal W} \}$ and ${\cal N}^-=\{\underline{w}, \forall w \in {\cal W} \}$. 
\end{itemize}
\end{rem}

The OD graph consists of four layers described below in the order from top to bottom, namely,  e-solo passenger flow, e-solo vehicle flow, e-pooling vehicle flow, and e-pooling passenger flow. An e-hailing platform solves optimal vehicle flow dispatching based on two vehicle flow layers, namely, solo and pooling vehicle flows; while driver-order matching is solved on two passenger flow layers, namely, solo and pooling passenger flows. Passenger orders requesting solo $q^{[e]}_{w}$ or pooling $q^{[p]}_{w}, \forall w\in {\cal W}$ service induce vehicle movement and cause traffic congestion. Thus, the coupling between one vehicle flow layer and its corresponding passenger flow layer is reflected in demand-supply constraints. For solo passenger orders, the demand-supply constraint is quite straightforward, which is the vehicle flow on one OD path needs to be no less than the passenger flow on that path. However, for pooling passenger orders, the demand-supply constraint is not straightforward to formulate, which requires to solve passenger-vehicle matching to be detailed in Sec.~\ref{subsec:match}. The algorithm to generate a layered OD graph is summarized in Algorithm \ref{alg:OD_generation_1}.

\begin{algorithm}[H]
	\caption{Layered OD graph generation}
	\label{alg:OD_generation_1}
	\begin{algorithmic}[1]
	\Require{The node pair set for e-hailing orders $\mathcal{W}$. Each $w \in \mathcal{W}$ represents an OD pair in the road network.} 
    \State{\textbf{Step 1}. Generate the vehicle flow subgraph for e-pooling service.} 
    
    \ \ \ \ \ Nodes: Origins  $\bar{w}_{[p]}$ and destinations $\underline{w}_{[p]},\forall w \in \mathcal{W}$.

    \ \ \ \ \ Edges: Origins are fully connected by edges $(\bar{w}_{[p]},\bar{k}_{[p]}), \forall w,k \in \mathcal{W}, w \neq k$.  
    
    \ \ \ \ \ \ \ \ \ \ \ \ \ \ \ \ \ Destinations are fully connected by edges  $(\underline{w}_{[p]},\underline{k}_{[p]}), \forall w,k \in \mathcal{W}, w \neq k$. 
    
    \ \ \ \ \ \ \ \ \ \ \ \ \ \ \ \ \ Origins and destinations are connected by edges $(\bar{w}_{[p]},\underline{k}_{[p]}), \forall w,k \in \mathcal{W}$.

    \State{\textbf{Step 2}. Generate the vehicle flow subgraph for e-solo service.}
    
    \ \ \ \ \ Nodes: Origins $\bar{w}_{[e]}$ and destinations $\underline{w}_{[e]},\forall w \in \mathcal{W}$.

    \ \ \ \ \ Edges: Origins and destinations are connected by edges $(\bar{w}_{[e]},\underline{w}_{[e]}), \forall w \in \mathcal{W}$.

    \State{\textbf{Step 3}. Link the e-pooling and e-solo vehicle flow subgraphs.}
    
     \ \ \ \ \ Nodes: Virtual origins $\bar{w}$ and virtual destinations $\underline{w}, \forall w \in \mathcal{W}$. 
     
     \ \ \ \ \ \ \ \ \ \ \ \ \ \ \ \ \ Origins $\bar{w}_{[m]}$ and destinations $\underline{w}_{[m]}, \forall w \in \mathcal{W}, m \in \{ e,p\}$. 
     
     \ \ \ \ \ Edges: Virtual origins and origins are connected by edges
      $(\bar{w},\bar{w}_{[m]}), \forall w \in \mathcal{W}, m \in \{ e,p\}$. 
      
      \ \ \ \ \ \ \ \ \ \ \ \ \ \ \ \ \ Destinations and virtual destinations are connected by edges
      $(\underline{w}_{[m]}, \underline{w}), \forall w \in \mathcal{W}, m \in \{ e,p\}$.

    \State{\textbf{Step 4}. Generate the vehicle subgraph for rebalancing.}
    
    \ \ \ \ \ Nodes: Virtual origins $\bar{w}$ and virtual destinations $\underline{w}, \forall w \in \mathcal{W}$.
    
    \ \ \ \ \ Edges: Virtual destinations and virtual origins are connected by edges $(\underline{w}, \bar{k}), \forall w,k \in \mathcal{W}$.
    
	\end{algorithmic}
\end{algorithm}

Based on the layered OD graph generated by Algorithm \ref{alg:OD_generation_1}, we then discuss vehicle and passenger flows sequentially.

\subsubsection{Vehicle flow on vehicle OD graph}
 
We first describe how e-haling vehicles flow within the OD graph. 
To ensure that the total number of vehicles is conserved according to Assumption (A3), e-hailing vehicles move within a closed, integrated e-solo and e-pooling vehicle OD graphs, denoted as ${\cal G}^{[e]}$, and ${\cal G}^{[p]}$, respectively. 
Let us start with vehicles waiting at an arbitrary drop-off node $\underline{k}$ after dropping off passengers of OD pair $k\in {\cal W}$. 
For vehicles that are matched to e-solo orders of OD pair $w\in {\cal W}, w\neq k$, these cars need to first rebalance themselves to the pick-up node $\bar{w}$ and then enter the e-solo vehicle OD graph ${\cal G}^{[e]}$ where they pick up e-solo passengers and move directly to drop-off nodes. 
For vehicles that are matched to e-pooling orders of OD pair $w\in {\cal W}, w\neq k$, these cars also need to rebalance themselves and then enter the e-pooling vehicle OD graph ${\cal G}^{[p]}$, where these cars need to visit a sequence of two pick-up nodes before moving on to drop-off nodes. 
After dropping off all passengers, vehicles move on to virtual drop-off nodes and wait to be matched again.  

Note that two vehicle OD graphs ${\cal G}^{[e]}$ and ${\cal G}^{[p]}$ are coupled both at virtual pick-up nodes and virtual drop-off nodes. 
In other words, two vehicle OD graphs form a closed system and vehicles flow through these two layers without exiting the graph.




\subsubsection{Passenger flow on passenger OD graph}

First, let us focus on how passengers of an arbitrary OD pair $w\in {\cal W}$ flow within the OD graph. Prior-trip, passengers determine to use e-solo or e-pooling service. Those who select e-solo service enter the e-solo OD graph ${\cal G}^{[\underline{e}]}$, move directly to the destination node, and then leave the graph. Those who select e-pooling service move to the e-pooling passenger OD graph ${\cal G}^{[\underline{p}]}$, where one passenger is the first or the second picked up by an e-haling vehicle and then move to drop-off nodes in sequence. After being dropped off, passengers leave the graph. 
Moreover, two passenger OD graphs ${\cal G}^{[\underline{e}]}$ and ${\cal G}^{[\underline{p}]}$ are coupled only at virtual pick-up nodes and left open at drop-off nodes.  

\subsubsection{Typical vehicle and passenger flows on OD graph}
\begin{figure}[H]
	\centering
	\includegraphics[scale=.6]{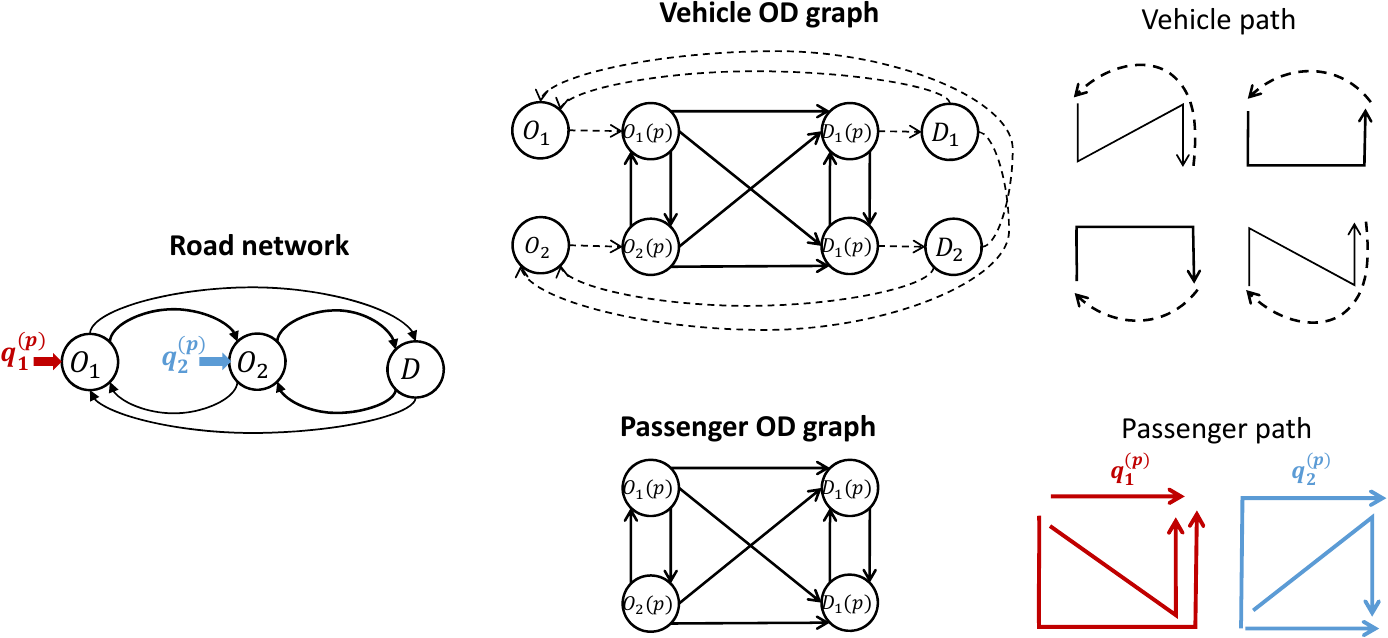}
	\caption{OD graph and path demonstration on a 3-node network with 2-OD pair}
	\label{fig:3_node_OD_exp}
\end{figure}

To further demonstrate how vehicles and passengers move over their respective OD graphs, we develop a simplified example on a 3-node network with 2-OD pair in Fig.~\ref{fig:3_node_OD_exp}, with only e-pooling demands and vehicles. The original road network, vehicle and passenger OD graphs, and feasible vehicle and passengers paths are indicated. In Fig.~\ref{fig:3_node_OD_exp}, we will demonstrate a typical e-pooling vehicle trip and a typical e-pooling passenger

A typical e-pooling vehicle trip is an enclosed circulation path. It starts from a destination where an empty car first repositions to an origin, after dropping off a past passenger. At the origin, it picks up a passenger and moves to the next origin. After carrying 2 groups of passengers from 2 origins, it moves to their respective destinations in sequence. After dropping off these passengers, it restarts the cycle, namely, repositioning, picking up passenger 1, picking up passenger 2, dropping off them one by one.

A typical e-pooling passenger (let us call it Passenger 1) can experience a trip of $O_1-D_1$, $O_1-O_2-D_1$, or $O_1-O_2-D_2-D_1$. The same is applied to Passenger 2 due to symmetry.
\begin{enumerate}
    \item $O_1-D_1$: When Passenger 1 is picked up, another passenger from a different origin is already on board. After Passenger 1 is picked up, she is the first one to be dropped off.
    \item $O_1-O_2-D_1$: When Passenger 1 is picked up, another passenger from a different origin is already on board. However, Passenger 1 is dropped off after the other passenger.
    \item $O_1-O_2-D_2-D_1$: Passenger 1 is the first to be picked up and the last to be dropped off.
\end{enumerate}

In summary, with vehicle and passenger OD graphs defined, solving vehicle and passenger flows is equivalent to solving network flow problems over corresponding OD graphs. That is why properly defining OD graphs for each entity is crucial. 
In the next section, we will elaborate on how to model our problems using the network flow formulation. 
Vehicle and passenger OD graphs will be revisited. 

\section{Module 1: E-haling platform choice on OD graphs}
\label{sec:1_OD}

Assuming that customers select their respective mode choices and place orders, the e-platform needs to first dispatch idle vehicles and then match them to orders. As mentioned before, Module 1 can be essentially transformed into variants of minimum-cost flow problems, when the customer choice is fixed and when travel cost is not flow-dependent. 
However, constructing Module 1 directly using network flow problems is non-trivial and can be tricky, because of the need for properly represented OD graphs (to be introduced in Secs.~\ref{subsubsec:vehOD},\ref{subsubsec:pasOD}), 
as well as the e-pooling constraints (to be introduced in Secs.~\ref{subsec:constr_M1.1},\ref{subsec:const_pass}).  
In the next two subsections, we will specify how to compute edge cost and the objective function, as well as how to formulate flow conservation and other constraints. 
\begin{rem}
\begin{enumerate}
    \item 
    In principle, e-hailing vehicle dispatching and vehicle-passenger matching are combinatorial problems. 
    The major contribution of this paper is that, these two problems can be reformulated as network flow problems, if isolated from other modules, in other words, when the modal demands and travel cost are given.
    
    \item To establish such an equivalence, the challenges lie in the construction and representation of a graph that encodes flow conservation for e-pooling vehicle and passenger flows. 
    We need an appropriate graph representation and carefully defined flow conservation constraints. 

   \item The overall unified model is not minimum-cost flow model for two reasons: (1) the elastic demand is subject to a customer choice model, and (2) edge cost is congestion-dependent, thus, depends on a network congestion model. 
\end{enumerate}
\end{rem}

\subsection{Module 1.1: Vehicle dispatching}
 
Finding an optimal vehicle dispatching plan is essentially equivalent to solving a minimum-cost circulation network flow problem. Analogously, an e-platform e-hailing vehicles circulate in a network continuously between origins and destinations. 
When vehicles move from origins to destinations, they are occupied by 1 or 2 passengers. 
When vehicles move from destinations to origins, which is the rebalancing phase, these vehicles are vacant and need to reposition to match new orders. 
In summary, e-hailing vehicle dispatching is a \emph{minimum-cost circulation network flow problem} (MCCNFP) (introduced in \ref{append:mccnfp}), provided with riders' requests and travel cost as input. 
 
Module 1.1 solves the optimal dispatch of idle vehicles (from destination to origin nodes) on the basis of vehicle OD graphs. 
Subsequently, we will introduce the graph, specify edge cost that leads to an objective function, and describe flow constraints, which ultimately complete the picture of the formulation of MCCNFP.  

\subsubsection{Vehicle OD graph}\label{subsubsec:vehOD}

Vehicle dispatching is essentially a minimum cost flow circulation problem over a vehicle OD graph, when edge cost is given. 
To formulate the vehicle dispatching problem as a circulation problem, we need to inspect the coupled e-solo and e-pooling vehicle OD subgraphs altogether, rather than on individual modal subgraph. 



 \subsubsection{Cost specification}
 \label{subsec:cost_M1.1}
 
Traversing edges incur travel time based cost. Profit is only earned when an order is dropped of at its destination $\underline{w}, \forall w\in {\cal W}$. Thus, cost on those edges that connect a destination node is the negative net profit, which is travel cost minus the fare. Below we define the net cost on each edge in the vehicle OD graph. 

\begin{subequations}\label{eq:opt_edge_cost}
	\begin{align}
	& \mbox{OD edge in e-solo:} & C^{[e]}_{\bar{w}\underline{w}} = -r^{[e]}_{w} + \beta_{(in-veh)} t_{\bar{w}\underline{w}} +c^{[e](se-veh)}_{\Bar{w}}, \forall w\in {\cal W},
	\label{subeq:c_od_s}\\
	& \mbox{OO edge in e-pooling:} & C^{[p]}_{\bar{w}\bar{k}} =  \beta_{(in-veh)} t_{\bar{w}\bar{k}}+ c^{[p](se-veh)}_{\Bar{w}}, \forall w,k\in {\cal W}, w\neq k, \label{subeq:c_o1o2_r}\\
	& \mbox{OD edge in e-pooling:} & C^{[p]}_{\bar{w}\underline{k}} = -r^{[p]}_{w} + \beta_{(in-veh)} t_{\bar{w}\underline{k}}+c^{[p](se-veh)}_{\Bar{w}}, \forall w,k\in {\cal W}, 
	\label{subeq:c_od_p}\\
	& \mbox{DD edge in e-pooling:} & C^{[p]}_{\underline{w}\underline{k}} = -r^{[p]}_{w} + \beta_{(in-veh)} t_{\underline{w}\underline{k}},
	\forall w,k\in {\cal W}, w\neq k. \label{subeq:c_d_r}\\
	& \mbox{Virtual source edge:} & C_{\bar{w} \bar{w}_{[p]}} = 0, \forall w\in {\cal W}, \label{subeq:c_o'o}\\
	& \mbox{Virtual sink edge:} & C_{\underline{w}_{[p]} \underline{w}} = 0, \forall w\in {\cal W},\label{subeq:c_dd'}\\
	& \mbox{Rebalancing edge:} & C^{\bar{w}}_{\underline{k}} =  \beta_{(in-veh)} t_{\underline{k}\bar{w}}, \forall w,k\in {\cal W}, \label{subeq:c_do}\\
	\end{align}
	\label{eq:edgecost_od}
\end{subequations}
where, $\beta_{(in-veh)}$ is a coefficient converting travel time to the value of time. 
$r^{[e]}_{w}, r^{[p]}_{w}$ are fares of using e-solo or e-pooling service for OD pair $w$, which will be defined subsequently. $c^{[e](se-veh)}_{\Bar{w}}, c^{[p](se-veh)}_{\Bar{w}}$ denote search friction of drivers at origins to be detailed in Sec.~\ref{subsec:nodal}.


In a ride-sourcing market, 
one widely used assumption about cost-sharing protocols between riders is that, trip fare should be proportional to the actual travel time of a single trip \citep{wang2018RS,chen2020dynamic}. In addition, e-pooling services usually can get a discount \citep{ke2020pricing}. Some studies use travel distances between users' origins and destinations to decide trip fares for each rider \citep{enzi2021ModelingAS}. Game theory has also been applied to split the cost among travellers:  Shapley values, representing riders' trip fares,  can be calculated based on vehicles'  travel distance and riders' priorities \citep{levinger2020shapley}. Auction-based prices offered by riders is another tool to split cost in on-demand systems \citep{bian2020mecha}. In our work, we assume that trip fare is proportional to travel time and travel distance and the fare for e-pooling is discounted by a multiplier.
Define the base fare for orders of OD pair $w\in {\cal W}$ as $R_{w}$, which consists of a fixed fare, time-based and distanced-based rates, 
\begin{equation}\label{eq:R_jw}
	R_{w} 
	= F^{}_{w} + \underbrace{\alpha_1(t_{\bar{w}\underline{w}} 
	- \hat{t}_{\bar{w}\underline{w}})}_{time-based} + \underbrace{\alpha_2 l_{\bar{w}\underline{w}}}_{distance-based},
\end{equation}
where, $\alpha_1$ and $\alpha_2$ are coefficients of time-based and distance-based fares, respectively. $t_{\bar{w}\underline{w}}$ is the travel time, $\hat{t}_{\bar{w}\underline{w}}$ is the free-flow travel time and $l_{\bar{w}\underline{w}}$ is the distance between node $\Bar{w}$ and $\underline{w}$. Assume that the fare for e-pooling is discounted by a multiplier, $\gamma_p\in \left(0,1\right)$.  
The fares for e-solo and e-pooling are computed as below:
\begin{subequations}\label{eq:opt_fare_discount}
	\begin{align}
& r^{[e]}_{w} = R_{w},\\
& r^{[p]}_{w} = \gamma_p R_{w}
	\end{align}
\end{subequations}

\begin{rem}
    We use disutility or cost instead of profit or utility. So the earnings are regarded as negative cost. 
\end{rem}

\subsubsection{Objective function}
 \label{subsec:obj_M1.1}
The objective function consists of two components:
the transport cost for idle vehicle dispatching (computed as the product of travel cost on rebalancing edges and the rebalancing flow),
and the total negative net profit earned by an e-hailing platform (computed as the product of a single ride's negative net profit and the occupied vehicle flow).
Mathematically,

\begin{subequations}\label{eq:opt_z2}
	\begin{align}	
	& \mbox{Idle vehicle dispatch:} & \sum_{w,k \in {\cal W}} 
	C_{\underline{k}}^{\bar{w}} z_{\underline{k}}^{\bar{w}}, \label{subeq:obj_transp}\\
	& \mbox{Passenger-carrying:} & \sum_{m\in {\cal M}}\sum_{u,v\in {\cal N}(m)} 
	C^{[m]}_{vu} z^{[m]}_{vu}, \label{subeq:obj_pick}
	\end{align}
\end{subequations}
\begin{figure}
	\centering
	\includegraphics[scale=.6]{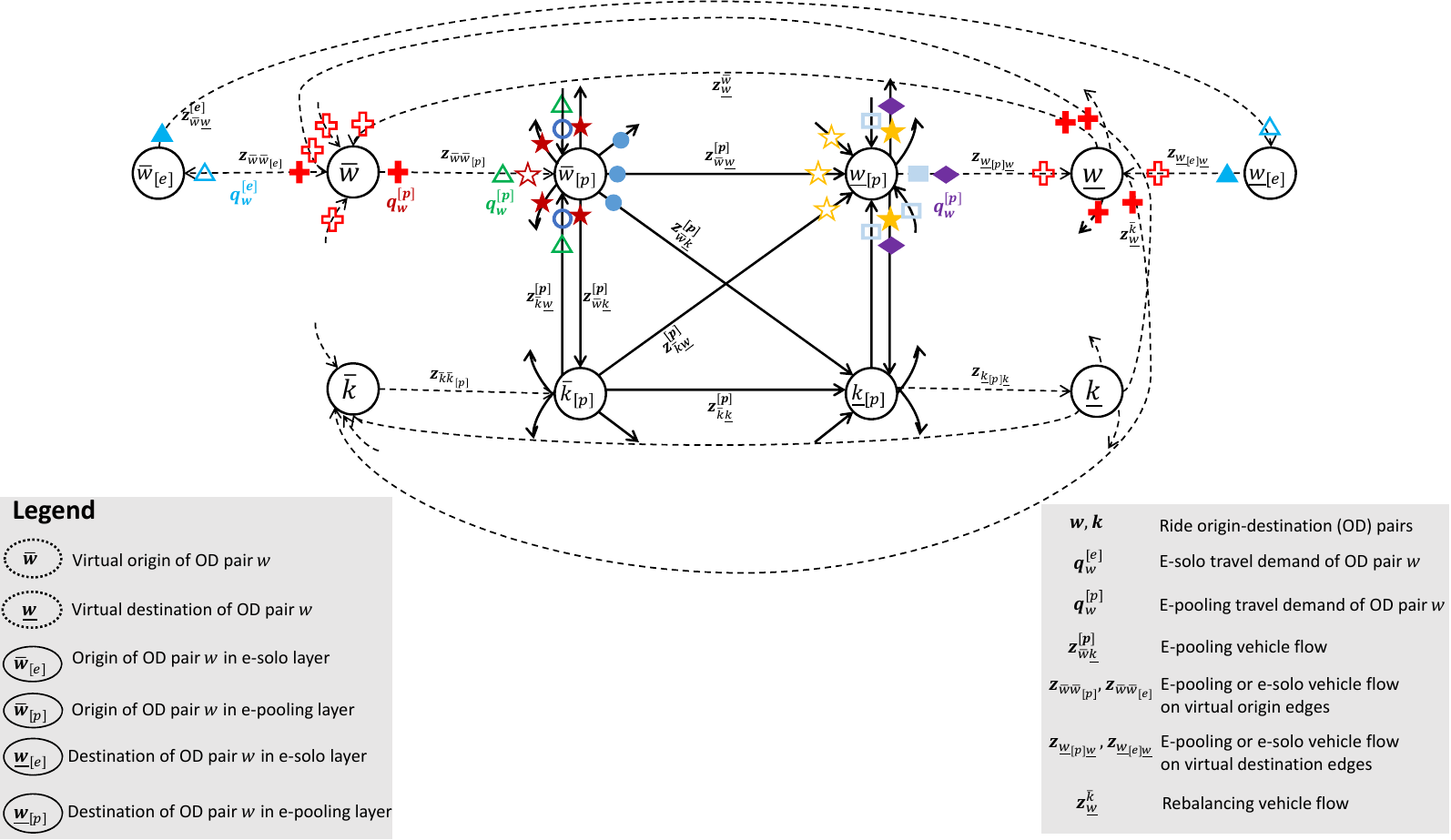}
	\caption{Constraints illustration on an OD vehicle graph} 
	\label{fig:ODnet_veh}					
\end{figure}

\subsubsection{Vehicle flow constraints}
  \label{subsec:constr_M1.1}
 
Fig.~\ref{fig:ODnet_veh} illustrates the constraints associated with vehicle flows on a vehicle OD graph. 
In this figure, we indicate links involved in each constraint with a specific symbol. 
The symbol type associated with each equation is indicated before each equation. 
When defining constraints, we will use $w$ as the primary OD pair index, and $k$ as the auxiliary OD pair index. 
As aforementioned, $\bar{w},\bar{k}\in {\cal N}^{+}(m)$ represent the origin nodes of OD $w,k$, respectively, and $\underline{w},\underline{k}\in {\cal N}^{-}(m)$ represent the destination nodes, respectively.
All the involved edges are marked with the same symbols.  To separate in- and out-edge, the in-edge is marked with a shallow symbol, while the out-edge with a solid symbol.

\paragraph{Flow conservation at origin from virtual origin edge to pick-up edges for e-pooling}

To ensure that e-pooling vehicles visit two origins, the e-pooling vehicle flow entering an origin node, $\bar{w}\in {\cal N}^{+}(p)$, can only be allowed to visit from $\bar{w}$ a second origin, i.e., $\forall\bar{k}\neq \bar{w}\in {\cal N}^{+}(p)$, before they head to a destination node. 
In other words, vehicles coming from virtual edges need to visit two origin nodes and are not allowed to directly go to drop-off edges.

\begin{eqnarray*}
    & (\textcolor{red}{\bigstar}) 
    & z_{\bar{w}\bar{w}_{[p]}} = \sum_{k\in {\cal W}, k\neq w} z^{[p]}_{\bar{w}\bar{k}}, 
	\forall w \in {\cal W}, \label{subeq:conserv_viro_o}
\end{eqnarray*}
    

\paragraph{Flow conservation at origin from pick-up edges to drop-off edges for e-pooling}  

Let us focus on the e-pooling vehicles that visit the origins of other OD pairs, i.e., $\forall \bar{k}\in {\cal N}^{+}(p), k\neq w$ prior to $\bar{w}\in {\cal N}^{+}(p)$.
After these vehicles visit origin node $\bar{w}$, they would need to visit the destinations of other ODs than $w$, i.e., $\forall \underline{k}\in {\cal N}^{-}(p), k\neq w$, as well as the destination $\underline{w}\in {\cal N}^{-}(p)$, regardless of the visiting sequence of destination nodes.

\begin{eqnarray*}
    & (\textcolor{blue}{\bigcirc})  & \sum_{k\in {\cal W}, k\neq w} z^{[p]}_{\bar{k}\bar{w}} 
    = \sum_{k\in {\cal W}, k\neq w}  z^{[p]}_{\bar{w}\underline{k}}
    + z^{[p]}_{\bar{w}\underline{w}}, 
	\forall w \in {\cal W}, \label{subeq:conserv_om_out}
\end{eqnarray*}    

\paragraph{Flow conservation at destination associated with drop-off edges for e-pooling}

All the e-pooling vehicle flows originating from origins into the destination of the OD pair $\underline{w}\in {\cal N}^{-}(p)$ need to visit the corresponding destinations of these OD pairs, i.e., $\forall \underline{k}\in {\cal N}^{-}(p), k\neq w$.

\begin{eqnarray*}
& (\textcolor{yellow}{\bigstar}) 
& z^{[p]}_{\bar{w}\underline{w}} 
	+ \sum_{k\in {\cal W},k\neq w} z^{[p]}_{\bar{k}\underline{w}} = 
    \sum_{k\in {\cal W},k\neq w}
    z^{[p]}_{\underline{w}\underline{k}},
	\forall w \in {\cal W},
\label{subeq:conserv_dm}
\end{eqnarray*} 

\paragraph{Flow conservation at destination from drop-off edges to virtual edge for e-pooling}
All the e-pooling vehicle flows originating from other destinations into the destination of OD pair $\underline{w}_{[p]}\in {\cal N}^{-}(p)$ end at its associated virtual destination node $\underline{w}\in {\cal N}^{-}$. 
In other words, when the e-pooling vehicle flows already visit one destination before visiting the destination of OD pair $\underline{w}_{[p]}\in {\cal N}^{-}(p)$, it means that $\underline{w}$ is the second destination to visit. According to our assumption A1 that only two OD pairs are allowed to pool, these vehicles should not visit a third destination and thus $\underline{w}$ is their final destination.

\begin{eqnarray*}
& (\hrectangle) 
& \sum_{k\neq w} 
    z^{[p]}_{\underline{k}\underline{w}}
	= z_{\underline{w}_{[p]}\underline{w}},
	\forall w\in {\cal W}, 
\label{subeq:conserv_dm_out}
\end{eqnarray*} 

\paragraph{Demand conservation at origin or destination for e-pooling.} The total e-pooling vehicle supply originating from the virtual origin of OD pair $\bar{w}\in {\cal N}^{+}$ needs to be no less than the total e-pooling demand for OD pair $\forall w\in {\cal W}$. This is the supply demand equilibrium condition described in Sec.~\ref{subsec:game}.
\begin{eqnarray*}
& (\triangle)
& z_{\bar{w}\bar{w}_{[p]}} 
	+ \sum_{k\in {\cal W},k\neq w} z^{[p]}_{\bar{k}\bar{w}}
	\geqslant q^{[p]}_{w},
	\forall w\in {\cal W}, 
	\label{subeq:supp_dem_Ow}
\end{eqnarray*}

The total e-pooling vehicle supply ending at the destination of the OD pair $\underline{w}\in {\cal N}^{-}(p)$ needs to be no less than the total e-pooling demand for the OD pair $\forall w\in {\cal W}$. 

\begin{eqnarray*}
& (\mdwhtdiamond)
& z_{\underline{w}_{[p]}\underline{w}} 
	+ \sum_{k\in {\cal W},k \neq w} z^{[p]}_{\underline{w}\underline{k}} 
	\geqslant q^{[p]}_{w}, 
	\forall w\in {\cal W}, 
	\label{subeq:supp_dem_r_Dw}
\end{eqnarray*}



\paragraph{Flow conservation at origin or destination for e-solo}
Demand conservation conditions for e-solo are relatively simple. 
Over the e-solo layer, at origin $\forall\bar{w}_{[e]}\in {\cal N}^{+}(e)$ of OD pair $w\in {\cal W}$, the inflow from the virtual origin $\bar{w}\in {\cal N}^{+}$ equals the outflow to the associated destination $\underline{w}_{[e]}\in {\cal N}^{-}(e)$.  
Similarly, at destination $\forall\underline{w}_{[e]}\in {\cal N}^{-}(e)$ of OD pair $w\in {\cal W}$, the inflow from its associated origin $\bar{w}_{[e]}\in {\cal N}^{+}(e)$ equals the outflow to the virtual destination $\underline{w}\in {\cal N}^{-}$.
\begin{eqnarray*}
& (\tco{\triangle})
	& z_{\bar{w} \bar{w}_{[e]}} = z^{[e]}_{\bar{w}\underline{w}},
	\forall w\in {\cal W},
	\label{subeq:conserv_o_e}\\
& (\tco{\triangle})
	& z^{[e]}_{\bar{w}\underline{w}} = z_{\underline{w}_{[e]}\underline{w}},
	\forall w\in {\cal W}, 
	\label{subeq:conserv_d_e}
\end{eqnarray*}

\paragraph{Demand conservation at origin for e-solo}
Over the e-solo layer, the inflow into the origin node $\forall\bar{w}_{[e]}\in {\cal N}^{+}(e)$ of OD pair $w\in {\cal W}$ needs to be no less than the total e-solo demand for OD pair $w\in {\cal W}$. 
\begin{eqnarray*}
& (\tco{\triangle})
	& z_{\bar{w}\bar{w}_{[e]}}
	\geqslant q^{[e]}_{w}, 
	\forall w\in {\cal W}, 
	\label{subeq:supp_dem_e}
\end{eqnarray*}

\paragraph{Demand conservation at virtual origin or destination}
At the virtual origin node $\forall\bar{w}\in {\cal N}$ of OD pair $w\in {\cal W}$, the inflow from all the destination nodes (for rebalancing) needs to be equal to the outflows to the corresponding origins across all travel modes. 
In other words, all the unoccupied vehicles that are rebalanced to origin $\bar{w}\in {\cal N}$ of OD pair $w\in {\cal W}$ need to pick up passengers of the same OD pair who select various travel modes. 
\begin{eqnarray*}
& (\tcr{+})
& \sum_{k\in {\cal W}} z^{\bar{w}}_{\underline{k}}
	=\sum_{m\in {\cal M}} z_{\bar{w}\bar{w}_{[m]}},
	\forall w \in {\cal W}, 
	\label{subeq:conserv_o}
\end{eqnarray*}

Similarly, at a virtual destination node $\underline{w}\in {\cal N}$, the inflow from its associated destination node across travel modes needs to be equal to the outflows to all virtual origins. 
\begin{eqnarray*}
& (\tcr{+})
& \sum_{m\in {\cal M}} z_{\underline{w}_{[m]}\underline{w}}
	=\sum_{k\in {\cal W}} z^{\bar{k}}_{\underline{w}},  
	\forall w \in {\cal W}, 
	\label{subeq:conserv_d}
\end{eqnarray*}

\paragraph{Total flow conservation on virtual links}
Vehicle flows on virtual origin links flowing into all origin nodes across all travel modes need to be equal to those on virtual destination links flowing out of all destination nodes across all travel modes.
\begin{eqnarray*}
& (\tcr{+})
    & \sum_{m\in {\cal M}} \sum_{w\in {\cal W}} z_{\bar{w}\bar{w}_{[m]}} 
	=\sum_{m\in {\cal M}} \sum_{w\in {\cal W}} z_{\underline{w}_{[m]}\underline{w}},
	\label{subeq:conserv_fleet}
\end{eqnarray*}

\paragraph{Fleet Sizing}
The total fleet hours, including both rebalancing and occupied fleets, should be no more than a predefined total fleet size.
\begin{eqnarray*}
	& \sum_{k,w \in {\cal W}} 
	z^{\bar{w}}_{\underline{k}} t_{\underline{k}\bar{w}}
	+ \sum_{m\in {\cal M}}\sum_{(v,u)\in {\cal L}(m)} z^{[m]}_{vu}t_{vu}
	\leqslant S, 
	\label{subeq:fleet}
\end{eqnarray*}

\subsubsection{Nonlinear complementarity formulation}

We reformulate vehicle dispatching as a nonlinear complementarity problem (NCP). We introduce $\perp$ as the orthogonal sign representing the inner product of two vectors. Mathematically




\begin{subequations}\label{subeq:NCP_M1.1}
\fontsize{9}{0}\selectfont
\begin{align}
	& [\textbf{M1.1}.\mbox{NCP}\mbox{-VehDispatch}] \nonumber\\
& \mbox{\textbf{Profit optimization for e-pooling}:} \nonumber\\
	& 0 \leqslant z^{\bar{w}}_{\underline{k}} \perp C^{\bar{w}}_{\underline{k}} 
	-\pi_{\bar{w}} +\pi_{\underline{k}} 
	+ \lambda_{fleet} 
	\geqslant 0, \forall w \in {\cal W}, \label{subeq:NCP_profit_vu}\\
	& 0 \leqslant z^{[p]}_{\bar{w} \bar{k}}
	\perp 
	 C^{[p]}_{\bar{w} \bar{k}} 
	+\pi^{[p]+}_{\bar{w}} -\pi^{[p]-}_{\bar{k}}
	- \phi^{[p]}_{\bar{k}} 
	- \lambda^{[p]}_{\bar{w}\bar{k}}
	+ \lambda_{fleet}
	\geqslant 0, \forall w,k\in {\cal W},k\neq w, \\
	& 0 \leqslant z^{[p]}_{\bar{k}\underline{w}} 
	\perp C^{[p]}_{\bar{k}\underline{w}} 
	+\pi^{[p]-}_{\bar{k}}
	-\pi^{[p]+}_{\underline{w}}
    -\lambda^{[p]}_{\bar{k}\underline{w}}
	+\lambda_{fleet} \geqslant 0, \forall w,k\in {\cal W}, k\neq w,\\
	& 0 \leqslant z^{[p]}_{\bar{w}\underline{w}} 
	\perp C^{[p]}_{\bar{w}\underline{w}} 
	+\pi^{[p]-}_{\bar{w}}
	-\pi^{[p]+}_{\underline{w}}
	-\lambda^{[p]}_{\bar{w}\underline{w}}
	+\lambda_{fleet} \geqslant 0, \forall w\in {\cal W},\\
	& 0 \leqslant z^{[p]}_{\underline{w}\underline{k}} 
	\perp C^{[p]}_{\underline{w}\underline{k}}  
	+\pi^{[p]+}_{\underline{w}}
	-\lambda^{[p]}_{\underline{w}\underline{k}}
	-\pi^{[p]-}_{\underline{k}}
	- \phi^{[p]}_{\underline{w}} 
	+\lambda_{fleet} \geqslant 0, \forall w,k\in {\cal W}, k\neq w, \label{subeq:NCP_profit_wv}\\
    & 0\leqslant z_{\Bar{w}\Bar{w}_{[p]}} \perp 
    C_{\Bar{w}\Bar{w}_{[p]}} 
    - \pi^{[p]+}_{\Bar{w}} 
    - \phi^{[p]}_{\Bar{w}} 
    - \pi_{\Bar{w}}
    -\lambda
    \geqslant 0, \forall w\in {\cal W}, 
	\label{subeq:NCP_x_1>0}\\
	& 0\leqslant z_{\underline{w}_{[p]}\underline{w}} \perp 
	C_{\underline{w}_{[p]}\underline{w}} 
	+ \pi^{[p]-}_{\underline{w}} 
	- \phi^{[p]}_{\underline{w}} 
	+ \pi_{\underline{w}} 
	+\lambda
	\geqslant 0, \forall w\in {\cal W}, 
	\label{subeq:NCP_x_2>0}	\\
    & \mbox{\textbf{Profit optimization for e-solo:}} \nonumber\\
	& 0 \leqslant z^{[e]}_{\bar{w}\underline{w}} 
	\perp C^{[e]}_{\bar{w}\underline{w}} 
	+ \pi^{[e]}_{\bar{w}} - \pi^{[e]}_{\underline{w}}
	-\lambda^{[e]}_{\bar{w}\underline{w}}
	+ \lambda_{fleet}
	\geqslant 0, \forall w \in {\cal W}, \\
	& 0\leqslant z_{\Bar{w}\Bar{w}_{[e]}} \perp 
    C_{\Bar{w}\Bar{w}_{[e]}} - \pi^{[e]}_{\Bar{w}} + \pi_{\Bar{w}}
    - \phi^{[e]}_{w} 
    -\lambda 
    \geqslant 0, \forall w\in {\cal W}, 
	\label{subeq:NCP_dummy_o_e}\\
	& 0\leqslant z_{\underline{w}_{[e]}\underline{w}} \perp 
	C_{\underline{w}_{[e]}\underline{w}} + \pi^{[e]}_{\underline{w}} - \pi_{\underline{w}} 
	+\lambda 
	\geqslant 0, \forall w\in {\cal W}, 
	\label{subeq:NCP_dummy_d_e}\\
	& \mbox{\textbf{Flow conservation at origins for e-pooling:}} \nonumber\\
	& 0=z_{\Bar{w}\Bar{w}_{[p]}} - \sum_{k\in {\cal W},k\neq w} z^{[p]}_{\Bar{w}\Bar{k}} \perp \pi^{[p]+}_{\Bar{w}} \mbox{ free}, 
	\forall w\in {\cal W}, 
	\label{subeq:NCP_conserv_om}\\
	& 0=\sum_{k\in {\cal W},k\neq w} z^{[p]}_{\Bar{k}\bar{w}} - \sum_{k\in {\cal W}, k \neq w} z^{[p]}_{\Bar{w}\underline{k}}-z^{[p]}_{\Bar{w}\underline{w}}
	\perp \pi^{[p]-}_{\Bar{w}} \mbox{ free}, 
	\forall w\in {\cal W}, 
	\label{subeq:NCP_conserv_om_out}\\
	& \mbox{\textbf{Flow conservation at destination for e-pooling:}} \nonumber\\
    & 0 = z^{[p]}_{\bar{w}\underline{w}} 
	+ \sum_{k \in {\cal W}, k\neq w} z^{[p]}_{\bar{k}\underline{w}} - 
    \sum_{k \in {\cal W}, k\neq w}
    z^{[p]}_{\underline{w}\underline{k}}
    \perp \pi^{[p]+}_{\underline{w}} \mbox{ free},
	\forall  w\in {\cal W},
	\label{subeq:NCP_conserv_dm}\\
	& 
	0 = \sum_{k\neq w} 
    z^{[p]}_{\underline{k}\underline{w}}
	- z_{\underline{w}_{[p]}\underline{w}}
	\perp \pi^{[p]-}_{\underline{w}} \mbox{ free}, 
	\forall w\in {\cal W},
	\label{subeq:NCP_conserv_dm_out}
	\\
	& \mbox{\textbf{Demand conservation at origin/destination for e-pooling:}} \nonumber\\
	& 0\leqslant z_{\bar{w}\bar{w}_{[p]}} 
	+ \sum_{k\in {\cal W},k\neq w} z^{[p]}_{\Bar{k}\bar{w}}
	-q^{[p]}_{w}
	\perp \phi^{[p]}_{\bar{w}} \geqslant 0, 
	\forall w\in {\cal W}, 
	\label{subeq:NCP_supp_dem_Ow}\\
	& 0\leqslant z_{\underline{w}_{[p]}\underline{w}} 
	+ \sum_{k\in {\cal W},k\neq w} z^{[p]}_{\underline{w}\underline{k}} - q^{[p]}_{w}
	\perp \phi^{[p]}_{\underline{w}} \geqslant 0, 
	\forall w\in {\cal W}, 
	\label{subeq:NCP_supp_dem_r_Dw}\\
& \mbox{\textbf{Flow conservation at origin/destination for e-solo:}} \nonumber\\
	& 0=z_{\bar{w} \bar{w}_{[e]}} - z^{[e]}_{\bar{w}\underline{w}}
	\perp \pi^{[e]}_{\bar{w}} \mbox{ free}, 
	\forall w\in {\cal W},
	\label{subeq:NCP_conserv_o_e}\\
	& 0=z^{[e]}_{\bar{w}\underline{w}} - z_{\underline{w}_{[e]}\underline{w}}
	\perp \pi^{[e]}_{\underline{w}} \mbox{ free}, 
	\forall w\in {\cal W}, 
	\label{subeq:NCP_conserv_d_e}\\
	& \mbox{\textbf{Demand conservation at origin for e-solo:}} \nonumber\\
	& 0\leqslant z_{\bar{w}\bar{w}_{[e]}}-q^{[e]}_{w} 
	\perp \phi^{[e]}_{w} \geqslant 0, 
	\forall w\in {\cal W}, 
	\label{subeq:NCP_supp_dem_e}\\
   & \mbox{\textbf{Demand conservation at virtual origin/destination:}} \nonumber\\
	& 0=\sum_{k\in {\cal W}} z^{\Bar{w}}_{\underline{k}}
	-\sum_{m\in {\cal M}} z_{\Bar{w}\Bar{w}_{[m]}}
	\perp \pi_{\Bar{w}} \mbox{ free},  
	\forall w\in {\cal W}, 
	\label{subeq:NCP_conserv_o}\\
	& 0=\sum_{m\in {\cal M}} z_{\underline{w}_{[m]}\underline{w}}
	-\sum_{k \in {\cal W}} z^{\Bar{k}}_{\underline{w}} 
	\perp \pi_{\underline{w}} \mbox{ free},  
	\forall w \in {\cal W}, 
	\label{subeq:NCP_conserv_d}\\
& \mbox{\textbf{Total flow conservation on virtual links:}} \nonumber\\
    & 0= \sum_{m\in {\cal M}} \sum_{w\in {\cal W}} z_{\bar{w}\bar{w}_{[m]}} 
	- \sum_{m\in {\cal M}} \sum_{w\in {\cal W}} z_{\underline{w}_{[m]}\underline{w}}\perp \lambda  \mbox{ free}, 
	\label{subeq:NCP_conserv_fleet}\\
& \mbox{\textbf{Fleet sizing:}} \nonumber\\
	& 0\leqslant 
	S
	-\sum_{k,w \in {\cal W}} 
	z^{\Bar{w}}_{\underline{k}} t_{\underline{k}\Bar{w}}
	-\sum_{m\in {\cal M}}\sum_{(v,u)\in {\cal L}(m)} z^{[m]}_{vu}t_{vu}
	\perp \lambda_{fleet} \geqslant 0, 
	\label{subeq:NCP_fleet}
\end{align}
\end{subequations}

Here we would like to discuss the economic interpretation of the multiplier associated with the demand conservation constraints defined in Equs.~(\ref{subeq:NCP_supp_dem_Ow},\ref{subeq:NCP_supp_dem_e}). 
The multiplier $\phi^{[e]}_{w}, \phi^{[p]}_{\Bar{w}}$ can be deemed as ``demand price" or ``surge price," which is a fee that an e-hailing passenger pays to drivers. 
In other words, when there is a surplus of e-hailing vehicles, 
the passengers do not need to wait;
when there is no surplus of e-hailing vehicles, 
the passengers have to wait. 
We summarize the above two conditions in Equ.~(\ref{eq:dif_dem_e_p}) for e-solo and e-pooling, respectively. 
Equ.~(\ref{eq:dif_dem_e_p}) will be revisited in Sec.~\ref{sec:3_mode} for the computation of passengers' waiting cost. 
\begin{subequations}\label{eq:dif_dem_e_p}
\begin{align}
& \begin{cases}
	& z_{\bar{w}\bar{w}_{[e]}} > q^{[e]}_{w} \Rightarrow \phi^{[e]}_{w} = 0, \mbox{ (e-solo passengers no waiting)}\\ 
	& z_{\bar{w}\bar{w}_{[e]}} = q^{[e]}_{w} \Rightarrow \phi^{[e]}_{w} >0. \mbox{ (e-solo passengers waiting)} 
\end{cases} \label{subeq:dif_dem_e}\\
& \begin{cases}
	&  z_{\bar{w}\bar{w}_{[p]}} 
	+ \sum_{k\in {\cal W},k\neq w} z^{[p]}_{\Bar{k}\bar{w}}
	> q^{[p]}_{w}
	\Rightarrow \phi^{[p]}_{\bar{w}} = 0, 
	\mbox{ (e-pooling passengers no waiting)}\\
	&  z_{\bar{w}\bar{w}_{[p]}} 
	+ \sum_{k\in {\cal W},k\neq w} z^{[p]}_{\Bar{k}\bar{w}}
	= q^{[p]}_{w}
	\Rightarrow \phi^{[p]}_{\bar{w}} > 0. 
	\mbox{ (e-pooling passengers waiting)}
\end{cases}\label{subeq:supp_dem_r_Ow}
\end{align}
\end{subequations}

Likewise, the economic meaning of the multiplier 
$\lambda_{fleet}$ associated with the fleet sizing constraint defined in Equ.~(\ref{subeq:NCP_fleet}) can be regarded as the incentive paid by the e-platform to drivers and imposed as an extra cost to the platform, shown in Equs~(\ref{subeq:NCP_profit_vu}-\ref{subeq:NCP_profit_wv}). 
When there are more drivers than needed, the e-platform does not have to invest. 
When there are fewer drivers, the e-platform pays an extra bonus to incentivize more drivers to join.
Accordingly, 

\begin{equation}\label{eq:surplus_w}
\begin{cases}
	& S > \sum_{k,w \in {\cal W}} 
	z^{\Bar{w}}_{\underline{k}} t_{\underline{k}\Bar{w}} + \sum_{m\in {\cal M}}\sum_{(u,v)\in {\cal L}(m)} z^{[m]}_{vu}t_{vu}
	\Rightarrow \lambda_{fleet} = 0, \mbox{ (underworking, e-platform no incentive)} 
	\\
	&  
	S = \sum_{k,w \in {\cal W}} 
	z^{\Bar{w}}_{\underline{k}} t_{\underline{k}\Bar{w}} + \sum_{m\in {\cal M}}\sum_{(v,u)\in {\cal L}(m)} z^{[m]}_{vu}t_{vu} \Rightarrow \lambda_{fleet} >0. \mbox{ (total work hours reached, e-platform pays)} 
\end{cases}
\end{equation}

To demonstrate how to apply the vehicle dispatching module to the 3-node network (Fig.~\ref{fig:3_node_OD_exp}), we lay out all the equations shown in \ref{append:exmp4.1}.

\subsection{Module 1.2: Vehicle-passenger matching}
 \label{subsec:match}

How would passengers move along with vehicle flows? 
Because passengers do not get to choose which e-hailing vehicles to take, the passenger flow is determined by the driver-passenger matching plan devised by the e-hailing platform, not chosen by passengers. 
Given an optimal vehicle dispatch plan $\{z^{[m]}_{uv}, \forall u,v\in {\cal N}^{+}(m)\cup{\cal N}^{-}(m), m\in {\cal M}\}\bigcup\{z_v^u, \forall u\in {\cal N}^{+}, v\in{\cal N}^{-}\}$, the e-platform needs to match driver-passenger and output passenger flows through a flow network. E-pooling vehicle-passenger matching is essentially a minimum-cost flow problem, provided with dispatched vehicle flow, riders' requests, and travel cost as input. 
In other words, the e-platform aims to find a matching strategy to send all passengers from their origins to destinations with the minimum cost over all edges on a passenger OD graph. Below we will introduce in sequence the structure of the passenger OD graph, the platform's objective, and passenger flow constraints. 

\subsubsection{Passenger OD graph}\label{subsubsec:pasOD}

\begin{figure}[H]
	\centering
	\includegraphics[scale=.6]{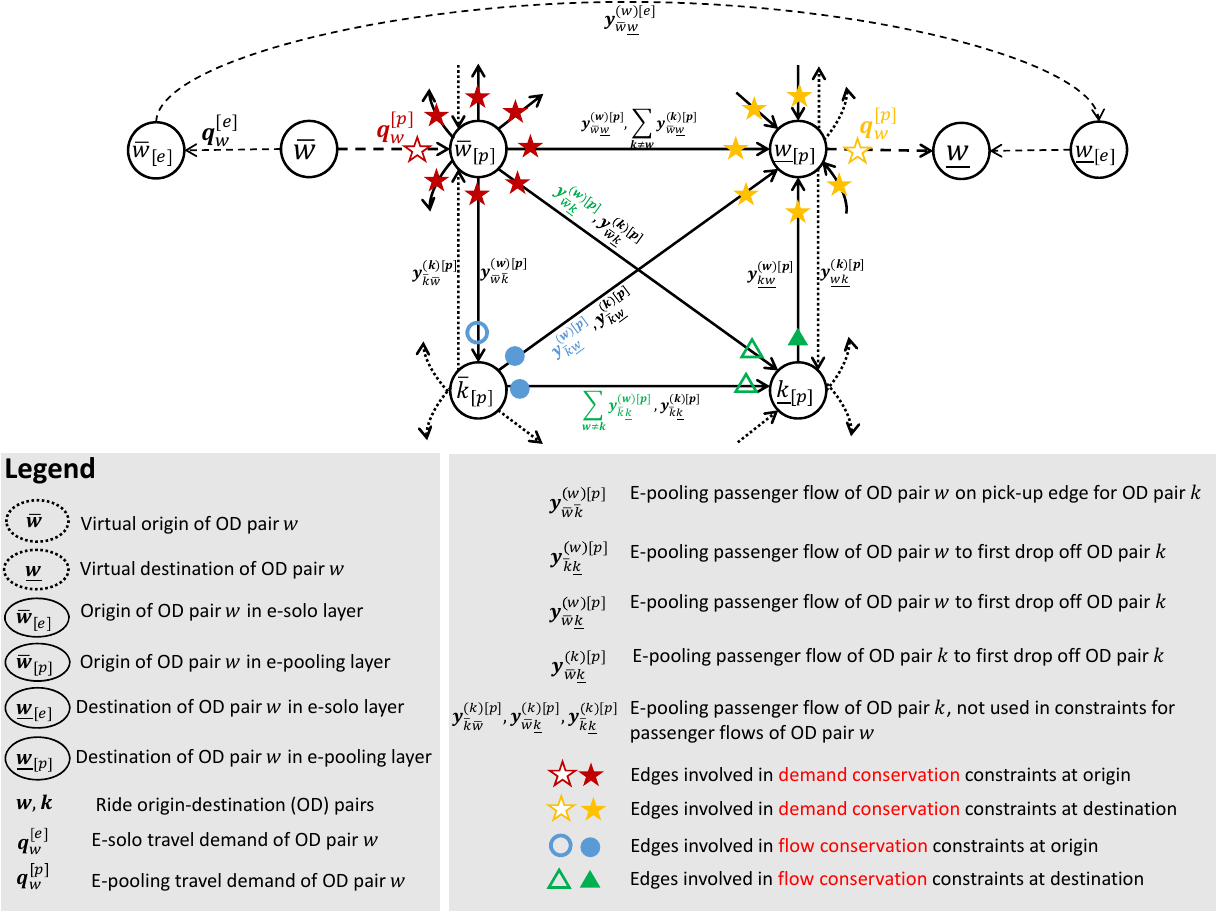}
	\caption{Passenger OD graph for OD pair $w\in {\cal W}$ and constraints illustration for the e-pooling passenger flow}
	\label{fig:ODnet_pas_ego}					
\end{figure}

The components of a passenger OD graph is illustrated in Fig.~\ref{fig:ODnet_pas_ego}.
A passenger OD graph is a subgraph of its vehicle OD graph counterpart.
The e-pooling layer resembles the topology of its associated e-pooling vehicle OD graph, except that there is no virtual origin nor virtual destination for a passenger OD graph. 
In the e-pooling layer, all the origins are aligned on the left-hand side and all the destinations are aligned on the right-hand side. 
All origins are connected with one another, and all destinations are connected with one another. 
Directed edges also emanate from all origins to all destinations. 
Different from the vehicle OD graph, however, the feasible path set on a passenger OD graph is smaller. 
First, there does not exist any e-pooling passenger flow of OD pair $\forall w\in {\cal W}$ prior to its origin $\bar{w}$ nor post its destination $\underline{w}$.
Second, there does not exist rebalancing flows for passengers and passenger flows end at their associated destinations. 
Thus, for passenger flow $y^{(w)[m]}_{uv}$ of OD pair $\forall w\in {\cal W}$, there only exist outgoing links from its origin $\bar{w}$ and only incoming links to its destination $\underline{w}$ from other destinations. 
Fig.~\ref{fig:ODnet_pas_ego} demonstrates the passenger OD graph from the perspective of OD pair $\forall w\in {\cal W}$.

\begin{rem}
Note that on each edge in the passenger OD graph, there exist passenger flows for at maximum two OD pairs. 
For example, on edge $(\bar{w}_{[p]},\underline{k}_{[p]})$, there only exist two types of passenger flows, namely, $y^{(w)[p]}_{\bar{w}\underline{k}},y^{(k)[p]}_{\bar{w}\underline{k}}$. 
In other words, there does not exist any other passenger flows associated with OD pairs than $w,k\in {\cal W}$ on this edge.
\end{rem}

\subsubsection{Passenger flow constraints}
 \label{subsec:const_pass}
 
Define $y^{(w)[m]}_{uv}, \forall u,v\in {\cal N}(m), w\in {\cal W}$ as the passenger flow on an edge $(u,v)$ in the passenger OD graph belonging to OD pair $w$ and mode $m$. 

The e-solo passenger flow only exists when $u=\bar{w}, v=\underline{w}$ and it should be equal to its demand. 
In addition, the driver-passenger matching simply satisfies the supply demand constraint that the e-solo passenger flow needs not to exceed its vehicle flow on the same edge. 
Mathematically,
\begin{subequations}\label{eq:y_s}
	\begin{align}
	& y^{(w)[e]}_{uv} 
	\begin{cases}
	= q^{[e]}_{w}, u=\bar{w}, v=\underline{w} 
	\nonumber \\
	=0, \forall u\neq\bar{w} \mbox{ and } v\neq\underline{w}
	\end{cases}, 
	\nonumber \\
	& y^{(w)[e]}_{\bar{w}\underline{w}} \leqslant z^{[e]}_{w},\forall w\in {\cal W}. \nonumber 
	\end{align}
\end{subequations}

However, e-pooling passengers of OD pair $w$ can visit nodes other than their own origins and destinations. In other words, it is likely that 
$y^{(w)[p]}_{uv}>0, \forall u\neq\bar{w} \mbox{ and } v\neq\underline{w}$. 
Thus, we need to further solve the vehicle-passenger matching problem, other than the demand-supply constraint defined in Equation~(\ref{subeq:supp_dem_Ow}). 
Subsequently, we will primarily focus on defining e-pooling passenger flow constrains.

Similar to how vehicle flow constraints are demonstrated in Fig.~\ref{fig:ODnet_veh}, 
the constraints associated with e-pooling passenger flows on a passenger OD graph are illustrated in Fig.~\ref{fig:ODnet_pas_ego}. 
The symbols used to represent constrains follow the same style as in Fig.~\ref{fig:ODnet_veh}.
 

\paragraph{Flow conservation at origin/destination}
If the e-pooling passenger flow of OD pair $\forall w\in {\cal W}$ visit a second origin of OD pair $\forall k\in {\cal W}, k\neq w$, they can only move to its own destination $\underline{w}$ or the associated destination $\underline{k}$. 
\begin{eqnarray*}
    & (\textcolor{blue}{\bigcirc}) 
	& y^{(w)[p]}_{\bar{w}\bar{k}}
	= y^{(w)[p]}_{\bar{k}\underline{w}} 
	+ y^{(w)[p]}_{\bar{k}\underline{k}}, 
	\forall w,k\in {\cal W}, 
	w\neq k,
	\label{subeq:ok_y}
\end{eqnarray*}

Similarly, if the e-pooling passenger flow of OD pair $\forall w\in {\cal W}$ move to its own destination $\underline{w}$ from a second destination of OD pair $\forall k\in {\cal W}, k\neq w$, at the associated destination $\underline{k}$, passenger flows can only originate from either its own origin $\bar{w}$ or the origin of the second OD pair, i.e., $\bar{k}$. 

\begin{eqnarray*}
    & (\textcolor{green}{\triangle}) 
	& y^{(w)[p]}_{\underline{k}\underline{w}}
	= y^{(w)[p]}_{\bar{w}\underline{k}}
	+ y^{(w)[p]}_{\bar{k}\underline{k}}, 
    \forall w,k\in {\cal W}, 
	w\neq k,
	\label{subeq:ko_y}
\end{eqnarray*}

\paragraph{Demand conservation at origin/destination}

E-pooling passengers of OD pair $w$ need to be all picked up at its origin. 
In other words, the sum of these passenger flows needs to be distributed to outgoing links from its origin. 

\begin{eqnarray*}
	& (\textcolor{red}{\bigstar}) 
	& y^{(w)[p]}_{\bar{w}\underline{w}}
	+ \sum_{k\in {\cal W}, k\neq w} 
	\left[y^{(w)[p]}_{\bar{w}\bar{k}} + y^{(w)[p]}_{\bar{w}\underline{k}}\right]
	= q^{[p]}_{w},
	\forall w\in {\cal W}, 
	\bar{w}\in {\cal N}^{+}(p).
	\label{subeq:conserv_om_y}
\end{eqnarray*}

Similarly, e-pooling passengers of OD pair $w$ need to be all dropped off at its destination. In other words, the sum of these passenger flows needs to be collected from incoming links to its destination.
\begin{eqnarray*}
    & (\textcolor{yellow}{\bigstar}) 
	& y^{(w)[p]}_{\bar{w}\underline{w}}
	+ \sum_{k\in {\cal W}, k\neq w} 
	\left[y^{(w)[p]}_{\bar{k}\underline{w}} + y^{(w)[p]}_{\underline{k}\underline{w}}\right]
	= q^{[p]}_{w},
	\forall w\in {\cal W},
	\underline{w}\in {\cal N}^{-}(p).
	\label{subeq:conserv_dm_y}
\end{eqnarray*}

\paragraph{Prior to an origin or post a destination}
The e-pooling passenger flow of OD pair $\forall w\in {\cal W}$ does not exist prior to its origin nor post its destination.
\begin{eqnarray*}
	& y^{(w)[p]}_{v\bar{w}} = 0, 
	\forall w\in {\cal W}, 
	v\in {\cal N}^{+}(p), v\neq \bar{w}, 
	\label{subeq:vo_y}\\
	& y^{(w)[p]}_{\underline{w}v} = 0, 
	\forall w\in {\cal W}, 
	v\in {\cal N}^{-}(p), v\neq \underline{w},
	\label{subeq:dv_y}
\end{eqnarray*}

\paragraph{Origin or destination edges}
The e-pooling passenger flow of OD pair $\forall w\in {\cal W}$ on an origin edge is zero if it does not start from its own origin. 
\begin{eqnarray*}
	& y^{(w)[p]}_{uv}
	= 0, 
	\forall w\in {\cal W}, 
	u\in {\cal N}^{+}(p), u\neq \bar{w},
	v\in {\cal N}^{+}(p),
	\label{subeq:conserv_no_o_y}
\end{eqnarray*}

Similarly, the e-pooling passenger flow of OD pair $\forall w\in {\cal W}$ on a destination edge is zero if it does not end at its own destination. 
\begin{eqnarray*}
	& y^{(w)[p]}_{vu}
	= 0, 
	\forall w\in {\cal W}, 
	v,u\in {\cal N}^{-}(p), u\neq \underline{w}.
	\label{subeq:conserv_no_d_y}
\end{eqnarray*}

\paragraph{Supply demand constraint}
The e-pooling passenger flow of OD pair $\forall w\in {\cal W}, v,u\in {\cal N}(m)$ on an edge should not exceed the vehicle flow on that edge and is always non-negative. 
Note that we assume each car has a capacity of two. Thus, we just need to ensure that the passenger flow from one OD is no greater than the vehicle flow on the same edge. Accordingly, the passenger flows from two ODs are automatically no greater than twice the vehicle flow, which is the vehicle capacity. Recall that on each edge in the passenger OD graph, there exist passenger flows for at maximum two OD pairs. Thus, we have
\begin{eqnarray*}
    & 0\leqslant 
    y^{(w)[p]}_{vu} \leqslant  z^{[p]}_{vu},
	\forall w\in {\cal W}, 
	v,u\in {\cal N}(p), 
	\label{subeq:cap+_y}
\end{eqnarray*}

\subsubsection{Objective function}
 \label{subsec:obj_pass}

In the passenger OD graph, the edge cost is the travel time of traversing that edge, $c^{[m]}_{vu} = t_{vu}$. The nodal cost is the waiting cost at a pick-up node $\bar{w},\forall w \in {\cal W}$, denoted as $c^{[m](wt)}_{\bar{w}}$, which will be detailed in Equs.~(\ref{eq:c_wait_do}) and (\ref{eq:c_wait_p}).

  
The platform determines a driver-passenger matching plan, with a goal of optimizing customer satisfaction, represented by minimizing passengers' total in-vehicle travel cost plus waiting cost. Mathematically, the objective function is the sum of the following two components:


\begin{subequations}\label{eq:obj_y}
	\begin{align}	
	& \mbox{Travel time:} & 
	\beta_{(in-veh)} \sum_{v,u\in {\cal N}(m), (v,u)\in {\cal L}}  t_{vu} \cdot\sum_{w\in {\cal W}}y^{(w)[m]}_{vu}, \label{subeq:obj_y_transp}\\
	& \mbox{Waiting cost:} & c^{[m](wt)}_{\bar{w}} \cdot \sum_{u\in {\cal N}(m)} y^{(w)[m]}_{vu}. \label{subeq:obj_y_wait}
	\end{align}
\end{subequations}

\begin{rem}
\begin{enumerate}
\item The first term of travel time is time-based and the second term of waiting cost is cost based. 
Thus, a coefficient $\beta_{(in-veh)}$ representing the value of time for passengers in-vehicle is multiplied in front of the first term.
\item In this paper, ``vehicle dispatching" and ``vehicle-passenger matching” are separated into two sequential modules. When the platform determines where to dispatch idle vehicles to meet demands, it is profit-driven. After vehicles are dispatched to reach passengers, the platform aims to move passengers to their destinations as fast as possible to increase passenger satisfaction, thus it is time-based. 
\end{enumerate}
\end{rem}


\subsubsection{Nonlinear complementarity formulation}

We formulate the minimum-cost flow problem to match e-pooling vehicles and passengers on the e-pooling passenger OD graph as an NCP. Mathematically,



\begin{subequations}\label{eq:opt_y_ncp}
\fontsize{9}{0}\selectfont
	\begin{align}
& \mbox{[\textbf{M1.2}.\mbox{NCP-VehPassMatch]}:} \nonumber\\
& \mbox{\textbf{Cost minimization for passengers}:} \nonumber\\
	& 0 \leqslant 
	y^{(w)[p]}_{\bar{w}\bar{k}}
	\perp 
	\beta_{(in-veh)} t_{\bar{w}\bar{k}} 
	+ c^{[p](wt)}_{\bar{w}}
	- \phi^{[p]+}_{w} - \pi^{(w)[p]}_{\bar{k}}
	+ \lambda^{(w)[p]}_{\bar{w}\bar{k}}
	\geqslant 0, 
	\forall \bar{w}, \bar{k}\in {\cal N}^{+}(p), k\neq w,
	\forall w\in{\cal W}, 
	\label{subeq:obj_y_owok_ncp}\\ 
	& 0 \leqslant 
	y^{(w)[p]}_{\bar{w}\underline{k}}
	\perp 
	\beta_{(in-veh)}t_{\bar{w}\underline{k}} 
	+ c^{[p](wt)}_{\bar{w}}
	- \phi^{[p]+}_{w} 
	+ \pi^{(w)[p]}_{\underline{k}}
	+ \lambda^{(w)[p]}_{\bar{w}\underline{k}}
	\geqslant 0, 
	\forall \bar{w}\in {\cal N}^{+}(p), \underline{k}\in {\cal N}^{-}(p),
	\forall w\in{\cal W}, 
	\label{subeq:obj_y_owdk_ncp}\\ 
	& 0 \leqslant 
	y^{(w)[p]}_{\underline{k}\underline{w}}
	\perp 
	\beta_{(in-veh)}t_{\underline{k}\underline{w}} - \phi^{[p]-}_{w} 
	- \pi^{(w)[p]}_{\underline{k}}
	+ \lambda^{(w)[p]}_{\underline{k}\underline{w}}
	\geqslant 0, 
	\forall \underline{w}, \underline{k}\in {\cal N}^{-}(p),
	\forall w\in{\cal W}, 
	\label{subeq:obj_y_dkdw_ncp}\\ 
	& 0 \leqslant 
	y^{(w)[p]}_{\bar{w}\underline{w}}
	\perp 
	\beta_{(in-veh)}t_{\bar{w}\underline{w}} - \phi^{[p]+}_{w} 
	- \phi^{[p]-}_{w}
	+ \lambda^{(w)[p]}_{\bar{w}\underline{w}}
	\geqslant 0, 
	\forall \bar{w}\in {\cal N}^{+}(p),
	\forall w\in{\cal W}, 
	\label{subeq:obj_y_w_ncp}\\ 
	& 0 \leqslant 
	y^{(w)[p]}_{\bar{k}\underline{w}}
	\perp 
	\beta_{(in-veh)}t_{\bar{k}\underline{w}} 
	- \phi^{[p]-}_{w}
	+ \pi^{(w)[p]}_{\bar{k}}
	+ \lambda^{(w)[p]}_{\bar{k}\underline{w}}
	\geqslant 0, 
	\forall \bar{k}\in {\cal N}^{+}(p),
	\underline{w}\in {\cal N}^{-}(p), 
	\forall w\in{\cal W}, 
	\label{subeq:obj_y_okdw_ncp}\\ 
	& 0 \leqslant 
	y^{(w)[p]}_{\bar{k}\underline{k}}
	\perp 
	\beta_{(in-veh)}t_{\bar{k}\underline{k}} 
	+ \pi^{(w)[p]}_{\bar{k}}
	+ \pi^{(w)[p]}_{\underline{k}}
	+ \lambda^{(w)[p]}_{\bar{k}\underline{k}}
	\geqslant 0, 
	\forall \bar{k}\in {\cal N}^{+}(p),
	\underline{k}\in {\cal N}^{-}(p), 
	\forall k\in{\cal W}, k\neq w, 
	\label{subeq:obj_y_okdk_ncp}\\ 
& \mbox{\textbf{Demand conservation at origin/destination}:} \nonumber\\
	& 0 = 
	y^{(w)[p]}_{\bar{w}\underline{w}}
	+ \sum_{k\in {\cal W}, k\neq w} 
	\left[y^{(w)[p]}_{\bar{w}\bar{k}} + y^{(w)[p]}_{\bar{w}\underline{k}}\right]
	- q^{[p]}_{w}
	\perp \phi^{[p]+}_{w} \mbox{ free}, 
	\forall w\in {\cal W}, 
	\bar{w}\in {\cal N}^{+}(p),
	\label{subeq:conserv_om_y_ncp}\\
	& 0 = 
	y^{(w)[p]}_{\bar{w}\underline{w}}
	+ \sum_{k\in {\cal W}, k\neq w} 
	\left[y^{(w)[p]}_{\bar{k}\underline{w}} + y^{(w)[p]}_{\underline{k}\underline{w}}\right]
	- q^{[p]}_{w}
	\perp \phi^{[p]-}_{w} \mbox{ free}, 
	\forall w\in {\cal W},
	\underline{w}\in {\cal N}^{-}(p), 
	\label{subeq:conserv_dm_y_ncp}\\
& \mbox{\textbf{Flow conservation at origin/destination}:} \nonumber\\
	& 0 = 
	y^{(w)[p]}_{\bar{w}\bar{k}}
	- y^{(w)[p]}_{\bar{k}\underline{w}} 
	- y^{(w)[p]}_{\bar{k}\underline{k}}
	\perp \pi^{(w)[p]}_{\bar{k}} \mbox{ free}, 
	\forall w,k\in {\cal W}, 
	w\neq k,
	\label{subeq:ok_y_ncp}\\
	& 0 =
	y^{(w)[p]}_{\underline{k}\underline{w}}
	- y^{(w)[p]}_{\bar{w}\underline{k}}
	- y^{(w)[p]}_{\bar{k}\underline{k}}
	\perp \pi^{(w)[p]}_{\underline{k}} \mbox{ free}, 
    \forall w,k\in {\cal W}, 
	w\neq k,
	\label{subeq:ko_y_ncp}\\
& \mbox{\textbf{Nominal constraints}:} \nonumber\\
	& y^{(w)[p]}_{v\bar{w}} = 0, 
	\forall w\in {\cal W}, 
	v\in {\cal N}^{+}(p), v\neq \bar{w}, 
	\label{subeq:vo_y_ncp}\\
	& y^{(w)[p]}_{\underline{w}v} = 0, 
	\forall w\in {\cal W}, 
	v\in {\cal N}^{-}(p), v\neq \underline{w},
	\label{subeq:dv_y_ncp}\\
	& y^{(w)[p]}_{vu}
	= 0, 
	\forall w\in {\cal W}, 
	v,u\in {\cal N}^{+}(p), v,u\neq \bar{w},
	\label{subeq:conserv_no_o_y_ncp}\\
	& y^{(w)[p]}_{vu}
	= 0, 
	\forall w\in {\cal W}, 
	v,u\in {\cal N}^{-}(p), v,u\neq \underline{w}.
	\label{subeq:conserv_no_d_y_ncp}\\
& \mbox{\textbf{Supply demand constraint}:} \nonumber\\
    & 0\leqslant 
    z^{[p]}_{vu} - y^{(w)[p]}_{vu}
    \perp \lambda^{(w)[p]}_{vu}\geqslant 0, 
	\forall w\in {\cal W}, 
	v,u\in {\cal N}(p). 
	\label{subeq:cap+_y_ncp}
	\end{align}
\end{subequations}

Following suit with the economic interpretation of the multiplier in Equ.~(\ref{eq:dif_dem_e_p}), here we discuss the multipliers $\lambda^{(w)[m]}_{vu}, \forall w\in {\cal W}, v,u\in {\cal N}(m), \forall m\in {\cal M}$ associated with the supply-demand constraints. 
For all $\forall w\in {\cal W}, v,u\in {\cal N}(m), 
\forall m\in {\cal M}$,
\begin{equation}\label{eq:surplus_p}
\begin{cases}
	y^{(w)[m]}_{vu} < z^{[m]}_{vu}
	\Rightarrow \lambda^{(w)[m]}_{vu} = 0, 
	\mbox{ (demand $<$ supply (oversupply), no extra fee)}\\
	y^{(w)[m]}_{vu} = z^{[m]}_{vu}
	\Rightarrow \lambda^{(w)[m]}_{vu} > 0. 
	\mbox{ (demand = supply (undersupply), fee paid by passengers to drivers))}
\end{cases}
\end{equation}

To demonstrate how to apply the vehicle-passenger matching module to the 3-node network (Fig.~\ref{fig:3_node_OD_exp}), we also write out the equations related to passenger flows in \ref{append:exmp4.1}.









\section{Module 2: E-hailing vehicle route choice on congested networks}
\label{sec:3_congest}

E-solo, e-pooling, and rebalancing vehicles contribute to traffic congestion. 
Compared to conventional traffic assignment problems, 
route choice for e-hailing vehicles depends on the corresponding vehicle flow. 
To define vehicle flows for each augmented OD pair, we first introduce a list of binary node-OD incidence indicators.

$\delta^{(o)}_{u \bar{w}} = 
\begin{cases}
0, \mbox{ node $u$ is not the origin node of OD pair $w$}, u\neq \bar{w},\\
1, \mbox{ node $u$ is the origin node of OD pair $w$}, u=\bar{w},
\end{cases}
\forall u\in {\cal N}, w\in {\cal W}.$

$\delta^{(d)}_{u \underline{w}} = 
\begin{cases}
0, \mbox{ node $u$ is not the destination node of OD pair $w$}, u=\underline{w}, \\
1, \mbox{ node $u$ is the destination node of OD pair $w$}, u\neq\underline{w},
\end{cases}
\forall u\in {\cal N}, w\in {\cal W}.$

Building on two indicators, we define a list of node-OD incidence matrices below:

$\delta^{(od)}_{(u,v) w} = 
\delta^{(o)}_{u \bar{w}} \delta^{(d)}_{v \underline{w}} 
=
\begin{cases}
0, \mbox{ node pair $(u,v)$ is not OD pair $w$}, u\neq\bar{w}, v\neq\underline{w},\\
1, \mbox{ node pair $(u,v)$ is OD pair $w$}, u=\bar{w}, v=\underline{w},
\end{cases}
\forall u,v\in {\cal N}, w\in {\cal W}.$

$\delta^{(oo)}_{(u,v) wk} = 
\delta^{(o)}_{u \bar{w}} \delta^{(o)}_{v \bar{k}} 
=
\begin{cases}
0, \mbox{ node pair $(u,v)$ is not OO pair ($\bar{w},\bar{k}$)}, u\neq\bar{w}, j\neq\bar{k},\\
1, \mbox{ node pair $(u,v)$ is OO pair ($\bar{w},\bar{k}$)}, u=\bar{w}, v=\bar{k},
\end{cases} 
\forall u,v\in {\cal N}, w,k\in {\cal W}.$

$\delta^{(od)}_{(u,v) wk} = 
\delta^{(o)}_{u \bar{w}} \delta^{(d)}_{v\underline{k}} 
=
\begin{cases}
0, \mbox{ node pair $(u,v)$ is not OD pair ($\bar{w},\underline{k}$)}, u\neq\bar{w}, v\neq\underline{k},\\
1, \mbox{ node pair $(u,v)$ is OD pair ($\bar{w},\underline{k}$)}, u=\bar{w}, v=\underline{k},
\end{cases} 
\forall u,v\in {\cal N}, w,k\in {\cal W}.$

$\delta^{(dd)}_{(u,v) wk} = 
\delta^{(d)}_{u \underline{w}} \delta^{(d)}_{v\underline{k}} 
=
\begin{cases}
0, \mbox{ node pair $(u,v)$ is not DD pair ($\underline{w},\underline{k}$)}, u\neq\underline{w}, v\neq\underline{k},\\
1, \mbox{ node pair $(u,v)$ is DD pair ($\underline{w},\underline{k}$)}, u=\underline{w}, v=\underline{k},
\end{cases} 
\forall u,v\in {\cal N}, w,k\in {\cal W}.$

$\delta^{(do)}_{(u,v) wk} = 
\delta^{(d)}_{u \underline{w}} \delta^{(o)}_{v\bar{k}} 
=
\begin{cases}
0, \mbox{ node pair $(u,v)$ is not DO pair ($\underline{w},\bar{k}$)}, u\neq\underline{w}, v\neq\bar{k},\\
1, \mbox{ node pair $(u,v)$ is DO pair ($\underline{w},\bar{k}$)}, u=\underline{w}, v=\bar{k},
\end{cases} 
\forall u,v\in {\cal N}, w,k\in {\cal W}.$

From now on, we will replace the OD pair $w=(u,v)$ by $(i,d)$ to align notations used in this module. 
With the above node-OD incidence indicators, the vehicle demand on an arbitrary node pair $(i,d)$ is computed as: 

\begin{eqnarray*}\label{eq:qid}
\fontsize{9}{0}\selectfont
q_{(i,d)} 
=
    \sum_{w\in{\cal W}} 
    \underbrace{\delta^{(od)}_{(i,d)w} q^{[e]}_{w}}_{\mbox{\tiny e-solo vehicles for OD pair}}
    +\sum_{w, k\in {\cal W}} 
    \left( 
    \underbrace{\delta^{(oo)}_{(i,d) wk} z^{[p]}_{\bar{w}\bar{k}}}_{\mbox{\tiny e-pool vehicles for OO pair}} 
    +
    \underbrace{\delta^{(od)}_{(i,d) wk} z^{[p]}_{\bar{w}\underline{k}}}_{\mbox{\tiny e-pool vehicles for OD pair}} 
    +
    \underbrace{\delta^{(dd)}_{(i,d) wk} z^{[p]}_{\underline{w}\underline{k}}}_{\mbox{\tiny e-pool vehicles for DD pair}}
   \right)
   +\sum_{w, k\in {\cal W}} 
	\underbrace{\delta^{(do)}_{(i,d) wk} z^{\bar{k}}_{\underline{w}}}_{\mbox{\tiny rebalancing vehicles for DO pair}}. 
\end{eqnarray*}

\subsection{Traffic assignment problem}


Traffic assignment problem can be path, link, link-OD, or link-node based. We will use link-node based formulation. 
In a road network, define $x^{d}_{ij}$ as the flow on road segment or link $(i,j)\in {\cal L}_{(net)}$ destined to node $d\in {\cal N}$. 
Then the link flow on link $(i,j)\in {\cal L}_{(net)}$ is denoted as $x_{ij} = \sum_{d\in{\cal N}_{(net)}} x^{d}_{ij}$.   
Accordingly, define $\tau^{d}_i\geq 0$ as node potentials.


At user equilibrium, on link $(i,j)\in {\cal L}_{(net)}$ destined to node $d\in {\cal N}$, no driver can reduce her travel time by unilaterally switching routes \citep{di2018link}.
The corresponding link-node based complementarity conditions for route choice are formulated as:

\begin{equation}\label{eq:RC_z}
	0 \leqslant x^{d}_{ij} \perp \tau^{d}_{j} + t_{ij}(\mathbf{x}) - \tau^{d}_{i} \geqslant 0, 
	\forall (i,j)\in {\cal L}_{(net)}, 
	d\in {\cal N}.  \nonumber
\end{equation}

The complementarity conditions of flow conservation at intermediate nodes are:

\begin{equation}\label{eq:FC_z}
	0 \leqslant 
	\sum_{j:(i,j)\in {\cal L}_{(net)}} x^{d}_{ij} - \sum_{k:(k,i)\in {\cal L}_{(net)}}  x^{d}_{ki} 
	\perp \tau^{d}_{i} \geqslant 0, \forall i\in {\cal N}_{(net)} \setminus {\cal N}, d\in {\cal N}. \nonumber
\end{equation}

Demand conservation conditions are satisfied at augmented origin nodes corresponding to $\tau^{d}_{i}, \forall i\in {\cal N}^{+}, v\in{\cal N}^{-}$:

\begin{equation}
0 \leqslant 
\sum_{j:(i,j)\in {\cal L}_{(net)}} x^{d}_{ij} - \sum_{k:(k,i)\in {\cal L}_{(net)}} x^{d}_{ki}  
- 
q_{(i,d)} 
	\perp \tau^{d}_{i} \geqslant 0, \forall i, d\in {\cal N}. \label{eq:DC} \nonumber
\end{equation} 


\subsection{Module summary}

Summarizing the route choice equilibrium conditions in Equ.~\ref{eq:RC_z} and the conservation constraints in Equs.~\ref{eq:FC_z}-\ref{eq:DC} and \ref{eq:qid}, the complementarity problem of Module 2 is denoted as [\textbf{M2}.\mbox{NCP}\mbox{-VehRoute}]:

\begin{subequations}\label{eq:NCP_vehRoute}
\fontsize{9}{0}\selectfont
	\begin{align}
& [\textbf{M2}.\mbox{NCP}\mbox{-VehRoute}] \nonumber\\
	& \mbox{\textbf{Route choice}:} \nonumber\\
	& 0 \leqslant x^{d}_{ij} \perp \tau^{d}_{j} + t_{ij}(\mathbf{x}) - \tau^{d}_{i} \geqslant 0, 
	\forall (i,j)\in {\cal L}_{(net)}, 
	d\in {\cal N}, \label{subeq:RC_z}\\
    & \mbox{\textbf{Flow conservation at intermediate Nodes}:} \nonumber\\
	& 0 \leqslant 
	\sum_{j:(i,j)\in {\cal L}_{(net)}} x^{d}_{ij} - \sum_{k:(k,i)\in {\cal L}_{(net)}} x^{d}_{ki} 
	\perp \tau^{d}_{i} \geqslant 0, \forall i\in {\cal N}_{(net)} \setminus {\cal N}, d\in {\cal N}, \label{subeq:FC_z}\\
    & \mbox{\textbf{Demand conservation at augmented origin nodes}:} \nonumber\\
	& 0 \leqslant 
	\sum_{j:(i,j)\in {\cal L}_{(net)}} x^{d}_{ij} - \sum_{k:(k,i)\in {\cal L}_{(net)}} x^{d}_{ki} - q_{(i,d)} 
	\perp \tau^{d}_{i} \geqslant 0, \forall i, d\in {\cal N}, \label{subeq:DC}\\
	& \mbox{\textbf{Vehicle demand on augmented OD pairs}:} \nonumber\\
	& q_{(i,d)} =
    \sum_{w\in{\cal W}^{aug}} 
    \delta^{(od)}_{(i,d)w} q^{[e]}_{w} + 
    \sum_{w, k\in {\cal W}^{aug}} 
    \left(\delta^{(oo)}_{(i,d) wk} z^{[p]}_{\bar{w}\bar{k}} 
    + \delta^{(od)}_{(i,d) wk} z^{[p]}_{\bar{w}\underline{k}} 
    + \delta^{(dd)}_{(i,d) wk} z^{[p]}_{\underline{w}\underline{k}} 
   \right) + \sum_{w, k\in {\cal W}^{aug}} 
	\delta^{(do)}_{(i,d) wk} z^{\bar{k}}_{\underline{w}}. \label{subeq:qid}
	\end{align}
\end{subequations} 
\vspace{-3mm}

\section{Module 3: TNC customer mode choice}
\label{sec:3_mode}

\subsection{Demand conservation}

A traveler faces two mode choices: 
taking e-hailing alone or pooling with another passenger.
Denote the traffic demand between OD pair $w=(i,d)$ as $q_w>0$. 
For the fixed travel demand case, the total travel demand $q_w$ is split into available modes, denoted as $q^{[m]}_{w}, \forall m\in {\cal M}$. 
The demand conservation condition needs to be satisfied for each OD pair $\forall w\in {\cal W}$ across modes $\forall m\in {\cal M}$, namely, 
\begin{equation}
    \sum_{m\in {\cal M}} q^{[m]}_{w} = q_{w}, 
\forall w\in {\cal W}. \label{eq:DC_M3}
\end{equation} 






\subsection{Mode choice}

The travel disutility of being e-solo passengers for OD pair $w=(\bar{w},\underline{w})\in {\cal W}$ is the sum of 
e-solo fare (sum of a fixed fare, a distance-based, and a time-based pricing), 
nodal cost (including waiting cost and matching friction), 
edge cost (i.e., supply-demand surplus cost),
and OD cost (including in-vehicle travel cost): 

\begin{equation}\label{eq:Ce}
U^{[e]}_{w} 
= \underbrace{-r^{[e]}_{w}}_{\small\mbox{e-solo fare}}
+ \underbrace{	c^{[e](nod)}_{\bar{w}} }_{\small\mbox{nodal cost}}
+ \underbrace{	c^{[e](surp-pas)}_{w} }_{\small\mbox{edge cost}}
+ \underbrace{	c^{(in-veh)}_{w} }_{\small\mbox{OD cost}}.
\end{equation}

The travel disutility of being e-pooling passengers for OD pair $w$ is the sum of 
e-pooling fare (sum of a fixed cost, a distance-based, and a time-based cost), 
nodal cost (including waiting cost and matching friction), 
edge cost (i.e., supply-demand surplus cost),
and OD cost (including en-route travel cost and inconvenience cost arising from sharing rides with others in an enclosed space): 

\begin{equation}\label{eq:Ce_p}
U^{[p]}_{w} = 
 \underbrace{ 
 - r^{[p]}_{w}}_{\small\mbox{e-pooling fare}}
+ \underbrace{	c^{[p](nod)}_{\bar{w}} }_{\small\mbox{nodal cost}}
+ \underbrace{	c^{[p](surp-pas)}_{w} }_{\small\mbox{edge cost}}
+ \underbrace{	c^{(in-veh)}_{w} + c^{[p](inc)}_{w} }_{\small\mbox{OD cost}}. 
\end{equation}

At an origin node $\forall \bar{w}\in {\cal N}^{+}(m)$, travelers who are desirable for e-hailing choose either taking e-hailing alone or pooling with another passenger. 

\subsubsection{Cost specification: Nodal cost}
\label{subsec:nodal}
 
The nodal cost denotes the total waiting time for passengers to be picked up at a node. It consists of two parts: the time drivers spend on reaching the node and the waiting time caused by drivers' search friction at the node. Mathematically, the nodal cost at origin $\forall \bar{w}\in {\cal N}^+$ for travel mode $m\in {\cal M}$ is formulated as below: 

\begin{equation}\label{eq:Cnodal_w}
c^{[m](nod)}_{\bar{w}}
= 
\underbrace{
c^{[m](wt)}_{\bar{w}}
}_{\small\mbox{waiting: vehicles reach the node}}
+ 
\underbrace{c^{[m](se-pas)}_{\bar{w}}}_{\small\mbox{search friction: vehicle-passenger match at the node}}, 
\end{equation}


 


\paragraph{Waiting cost} 

An e-solo and an e-pooling passenger could experience different waiting time to be picked up by an e-hailing vehicle. Thus, we will compute waiting cost for two types of passengers separately. 

\noindent\textbf{E-solo.}
For an e-solo passenger of OD pair $w\in {\cal W}$, her waiting time to be picked up is the vacant vehicles' average travel time to the origin $\bar{w}$ from any drop-off node $\forall \underline{v}\in {\cal N}^{-}, v\in {\cal W}$, which is formulated as: 

\begin{equation}\label{eq:c_wait_do}
c^{[e](wt)}_{\bar{w}}
= \beta^{[e]}_{(wt)} \cdot
\underbrace{
\sum_{\underline{v}\in {\cal N}^{-}}
\frac{
z^{\bar{w}}_{\underline{v}} 
}
{
\sum_{\underline{v}'\in {\cal N}^{-}} z^{\bar{w}}_{\underline{v}'}
}
t_{\underline{v}\bar{w}}
}_{\small\mbox{vacant vehicles' average travel time to $\bar{w}$, $c_{\bar{w}}$}} 
\triangleq 
\beta^{[e]}_{(wt)} \cdot \sum_{\underline{v}\in {\cal N}^{-}}
o_{\underline{v}\bar{w}}
t_{\underline{v}\bar{w}}, 
\end{equation}
where $\beta^{[e]}_{(wt)}$ is the weight of waiting cost. We define $o_{\underline{v}\bar{w}} 
\equiv 
\frac{
z^{\bar{w}}_{\underline{v}} 
}
{
\sum_{\underline{v}'\in {\cal N}^{-}} z^{\bar{w}}_{\underline{v}'}
}
\geq 0$
as the vacant vehicle percentage flowing into the virtual origin node $\bar{w}\in {\cal N}^+$ from a specific virtual destination node $\underline{v}$, and it satisfied that: 
$\sum_{\underline{v}\in {\cal N}^{-}}
o_{\underline{v}\bar{w}}=1$. The complementarity conditions associated with $o_{\underline{v}\bar{w}}$ are:

\begin{subequations}\label{eq:o_vw}
	\begin{align}
	& 0 \leqslant o_{\underline{v}\bar{w}} 
	\perp 
	\sum_{\underline{v}'\in {\cal N}^{-}} z^{\bar{w}}_{\underline{v}'} \cdot
	o_{\underline{v}\bar{w}} 
    - z^{\bar{w}}_{\underline{v}} 
	+ \eta^{-}_{\underline{v}\bar{w}} 
	\geqslant 0,
	\forall \underline{v}\in {\cal N}^{-}, \bar{w}\in {\cal N}^{+}, \label{subeq:o_vw.c_vw_1} \\
	& 0 \leqslant \eta^{-}_{\underline{v}\bar{w}} \perp 1-o_{\underline{v}\bar{w}} \geqslant 0, 
	\forall \underline{v}\in {\cal N}^{-}, \bar{w}\in {\cal N}^{+}.
	\label{subeq:o_vw.eta}
	\end{align}
\end{subequations}


\noindent Derivation of complementarity conditions in Equ.~\ref{eq:o_vw}
is shown in \ref{append:KKT}.

\noindent \textbf{E-pooling.}
There are two scenarios to compute the waiting cost of an e-pooling passenger. For an e-pooling passenger of OD pair $w\in {\cal W}$, if she is the first passenger to be picked up, her waiting time is the vacant vehicles' average travel time to the origin $\bar{w}$ from any drop-off node $\forall \underline{v}\in {\cal N}^{-}, v\in {\cal W}$. If she is the second passenger to be picked up, her waiting time consists of two parts, namely, the vacant vehicles' average travel time to an origin $\bar{k}\in {\cal N}^{+}, k\in {\cal W}$ from a drop-off node $\forall \underline{v}\in {\cal N}^{-}, v\in {\cal W}$ and the 1-passenger occupied vehicles' average travel time to her pick-up node $\bar{w}$ from a previous pick-up location. Considering the 
fact that an e-pooling passenger can be picked up as either the first passenger or the second passenger, we calculate the weighted average waiting time according to the probability of being the first or second passenger to get picked up. Mathematically, $\forall w\in {\cal W}$:

{\small 
\begin{align}\label{eq:c_wait_p}
& c^{[p](wt)}_{\bar{w}}
= \beta_{(wt)}^{[p]} \cdot \Bigg\{
\frac{\sum_{k\in \mathcal{W},k \neq w} y^{(w)[p]}_{\Bar{w}\Bar{k}}}{\sum_{k\in \mathcal{W},k \neq w} y^{(w)[p]}_{\Bar{w}\Bar{k}}+\sum_{k\in \mathcal{W}} y^{(w)[p]}_{\Bar{w}\underline{k}}}\cdot \underbrace{
\sum_{\underline{v}\in {\cal N}^{-}}
\frac{
z^{\bar{w}}_{\underline{v}} 
}
{
\sum_{\underline{v}'\in {\cal N}^{-}} z^{\bar{w}}_{\underline{v}'}
}
t_{\underline{v}\bar{w}}
}_{\small\mbox{vacant vehicles' average travel time to $\bar{w}$, $c_{\bar{w}}$}} 
\nonumber \\
& + \frac{y^{(w)[p]}_{\Bar{w}\underline{w}}}{\sum_{k\in \mathcal{W},k \neq w} y^{(w)[p]}_{\Bar{w}\Bar{k}}+\sum_{k\in \mathcal{W}} y^{(w)[p]}_{\Bar{w}\underline{k}}}\cdot\sum_{\bar{k}\in {\cal N}^{+}(p), k\in {\cal W}, k\neq w} 
\Bigg\{
\underbrace{
\sum_{\underline{v}\in {\cal N}^{-}}
\frac{
z^{\bar{k}}_{\underline{v}}
}
{
\sum_{\underline{v}'\in {\cal N}^{-}} z^{\bar{k}}_{\underline{v}'}
}
t_{\underline{v}\bar{k}}
}_{\small\mbox{vacant vehicles' average travel time to $\bar{k}$, $c_{\underline{v}\bar{k}}$}}
+ 
t_{\bar{k}\bar{w}}
\Bigg\} , 
\nonumber \\
& + \sum_{\bar{k}\in {\cal N}^{+}(p), k\in {\cal W}, k\neq w} \frac{y^{(w)[p]}_{\Bar{w}\underline{k}}}{\sum_{k\in \mathcal{W},k \neq w} y^{(w)[p]}_{\Bar{w}\Bar{k}}+\sum_{k\in \mathcal{W}} y^{(w)[p]}_{\Bar{w}\underline{k}}}\cdot
\Bigg\{
\underbrace{
\sum_{\underline{v}\in {\cal N}^{-}}
\frac{
z^{\bar{k}}_{\underline{v}}
}
{
\sum_{\underline{v}'\in {\cal N}^{-}} z^{\bar{k}}_{\underline{v}'}
}
t_{\underline{v}\bar{k}}
}_{\small\mbox{vacant vehicles' average travel time to $\bar{k}$, $c_{\underline{v}\bar{k}}$}}
+ 
t_{\bar{k}\bar{w}}
\Bigg\} \Bigg\} \nonumber \\
&\triangleq  \beta_{(wt)}^{[p]} \cdot \Bigg\{ \sum_{k,k\neq w}  o^{(w)[p]}_{\bar{w}\bar{k}} \sum_{v} o_{\underline{v}\bar{w}}t_{\underline{v}\bar{w}}+ o^{(w)[p]}_{\bar{w}\underline{w}} \sum_{k,k\neq w}\Bigg\{ \sum_{v} o_{\underline{v}\bar{k}}t_{\underline{v}\bar{k}}+t_{\bar{k}\bar{w}}\Bigg\}+ \sum_{k,k\neq w}o^{(w)[p]}_{\bar{w}\underline{k}} \Bigg\{ \sum_{v} o_{\underline{v}\bar{k}}t_{\underline{v}\bar{k}}+t_{\bar{k}\bar{w}}\Bigg\} \Bigg\}
\end{align}
}

\noindent where $\beta^{[p]}_{(wt)}$ is the weight of waiting cost. The vacant vehicles' average travel time to the passenger's pick-up node $\bar{w}\in {\cal N}^{+}(p)$ or a different pick-up node $\bar{k}\in {\cal N}^{+}$, is similar to that defined in Equation~(\ref{eq:c_wait_do}). 
Following suit, we can introduce new variables $o_{\underline{v}\bar{w}},o_{\underline{v}\bar{k}}$ to reformulate these two terms. We introduce two new variables, namely, {\small$o^{(w)[p]}_{\Bar{w}\Bar{k}}=\frac{y^{(w)[p]}_{\Bar{w}\Bar{k}}}{\sum_{k\in \mathcal{W},k \neq w} y^{(w)[p]}_{\Bar{w}\Bar{k}}+\sum_{k\in \mathcal{W}} y^{(w)[p]}_{\Bar{w}\underline{k}}}$} as the proportion of the first passenger picked up at origin $\Bar{w}$,  {\small $o^{(w)[p]}_{\Bar{w}\underline{k}}=\frac{y^{(w)[p]}_{\Bar{w}\underline{k}}}{\sum_{k\in \mathcal{W},k \neq w} y^{(w)[p]}_{\Bar{w}\Bar{k}}+\sum_{k\in \mathcal{W}} y^{(w)[p]}_{\Bar{w}\underline{k}}}$} as the proportion of the second passenger picked up at origin $\Bar{w}$. Their complementarity conditions are:

\begin{subequations}\label{eq:prop_passenger}
	\begin{align}
	& 0 \leqslant o^{(w)[p]}_{\bar{w}\bar{k}}
	\perp 
	(\sum_{k\in \mathcal{W},k \neq w} y^{(w)[p]}_{\Bar{w}\Bar{k}}+\sum_{k\in \mathcal{W}} y^{(w)[p]}_{\Bar{w}\underline{k}}) \cdot 
	o^{(w)[p]}_{\bar{w}\bar{k}}
    - y^{(w)[p]}_{\bar{w}\bar{k}}
	+ \eta^{(w)[p]-}_{\bar{w}\bar{k}} 
	\geqslant 0, 
	\forall \bar{k}, \bar{w}\in {\cal N}^{+}(p), k\neq w\in {\cal W}, \label{subeq:prop_passenger_1}\\
	& 0 \leqslant \eta^{(w)[p]}_{\bar{w}\bar{k}} \perp 1-o^{(w)[p]}_{\bar{w}\bar{k}} \geqslant 0, 
	\forall \bar{k}, \bar{w}\in {\cal N}^{+}, k\neq w\in {\cal W}, \label{subeq:prop_passenger_2}\\
	& 0 \leqslant o^{(w)[p]}_{\bar{w}\underline{k}}
	\perp 
	(\sum_{k\in \mathcal{W},k \neq w} y^{(w)[p]}_{\Bar{w}\bar{k}}+\sum_{k\in \mathcal{W}} y^{(w)[p]}_{\Bar{w}\underline{k}}) \cdot 
	o^{(w)[p]}_{\bar{w}\underline{k}}
    - y^{(w)[p]}_{\bar{w}\underline{k}}
	+ \eta^{(w)[p]-}_{\bar{w}\underline{k}} 
	\geqslant 0, 
	\forall \bar{w}\in {\cal N}^{+}(p), \bar{k} \in {\cal N}^{-}(p), \label{subeq:prop_passenger_3}\\
	& 0 \leqslant \eta^{(w)[p]}_{\bar{w}\underline{k}} \perp 1-o^{(w)[p]}_{\bar{w}\underline{k}} \geqslant 0, 
	\forall \bar{w}\in {\cal N}^{+}(p), \bar{k} \in {\cal N}^{-}(p).\label{subeq:prop_passenger_4}
	\end{align}
\end{subequations}

\paragraph{Search friction} 
Search friction is incurred when e-hailing drivers seek passengers at origins. We use $c^{[m](se-veh)}_{\bar{w}}$ to denote the time drivers spend seeking passengers and $\Tilde{c}^{[m](se-pas)}_{\bar{w}}$ to denote the time passengers spend waiting to be picked up by drivers. Following the Douglas form of matching functions \citep{yang2010equilibria}, we have  {\small $A(\Tilde{c}^{[e](se-pas)}_{\bar{w}}\cdot q_w^{[e]})^{\alpha_1}(c^{[e](se-veh)}_{\bar{w}}\cdot z^{[e]}_{\bar{w}\underline{w}})^{\alpha_2}=q_w^{[e]}$} and {\small $A(\Tilde{c}^{[p](se-pas) }_{\bar{w}}\cdot q_w^{[p]})^{\alpha_1}(c^{[p](se-veh)}_{\bar{w}}\cdot \sum_{u \in {\cal N^+}(p) \cup {\cal N^+}}z^{[p]}_{u\bar{w}})^{\alpha_2}=q_w^{[p]}$}, where $\Tilde{c}^{[m](mf)}_{\bar{w}}, \forall w \in {\cal W}, m \in {\cal M}$ is waiting time of passengers at origins when matched with drivers. $z^{[e]}_{\bar{w}\underline{w}}$ and $\sum_{u \in {\cal N^+}(p) \cup {\cal N^+}}z^{[p]}_{u\bar{w}}$ represent the e-solo and e-pooling vehicle flow at origin $\bar{w}$, respectively. They fulfill the travel demand of OD pair $w, w \in \mathcal{W}$. $\alpha_1$ and $\alpha_2$ are matching elasticity and $A$ is a constant representing properties of a matching zone. Accordingly, we can calculate search friction of drivers,
\begin{subequations}
    \begin{align}
    &c^{[e](se-veh)}_{\bar{w}}= \beta_{(se-veh)} \cdot A^{-\frac{1}{\alpha_2}} (z^{[e]}_{\bar{w}\underline{w}})^{-1} (q^{[e]}_{w})^{\frac{1-\alpha_1}{\alpha_2}} (\Tilde{c}^{[e](se-pas)}_{\bar{w}} )^{-\frac{\alpha_1}{\alpha_2}}.\\
    &c^{[p](se-veh)}_{\bar{w}}= \beta_{(se-veh)} \cdot A^{-\frac{1}{\alpha_2}} (\sum_{u \in {\cal N^+}(p) \cup {\cal N^+}}z^{[p]}_{u\bar{w}})^{-1} (q^{[p]}_{w})^{\frac{1-\alpha_2}{\alpha_2}}  (\Tilde{c}^{[p](se-pas)}_{\bar{w}} )^{-\frac{\alpha_1}{\alpha_2}},
    \label{equ:waiting_time_one_to_one_1}
    \end{align} 
\end{subequations}
where, $\beta_{(se-veh)}$ is the cost coefficient. Note that $c^{[m](se-veh)}_{\bar{w}}$ and $\Tilde{c}^{[m](se-pas)}_{\bar{w}}$ depend on each other, meaning that to compute one we need to specify the other first. In this paper,  we assume passengers' waiting time to be matched with drivers $\Tilde{c}^{[m](se-pas)}_{\bar{w}}$ as constants. We have
\begin{align}
    c^{[m](se-pas)}_{\bar{w}}=\beta_{(se-pas)} \cdot [ \Tilde{c}^{[m](se-pas)}_{\Bar{w}}+\phi^{[m]}_{\Bar{w}} ]
\end{align}
where, $\beta_{(se-pas)}$ is the cost coefficient and $\phi^{[m]}_{\Bar{w}}$ is the multiplier associated with the demand constraints, denoting the extra waiting time of passengers (Equ.~\ref{eq:dif_dem_e_p}).

\subsubsection{Cost specification: Edge cost}

\paragraph{Supply-demand surplus for passenger flows}

Supply-demand surplus cost for passenger flows is a fee depicting the gap between the demand and supply of passengers, according to the economic interpretation in Equ.~(\ref{eq:surplus_p}). 
Thus, passenger flow supply-demand surplus cost for OD pair $w\in {\cal W}$, denoted as $c^{[m](surp-pas)}_{w}$, is proportional to the multiplier $\lambda^{(w)[m]}_{vu}, \forall w\in {\cal W}, v,u\in {\cal N}(m), \forall m\in {\cal M}$. 
Mathematically,
\begin{equation}
c^{[m](surp-pas)}_{w} = 
\beta^{[m]}_{(surp-pas)}
\sum_{(u,v)\in {\cal L}(m)} \sum_{h\in {\cal P}^w}
\delta^{(ed-pa)}_{(u,v)h}
\delta^{(pa-od)}_{hw} \lambda^{(w)[m]}_{vu}.
\end{equation}
where $\beta^{[m]}_{(surp-pas)}$ is the weight of edge cost.





\subsubsection{Cost specification: OD cost}

 \paragraph{In-vehicle travel cost}

The in-vehicle travel cost is the value of time for passengers to stay within an e-hail vehicle:  
\vspace{-2mm}
\begin{equation}
c^{(in-veh)}_{w} = 
\beta_{(in-veh)} t_{\bar{w}\underline{w}}.
\end{equation}
where $\beta_{(in-veh)}$ is the weight of in-vehicle travel cost.

 

 \paragraph{Inconvenience cost}
The inconvenience cost only exists for e-pooling customers. 
It is proportional to the travel  time and distance and can be defined as:  
\begin{equation}
c^{[p](inc)}_{w} = 
\beta_{1(inc)} t_{\bar{w}\underline{w}} + \beta_{2(inc)} l_{\bar{w}\underline{w}}.
\end{equation}
where $\beta_{1(inc)}$ and $\beta_{2(inc)}$ are  weights of time-based and distance-based inconvenience costs, respectively.


\subsubsection{Mode choice condition}

Define $\mu_{w}, \forall w\in {\cal W}$ as the minimum mode disutility for OD pair $\forall w\in {\cal W}$. At equilibrium, no traveler (i.e., e-solo, e-pooling) can reduce her modal disutility by unilaterally switching travel modes, i.e., 
the cost of utilized mode is equal, less than that of unused modes:  

\begin{subequations}\label{eq:condition_mode}
	\begin{align}
	q^{[m]}_{w} 
	\begin{cases}
	>0 \Rightarrow U^{[m]}_{w} = \mu_{w}  \nonumber\\
	=0 \Rightarrow U^{[m]}_{w} > \mu_{w} \nonumber
	\end{cases}, \forall w\in {\cal W}, m\in {\cal M}.
	\end{align}
\end{subequations}

This equilibrium condition can be formulated as a complementarity problem:
\vspace{-2mm}
\begin{equation}\label{eq:MC_p}
    0 \leqslant 
	U^{[m]}_{w} - \mu_{w} 
	\perp 
	q^{[m]}_{w} 
	\geqslant 0,  
	\forall w\in {\cal W}, m \in {\cal M}. 
\end{equation}

\subsection{Module summary}
Summarizing the demand conservation Equ.~\ref{eq:DC_M3} and the mode choice Equs.~\ref{eq:MC_p}, the complementarity problem of Module 3 is denoted as [\textbf{M3}.\mbox{NCP}\mbox{-CustChoice}]:

\begin{subequations}\label{eq:NCP_CustChoice_M3}
\fontsize{9}{0}\selectfont
\begin{align}
& [\textbf{M3}.\mbox{NCP}\mbox{-CustChoice}] \nonumber\\
    & \mbox{\textbf{Demand conservation}:} \nonumber\\
    & 0 =
    \sum_{m\in {\cal M}} q^{[m]}_{w} - q_{w}
    \perp 
    \mu_{w} \ \text{free}, 
    \forall w\in {\cal W}. \label{subeq:DC_M3}\\
    & \mbox{\textbf{Mode choice}:} \nonumber\\
	& 0 \leqslant 
	U^{[m]}_{w} - \mu_{w} 
	\perp 
	q^{[m]}_{w} 
	\geqslant 0,  
	\forall w\in {\cal W}, m \in {\cal M}. \label{eq:NCP_MC_p}\\
	& \mbox{\textbf{Travel disutility}:} \nonumber\\
	& U^{[e]}_{w} 
= -r^{[e]}_{w}+\beta^{[e]}_{(wt)} \cdot \sum_{\underline{v}\in {\cal N}^{-}}
o_{\underline{v}\bar{w}}
t_{\underline{v}\bar{w}}+\beta_{(se-pas)} \cdot [ \Tilde{c}^{[e](se-pas)}_{\Bar{w}}+\phi^{[e]}_{\Bar{w}} ]
+ \beta^{[e]}_{(surp-pas)}
\sum_{(u,v)\in {\cal L}(e)} \sum_{h\in {\cal P}^w}
\delta^{(ed-pa)}_{(u,v)h}
\delta^{(pa-od)}_{hw} \lambda^{(w)[e]}_{vu}
+ 	\beta_{(in-veh)} t_{\bar{w}\underline{w}}, \\
    & U^{[p]}_{w} = - r^{[p]}_{w}+ 	
\beta^{[p]}_{(wt)} \cdot \beta_{(wt)}^{[p]} \cdot \Bigg\{ \sum_{k,k\neq w}  o^{(w)[p]}_{\bar{w}\bar{k}} \sum_{v} o_{\underline{v}\bar{w}}t_{\underline{v}\bar{w}}+ o^{(w)[p]}_{\bar{w}\underline{w}} \sum_{k,k\neq w}\Bigg\{ \sum_{v} o_{\underline{v}\bar{k}}t_{\underline{v}\bar{k}}+t_{\bar{k}\bar{w}}\Bigg\}+ \sum_{k,k\neq w}o^{(w)[p]}_{\bar{w}\underline{k}} \Bigg\{ \sum_{v} o_{\underline{v}\bar{k}}t_{\underline{v}\bar{k}}+t_{\bar{k}\bar{w}}\Bigg\} \Bigg\}
\nonumber\\
&+\beta_{(se-pas)} \cdot [ \Tilde{c}^{[p](se-pas)}_{\Bar{w}}+\phi^{[p]}_{\Bar{w}} ]
+\beta^{[p]}_{(surp-pas)}
\sum_{(u,v)\in {\cal L}(p)} \sum_{h\in {\cal P}^w}
\delta^{(ed-pa)}_{(u,v)h}
\delta^{(pa-od)}_{hw} \lambda^{(w)[p]}_{vu}
+\beta_{(in-veh)} t_{\bar{w}\underline{w}}
+ 	\beta_{1(inc)} t_{\bar{w}\underline{w}} + \beta_{2(inc)} l_{\bar{w}\underline{w}}. \\
	& \mbox{\textbf{Nodal cost:}} \nonumber\\
	& 0 \leqslant o_{\underline{v}\bar{w}} 
	\perp 
	\sum_{\underline{v}'\in {\cal N}^{-}} z^{\bar{w}}_{\underline{v}'} \cdot
	o_{\underline{v}\bar{w}} 
    - z^{\bar{w}}_{\underline{v}} 
	+ \eta^{-}_{\underline{v}\bar{w}} \geqslant 0, 
	\forall \underline{v}\in {\cal N}^{-}, \bar{w}\in {\cal N}^{+}, \label{subeq:o_vw.c_vw}\\
	& 0 \leqslant \eta^{-}_{\underline{v}\bar{w}} \perp 1-o_{\underline{v}\bar{w}} \geqslant 0, 
	\forall \underline{v}\in {\cal N}^{-}, \bar{w}\in {\cal N}^{+}. \label{subeq:o_vw_e.eta_e}\\
	& 0 \leqslant o^{(w)[p]}_{\bar{w}\bar{k}}
	\perp 
	(\sum_{k\in \mathcal{W},k \neq w} y^{(w)[p]}_{\Bar{w}\Bar{k}}+\sum_{k\in \mathcal{W}} y^{(w)[p]}_{\Bar{w}\underline{k}}) \cdot 
	o^{(w)[p]}_{\bar{w}\bar{k}}
    - y^{(w)[p]}_{\bar{w}\bar{k}}
	+ \eta^{(w)[p]-}_{\bar{w}\bar{k}} 
	\geqslant 0, 
	\forall \bar{k}, \bar{w}\in {\cal N}^{+}(p), k\neq w\in {\cal W}, \label{subeq:prop_passenger_1_1}\\
	& 0 \leqslant \eta^{(w)[p]}_{\bar{w}\bar{k}} \perp 1-o^{(w)[p]}_{\bar{w}\bar{k}} \geqslant 0, 
	\forall \bar{k}, \bar{w}\in {\cal N}^{+}, k\neq w\in {\cal W}, \label{subeq:prop_passenger_2_1}\\
	& 0 \leqslant o^{(w)[p]}_{\bar{w}\underline{k}}
	\perp 
	(\sum_{k\in \mathcal{W},k \neq w} y^{(w)[p]}_{\Bar{w}\bar{k}}+\sum_{k\in \mathcal{W}} y^{(w)[p]}_{\Bar{w}\underline{k}}) \cdot 
	o^{(w)[p]}_{\bar{w}\underline{k}}
    - y^{(w)[p]}_{\bar{w}\underline{k}}
	+ \eta^{(w)[p]-}_{\bar{w}\underline{k}} 
	\geqslant 0, 
	\forall \bar{w}\in {\cal N}^{+}(p), \bar{k} \in {\cal N}^{-}(p), \label{subeq:prop_passenger_3_1}\\
	& 0 \leqslant \eta^{(w)[p]}_{\bar{w}\underline{k}} \perp 1-o^{(w)[p]}_{\bar{w}\underline{k}} \geqslant 0, 
	\forall \bar{w}\in {\cal N}^{+}(p), \bar{k} \in {\cal N}^{-}(p).\label{subeq:prop_passenger_4_1}
\end{align}
\end{subequations}

\section{Equilibrium properties}
\label{sec:sol}

Summarizing the three modules specified in Sections~\ref{sec:1_OD}-\ref{sec:3_mode}, 
including 
[\textbf{M1.1}.\mbox{NCP-VehDispatch}] (corresponding to Equs.~\ref{subeq:NCP_M1.1}), 
[\textbf{M1.2}.\mbox{NCP-VehPassMatch}] (corresponding to Equs.~\ref{eq:opt_y_ncp}), 
[\textbf{M2}.\mbox{NCP-VehRoute}] (corresponding to Equs.~\ref{eq:NCP_vehRoute}), 
[\textbf{M3}.\mbox{NCP-CustChoice}] (corresponding to Equs.~\ref{eq:NCP_CustChoice_M3}). 
we have a system of coupled NCPs, denoted as [\textbf{All}.\mbox{NCP}\mbox{-ESys}], that describe the equilibrium state where e-platform optimizes vehicle dispatching and vehicle-passenger matching, customers select travel modes, and vehicles select best routes.


We will introduce two propositions to prove the existence of equilibrium \citep{ban2019general}. Proposition \ref{prop:bounded} demonstrates conditions regarding the boundedness of feasible set in [\textbf{All}.\mbox{NCP}\mbox{-ESys}]. In Proposition \ref{prop:VI_penalty}, we first construct a variational inequality (VI) with an unbounded feasible set based on [\textbf{All}.\mbox{NCP}\mbox{-ESys}], and then demonstrate conditions for the solution existence of the proposed VI. The proof of Proposition \ref{prop:bounded} and \ref{prop:VI_penalty} is detailed in \ref{sec:appendix_exist}.

\begin{prop}\label{prop:bounded}
If $\exists \ \zeta>1, s.t. \ \sum_{k,k\neq w} z^{[m]}_{\bar{k}\bar{w}}+\sum_{k,k\neq w} z^{[m]}_{\bar{w}\bar{k}} \leqslant \zeta \cdot  q_w^{[m]}, \forall w \in \mathcal{W}, m \in \mathcal{M}$ and the travel time function on a road network $t_{ij}(\mathbf{x})> 0$ is continuous and monotone, then $x_{ij}, (i,j) \in {\cal L}_{net}$ and $\tau_i^d, i \in {\cal N}_{net}, d \in {\cal N}$ are bounded. 
\end{prop}

\begin{prop}\label{prop:VI_penalty}
If $\exists \eta>0,\ s.t.\ \forall \mathbf{q}\geqslant 0$ satisfying (\ref{subequ:poly_3}) and $\forall \mathbf{\tau}, \ \exists \ \mathbf{z}, \ s.t. \ (\mathbf{z},\mathbf{q}, \mathbf{\tau})$ satisfies (\ref{subequ:poly_1}, \ref{subequ:poly_4}) and $S-\sum_{(u,v)\in {\cal L}} z_{uv}t_{uv}\geqslant \eta$, a solution of $[\textbf{All}.\mbox{NCP}\mbox{-ESys}]$ under assumptions in Proposition \ref{prop:bounded} is then recovered by a variational inequality $V(\mathbf{H}^{\epsilon},\mathcal{F})$ when $\epsilon \rightarrow \infty$. The VI is constructed as: \begin{align}
    & [V(\mathbf{H}^{\epsilon},\mathcal{F})] \nonumber\\
    & \mathbf{H}^{\epsilon}(F) =\begin{Bmatrix}
    & C_{uv}+\epsilon_z t_{uv}\left[ \sum_{(u,v)}z_{uv}t_{uv} -S\right]_{+}, \forall (u,v) \in \mathcal{L} \nonumber \\
    &C^{(w)}_{uv} +\epsilon_y \left[ \sum_{v} y^{(w)}_{uv}-\sum_{w}\delta^{o}_{u\Bar{w}_{[m]}}q_w^{[m]} \right]_+ ,\forall w \in \mathcal{W}, (u,v) \in {\cal L}^{+}(m) \cup {\cal L}^{-}(m) \nonumber \\
    & U_w^{[m]}, \forall w \in \mathcal{W}, m \in \mathcal{M} \nonumber \\
    & t_{ij}(x)-(\tau_i^d-\tau_j^d), \forall (i,j) \in \mathcal{L}_{(net)}, d \in \mathcal{N} \nonumber \\
    & \sum_{j:(i,j)\in {\cal L}_{(net)}} x^{d}_{ij} - \sum_{k:(k,i)\in {\cal L}_{(net)}} x^{d}_{ki}  - q_{(i,d)}, \forall i \in \mathcal{N}_{net}, d \in \mathcal{N} \nonumber \\
    & \sum_{u} z_{uv} \cdot o_{uv}- z_{uv}, \forall (u,v) \in \mathcal{L},  u \in  {\cal N}^{-} \nonumber \\
    & \sum_{u} y^{(w)}_{uv} \cdot o^{(w)}_{uv}- y^{(w)}_{uv} , 
	\forall w \in {\cal W}, (u,v) \in \mathcal{L}, u \in {\cal N}^{+}(m) \nonumber
    \end{Bmatrix}
\end{align}
$\mathcal{F}$ is the feasible set of $F=(\mathbf{z,y,q,x,\tau,o})$ and $(\mathbf{z,y,q,o})\geqslant 0$ satisfy:
\begin{align}
    &\sum_{v'} z_{v'u}=\sum_{v} z_{uv}, \forall u \in \mathcal{N}  \label{subequ:poly_1} \\
    & \sum_{v'} y^{(w)}_{v'u}-\sum_{v} y^{(w)}_{uv}=0, \forall w \in \mathcal{W}, u \in {\cal N}^{+}(m) \cup {\cal N}^{-}(m) \label{subequ:poly_2} \\
    & \sum_{m\in {\cal M}} q^{[m]}_{w} = q_{w}, \forall w \in \mathcal{W} \label{subequ:poly_3} \\
    & 	\sum_{v'} z_{v'u}  \geqslant \sum_{w}\delta^{o}_{u\Bar{w}_{[m]}}q_w^{[m]}
	 , \forall u \in {\cal N}^+(m)  \label{subequ:poly_4} \\
    &y^{w}_{uv} \leqslant z_{uv}, \forall w\in {\cal W},
	(u,v) \in {\cal L}^{+}(m) \cup {\cal L}^{-}(m) \label{subequ:poly_5}  \\ 
    &o_{uv} \leqslant 1, \forall (u,v) \in \mathcal{L}, u \in {\cal N}^{-} \label{subequ:poly_6} \\
     &o^{(w)}_{uv} \leqslant 1, \forall w \in {\cal W}, (u,v) \in \mathcal{L}, u \in {\cal N}^{+}(m). \label{subequ:poly_7}
\end{align}
\end{prop}

\begin{rem}
It is challenging to prove the uniqueness of the equilibrium in [\textbf{All}.\mbox{NCP}\mbox{-ESys}] due to the nonlinearity in the product of decision variables. There are many ways dealing with multiple equilibria. Characterizing a solution set  \citep{di2013boundedlyTRB, di2016boundedly} may not work when there exist nonlinearity in the NCP. Another way is to utilize a bi-level model \citep{ban2009risk,di2014boundedly,di2016second,chen2021RUE} where the upper level is to select a risk-averse or risk-neutral equilibrium and the lower level is [\textbf{All}.\mbox{NCP}\mbox{-ESys}], which will be left for future research.
\end{rem}

\section{Numerical examples}\label{sec:exp}


In this section, we will apply [\textbf{All}.\mbox{NCP-ESys}] to the small network (3-node network 2-OD pair) and the Sioux Falls network to demonstrate our model. The numerical results on the small network are in \ref{append:parameters_small}.

The Sioux Falls network includes $24$ nodes and $76$ links. The topology of the network, node indexes, link indexes, and the link performance functions are downloadable from the github (\url{https://github.com/bstabler/TransportationNetworks}). Parameter values for Sioux Falls network are in \ref{append:sioux}. To solve the coupled system - [\textbf{All}.\mbox{NCP}\mbox{-ESys}], we first generate a layered OD graph in Python3.6 (Algorithm \ref{alg:OD_generation_1}) based on OD pairs on road networks. We then combine all the three modules as a large system of NCPs and solve it in GAMS \citep{ferrisgams}.

We would like to investigate three scenarios shown in Table~(\ref{tab:SF_scenario}). 
For Scenario~1 (2 ODs), we investigate two cases: travel time is congestion-independent (Case~1) and congestion-dependent (Case~2). 
For Scenario~2 (6 ODs), we also focus on two cases: search friction is small (Case~1) or large (Case~2). For Scenario~3 (more ODs), we look into the computation time and problem scale regarding the number of OD pairs. 


\begin{table}[H]\centering
    \fontsize{9}{0}\selectfont
	\caption{Three scenarios: 2, 6 \& more ODs}\label{tab:SF_scenario}
	\begin{tabular}{c|c|c|c|p{6cm}||p{3cm}}
		\hline
		Scenario & Origin & Destination  & Case & Description 
		& Parameter \\ \hline \hline
		\multirow{2}{*}{2 ODs} & \multirow{2}{*}{$\{1,2 \}$} & \multirow{2}{*}{$\{20 \}$}  & 1 & Link capacity carries the original values & \emph{CAPACITY} 
		\\ 
		& & & 2& Link capacity discounted by $50$ & \emph{CAPACITY}$/50$\\ \cline{1-6} 
	\multirow{2}{*}{6 ODs} & \multirow{2}{*}{$\{1,2,3 \}$} & \multirow{2}{*}{$\{18,20 \}$}  & 1 & Small search friction &  $\beta_{(se)}=1$ \\ 
	& & & 2 & Large search friction & $\beta_{(se)}=10$
	\\ \hline
	more ODs & - & - & - & Computation time \& Problem Scale & - \\ \hline
	\end{tabular}
\end{table}

\subsection{2 ODs}	

For 2 OD pairs (1,20) (indexed as OD~1) and (2,20) (indexed as OD~2), we aim to compare 2 cases: Case~1 where the network links have large capacities and Case~2 where the network links have smaller capacities and thus link travel times are more sensitive to link traffic flow. 
We deliberately assign different OD demands to these two OD pairs (i.e., $q_{(1,20)}=500, q_{(2,20)}=200$), so that readers can easily distinguish quantities of these two ODs. 
In addition, it represents an asymmetric demand scenario when OD demands of two ODs are different and matching all the e-pooling orders would not be guaranteed. 

In both cases, we vary the fare of the e-pooling service while fixing that of the e-solo service. 
Thus, we define the pooling-solo fare ratio as the ratio of e-pooling relative to e-solo pricing. 
The higher, the closer the e-pooling fare to the e-solo fare,  the fewer orders select e-pooling, 
and the more vehicles are needed to move the demands.

\begin{figure}[H]
	\centering
	\subfloat[OD demand fulfilled by e-pooling]{\includegraphics[scale=.30]{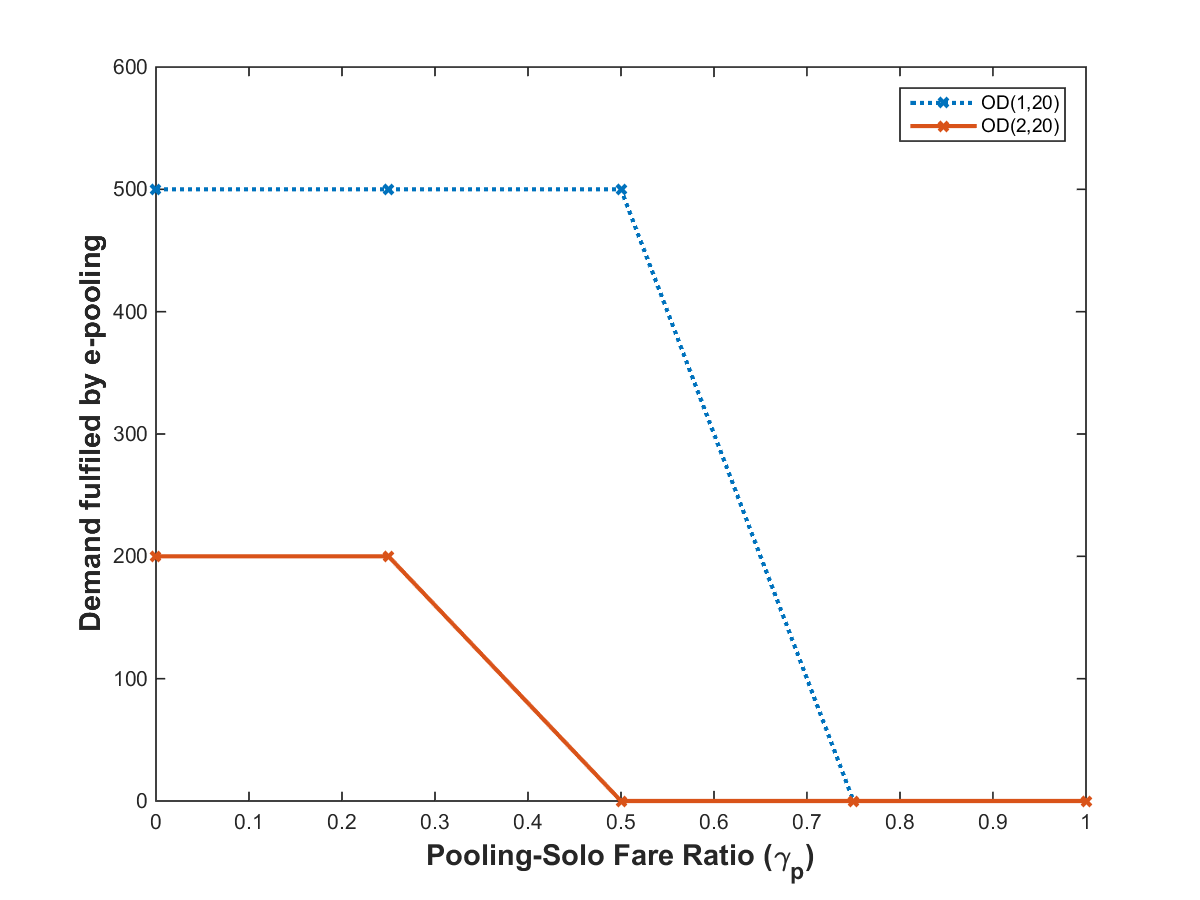}\label{subfig:SF_coeff_dm}}~
	\subfloat[Demand proportion by mode]{\includegraphics[scale=.30]{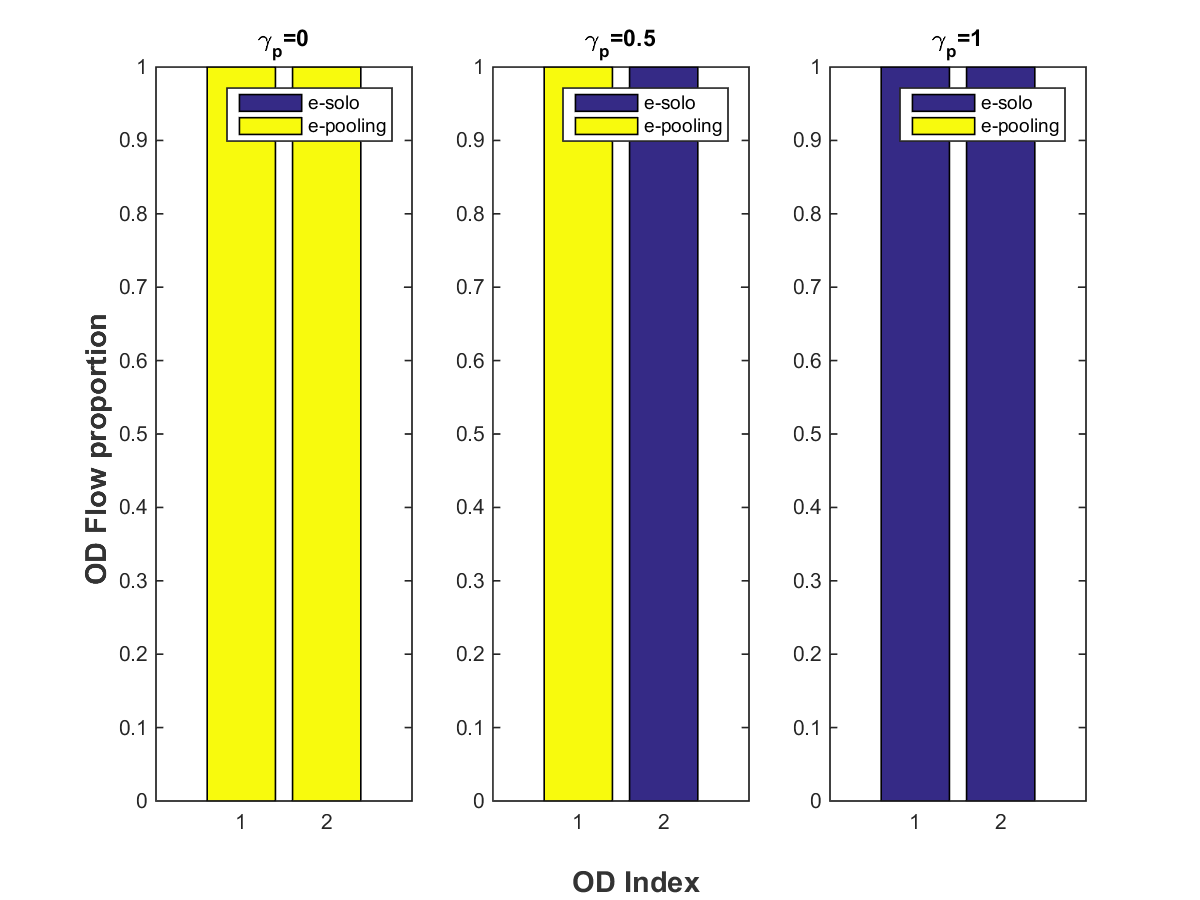}\label{subfig:SF_bar_dm}}
	\caption{OD demands for Scenario~1 Case~1 on Sioux Falls}
	\label{fig:SF_2OD_cap1_od}
\end{figure}

\paragraph{Case 1}

We plot the OD demand in Fig.~\ref{fig:SF_2OD_cap1_od}. 
In Figs.~\ref{subfig:SF_coeff_dm}, x-axis is the e-pooling fare coefficient increasing from 0 to 1, 
while y-axis represents the OD demands fulfilled by e-pooling for OD (1,20) in blue dashed line and OD (2,20) in red line, respectively.
Note that in the fixed total demand case, a decrease in the e-pooling demand indicates an increase in the e-solo demand. That is why we only plot the e-pooling demand. 
As the pooling-solo fare ratio increases, the number of e-pooling orders decrease to zero step-size. 
All orders are e-pooling when its fare is lower than 75\% of e-solo for OD (1,20) and 25\% of e-solo for OD (2,20), respectively.  
That being said, if the e-hailing company wants to promote e-pooling, the e-pooling fare discount should be higher than 25\% for OD (1,20) or 75\% for OD (2,20). 
Note that the e-pooling fare for OD pair (2,20) is much more sensitive to a discount than that for OD pair (1,20). Thus targeting the origin node 2 is the key. An option is to offer differential discounts to orders originating from nodes 1 and 2. 
In Fig.~\ref{subfig:SF_bar_dm}, the y-axis indicates the OD demand proportion between e-pooling (in yellow bar) and e-solo (in blue bar) for OD (1,20) (left) and OD (2,20) (right), respectively. 
Across the x-axis is demands sampled at three e-pooling fare ratios of 0,0.5,1, meaning that the e-pooling fare is 0, half of e-solo, and same as e-solo, respectively. 
When the ratio is zero, it represents an extreme case that e-pooling service is free. Thus all orders select e-pooling. 
When the ratio is 0.5, it indicates that e-pooling is half price of e-solo service. All orders from OD (2,20) abandon e-pooling and switch to e-solo entirely. This is because of the high inconvenience cost to pool rides. 
For OD (1,20), all the orders still select e-pooling due to the half fare despite of a higher pooling cost.
Undoubtedly, When the ratio is one, in other words, when e-pooling has no difference from e-solo, all customers switch to e-solo due to high inconvenience cost of sharing rides with others.

\begin{figure}[H]
	\centering
	\subfloat[Modal disutility]{\includegraphics[scale=.29]{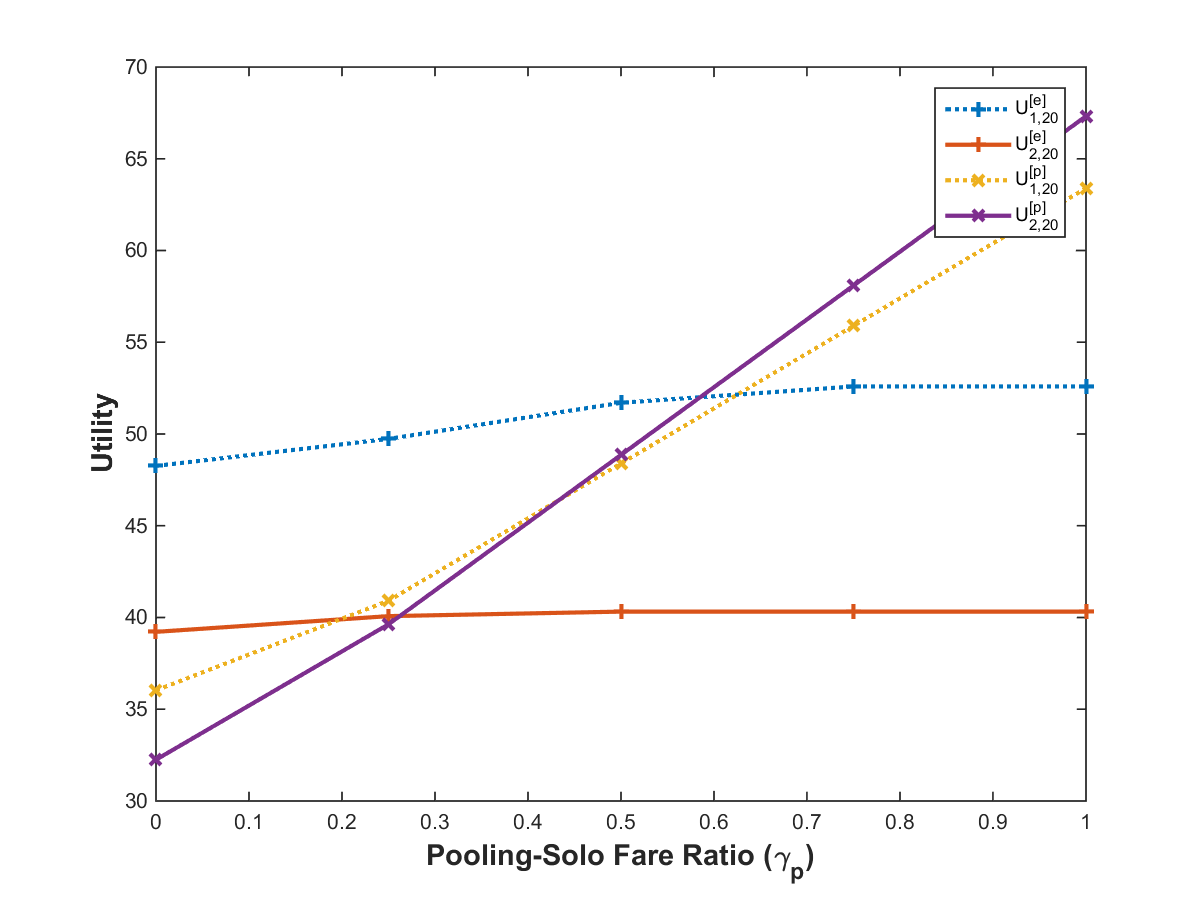}\label{subfig:SF_coeff_u}}~\hspace{-6mm}
	\subfloat[Fare by OD and mode]{\includegraphics[scale=.29]{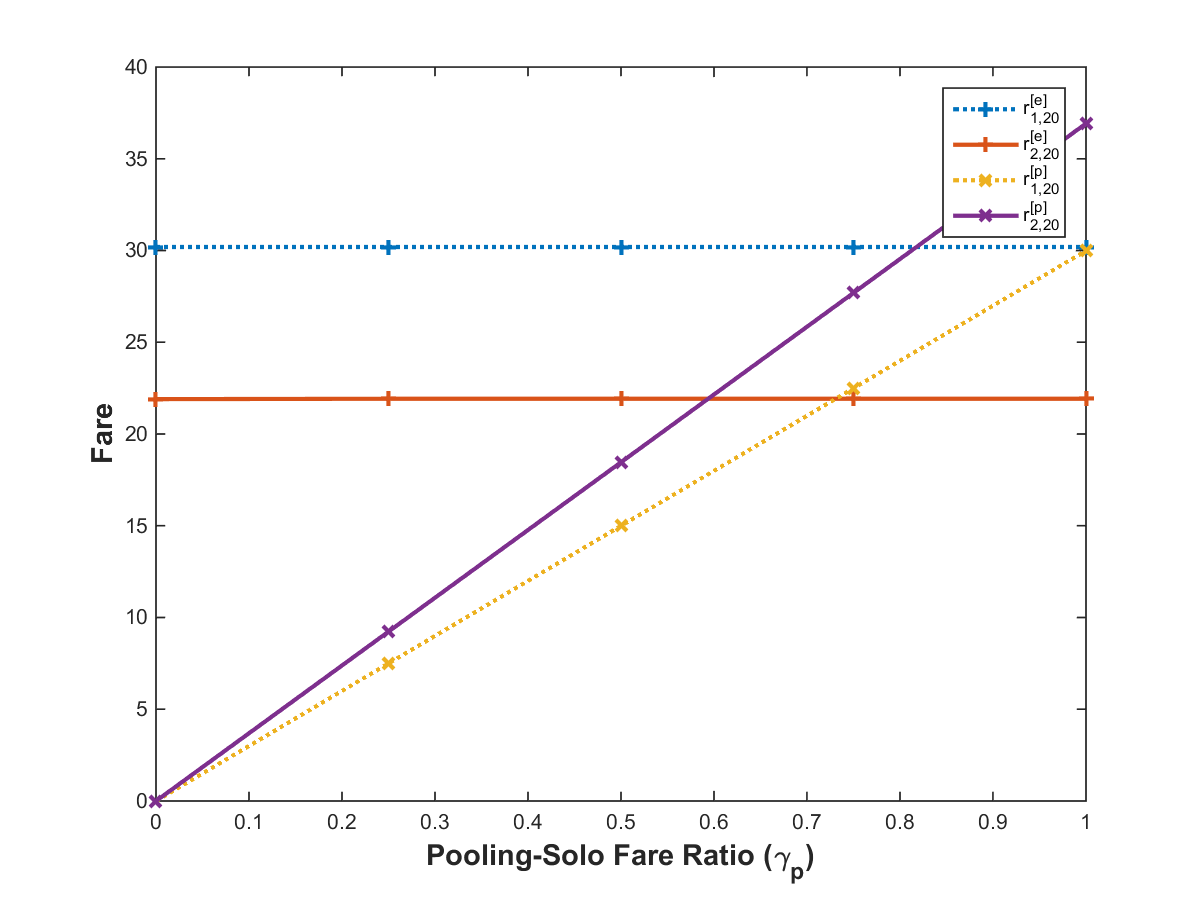}\label{subfig:SF_coeff_fare}}~\hspace{-6mm}
   \subfloat[Vehicle surplus \\(supply-demand surplus cost)]{\includegraphics[scale=.29]{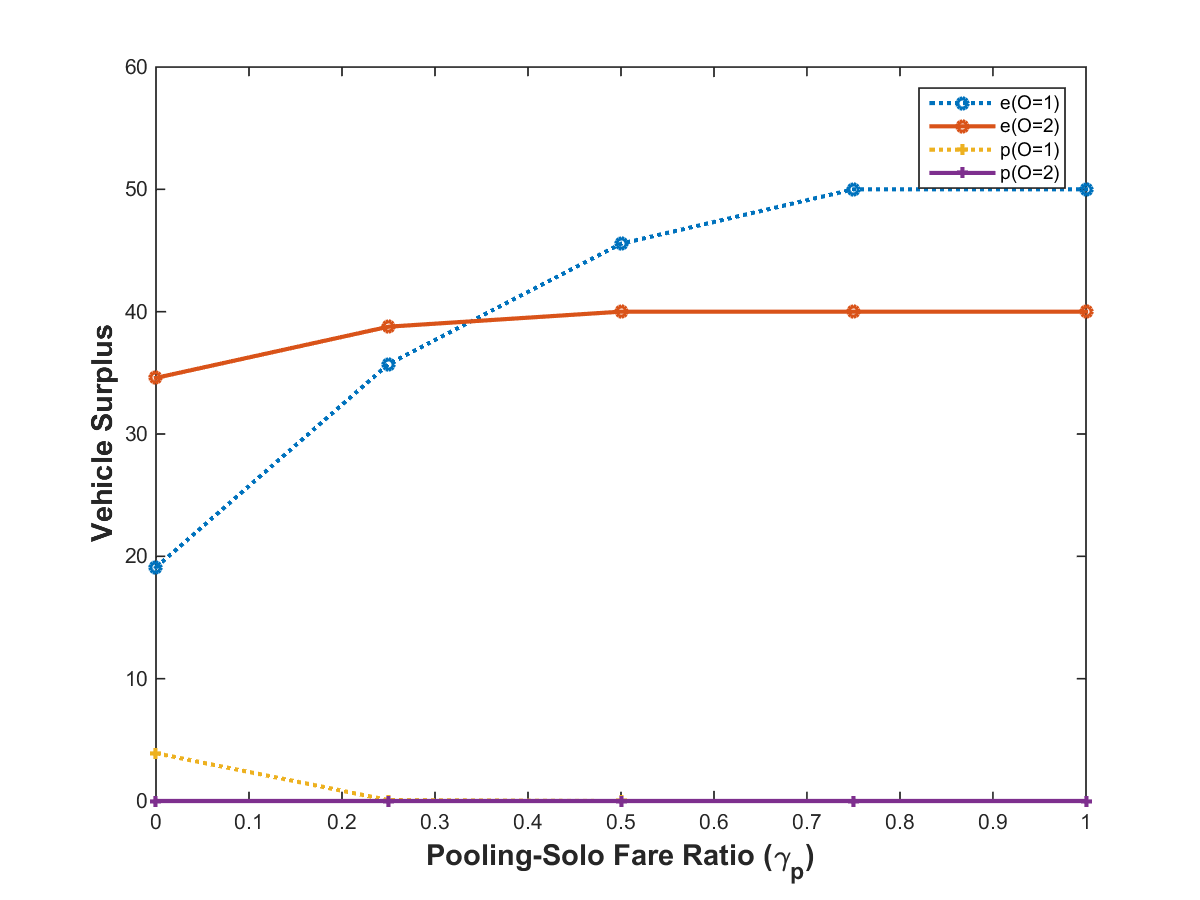}\label{subfig:SF_coeff_surplus}}
	\caption{Modal disutilities for Scenario~1 Case~1 on Sioux Falls}
    \label{fig:SF_2OD_cap1_u}
\end{figure}

\begin{figure}[H]
	\centering
	\subfloat[Performance measures]{\includegraphics[scale=.30]{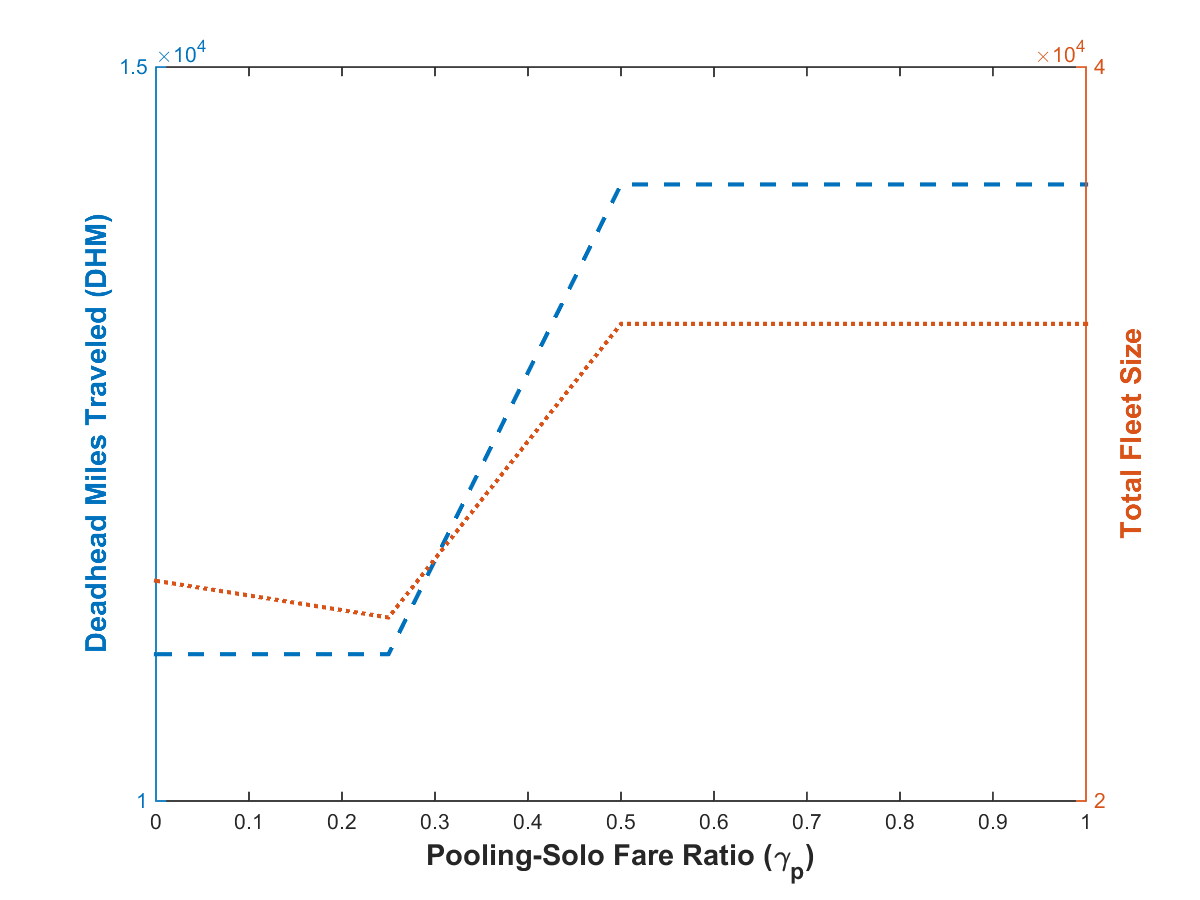}\label{subfig:SF_coeff_perf}}~
	\subfloat[Minimum travel time by OD]{\includegraphics[scale=.30]{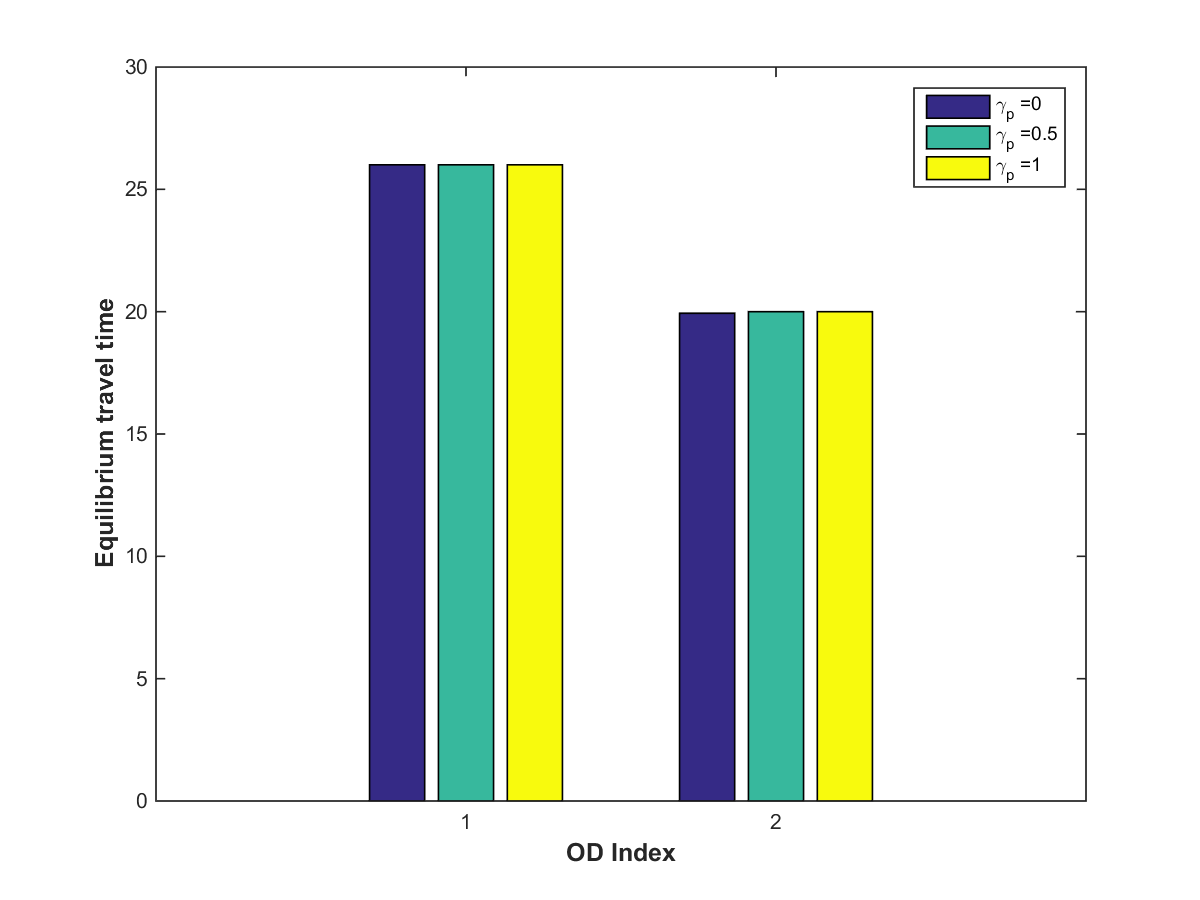}\label{subfig:SF_bar_minc}}
	\caption{Performance measures for Scenario~1 Case~1 on Sioux Falls}
    \label{fig:SF_2OD_cap1_perf}
\end{figure}

To compare the total and itemized disutility of two travel modes, we plot the modal disutility, fare, and supply-demand surplus cost 
in Figs.~\ref{fig:SF_2OD_cap1_u}.
In Fig.~\ref{subfig:SF_coeff_u}, the x-axis is the pooling-solo fare ratio, and the y-axis is the customer disutility. 
The disutility of selecting e-solo for OD (1,20) is represented by a dotted blue line with plus markers, 
that of selecting e-pooling for OD (1,20) by a dotted orange line with cross markers,
that of selecting e-solo for OD (2,20) by a solid red line with plus markers, 
and that of selecting e-pooling for OD (2,20) in a solid purple line with cross markers. 
As the pooling-solo fare ratio increases, disutilities of travel modes for both OD pairs increase generally. 
In particular, the disutilities of e-pooling increase faster than that of e-solo, because the pooling-solo fare ratio impacts the e-pooling fare directly. 
The disutility of e-solo also rises because of the indirect effect of the increasing number of vehicles on roads.
For OD (1,20), the e-pooling disutility is lower than the e-solo one when the pooling-solo fare ratio is less than 0.75. 
Similarly, for OD (2,20), the e-pooling disutility is lower than that of e-solo when the ratio is less than 0.25. 
These trends are consistent with those in modal demands analyzed above. 
Fig.~\ref{subfig:SF_coeff_fare} demonstrates fare for each mode.
For OD (1,20), the e-solo fare is indicated in a dotted blue line with plus markers and that of e-pooling is in a dotted orange line with cross markers, respectively. 
For OD (2,20), the e-solo fare is in a solid red line with plus markers and e-pooling is shown in a solid purple line with cross markers, respectively. 
As the pooling-solo fare ratio grows, the fares of e-pooling for both OD pairs go up, 
while those of e-solo remain constant. 
In other words, the e-pooling service becomes more and more expensive as the ratio increases and loses its competitiveness over the e-solo service, which repels more passengers to e-solo. In Fig.~\ref{subfig:SF_coeff_surplus}, we examine the extra waiting time of e-pooling riders to be matched with drivers, represented by the supply-demand surplus that constitutes part of the mode choice disutility. For e-solo, the surplus costs at origins 1 and 2 are represented by blue dotted and red solid lines with circular markers, respectively. 
The costs increase as the pooling-solo fare ratio increases, and that at node 1 increases at a faster rate than that at node 2. 
This is because more passengers select e-solo, and the e-solo supply is less than the demand, and thus the ``surge price" increases for the e-solo service.
For e-pooling, the surplus costs at origins 1 and 2 are represented by orange dotted and purple solid lines with cross markers, respectively. 
The surplus cost at origin 2 is always zero, meaning that the e-pooling vehicle size is always larger than the demand. 
When the pooling-solo fare ratio is lower than 0.25, surplus cost is non-zero and is reduced to zero when the ratio bypasses 0.25. 
This is because of the drop in the e-pooling demand relative to the supply.

To examine the systematic impact of pricing, we inspect DHM and total fleet size in Fig.~\ref{subfig:SF_coeff_perf} and equilibrium travel time in 
Fig.~\ref{subfig:SF_bar_minc}.
In Fig.~\ref{subfig:SF_coeff_perf}, 
the dashed blue line with plus markers represents DHM (the left y-axis) and the dotted red line represents total fleet size (the right y-axis). 
Both increase step-wise, the first one at 0.25 and the second one at 0.75, as the pooling-solo fare ratio increases. 
This is because the passengers' modal switch from e-pooling to e-solo. 
The more vehicles on road, the higher DHM and the larger fleet size are. 
In Fig.~\ref{subfig:SF_bar_minc}, the x-axis is two OD indices and the y-axis is the equilibrium travel time. Each bar represents the equilibrium travel time at one selected pooling-solo fare ratio. 
The equilibrium travel time reflects the traffic congestion level. All scenarios with three fare ratios share the same equilibrium travel time for the same OD. 
That means that even when every passenger uses e-solo service, it does not worsen the network traffic condition and traffic still runs at free flow. 
That leads us to move on to Case 2 when all the link capacities are compromised by a factor of 50, which makes modal utilities more sensitive to traffic congestion and travel time.

	

\paragraph{Case 2}

We discount the capacity of each link by 50 and see how a congestion prone road network would impact people's mode choice and performances. 
A congestion prone road network refers to one that has limited capacity and is thus sensitive to congestion. In other words, a few additional vehicles on a link can trigger a much higher delay. 
Our hypothesis is that due to the longer travel time on roads, passengers are more sensitive to traffic congestion and are thus prone to pool rides. 

\begin{figure}[H]
	\centering
	\subfloat[OD demand]{\includegraphics[scale=.30]{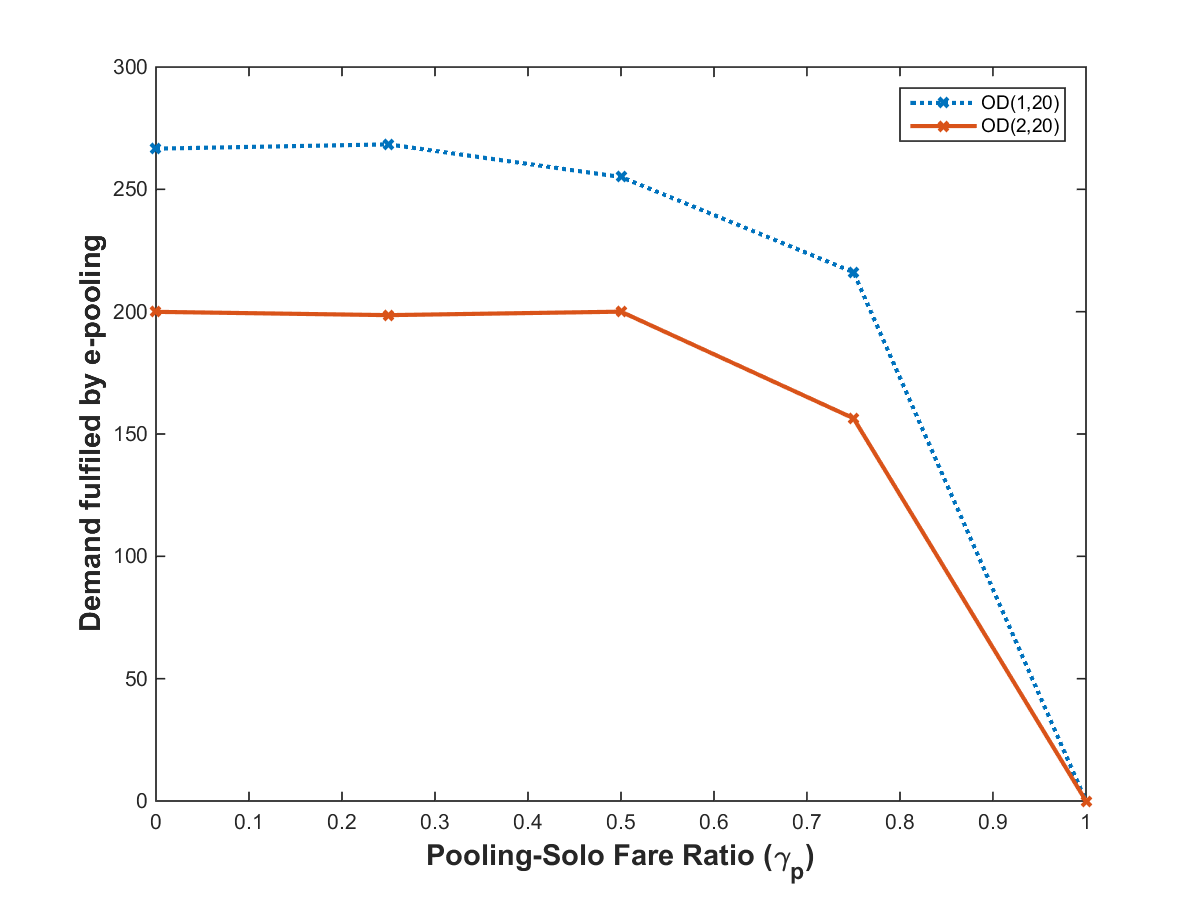}\label{subfig:SF_coeff_dm50}}~
	\subfloat[OD demand proportion by mode]{\includegraphics[scale=.30]{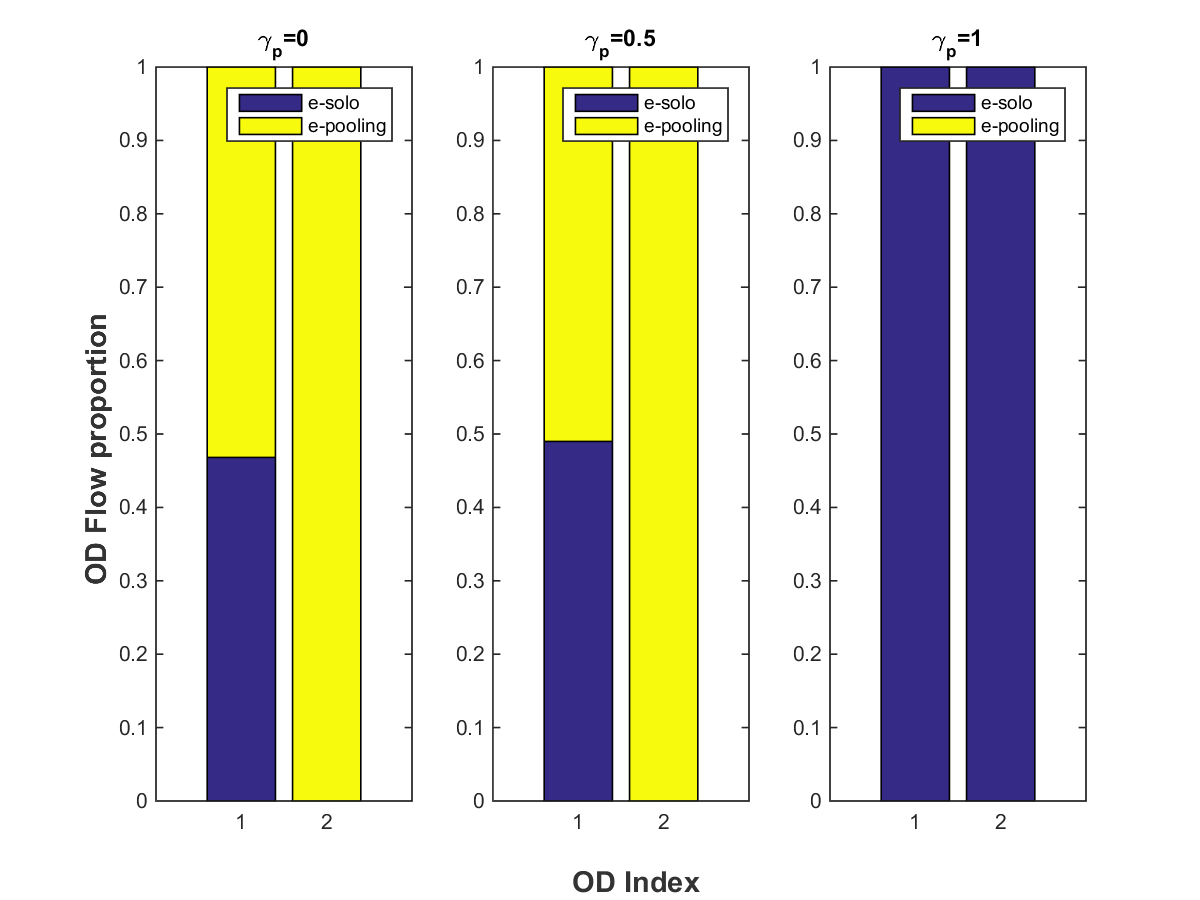}\label{subfig:SF_bar_dm50}}
	\caption{OD demands for Scenario~1 Case~2 on Sioux Falls}
	\label{fig:SF_2OD_cap50_od}
\end{figure}

\begin{figure}[H]
	\centering
	\subfloat[Modal disutility]{\includegraphics[scale=.29]{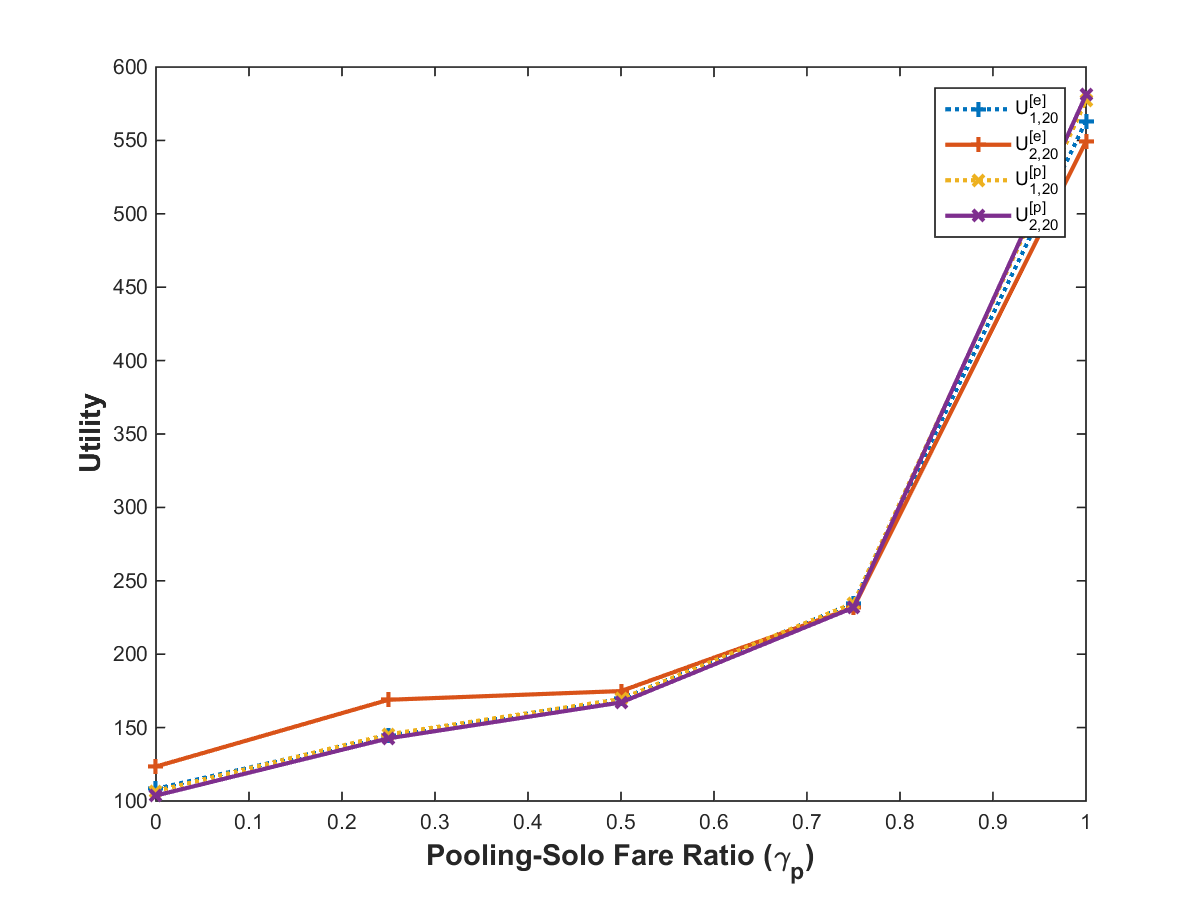}\label{subfig:SF_coeff_u50}}~\hspace{-6mm}
	\subfloat[Fare by OD and mode]{\includegraphics[scale=.29]{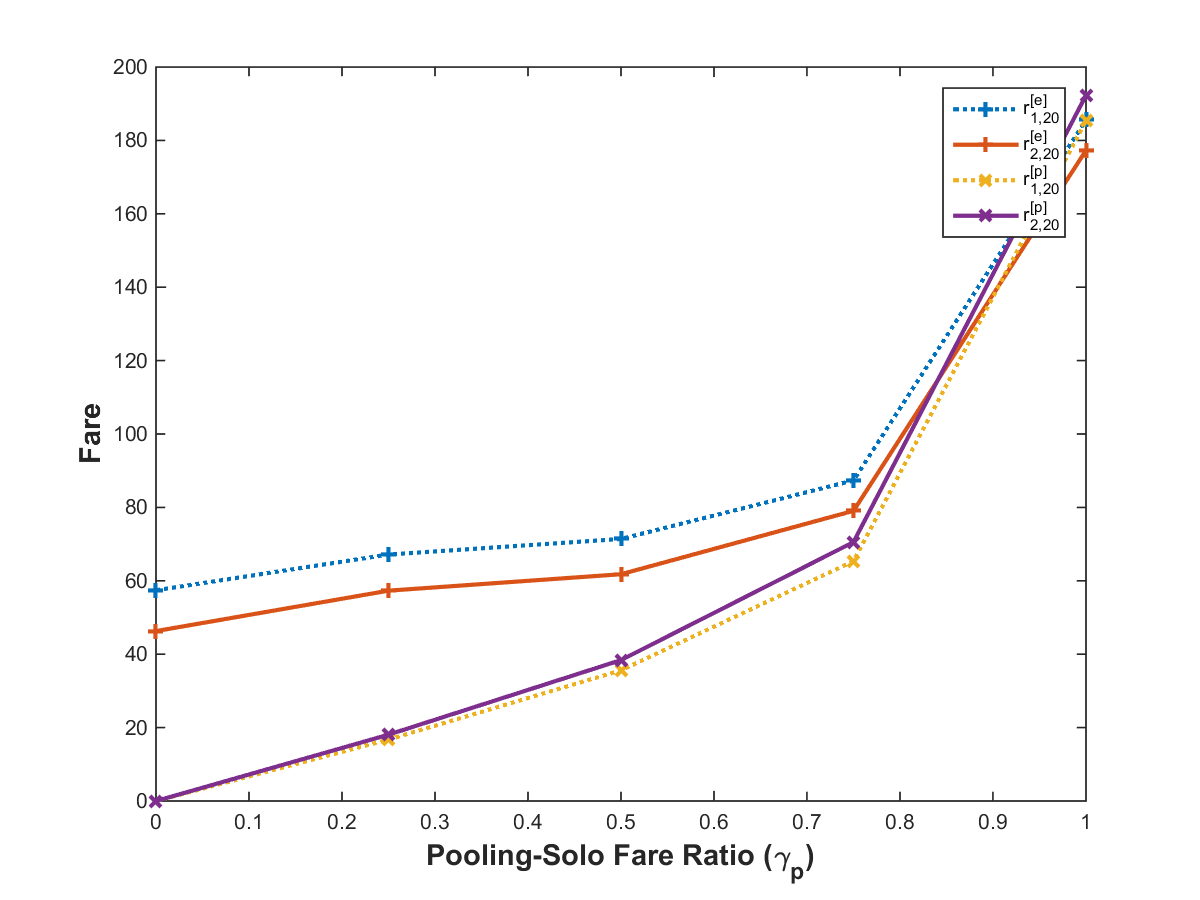}\label{subfig:SF_coeff_fare50}}~\hspace{-6mm}
    \subfloat[Vehicle surplus (supply-demand surplus cost)]{\includegraphics[scale=.29]{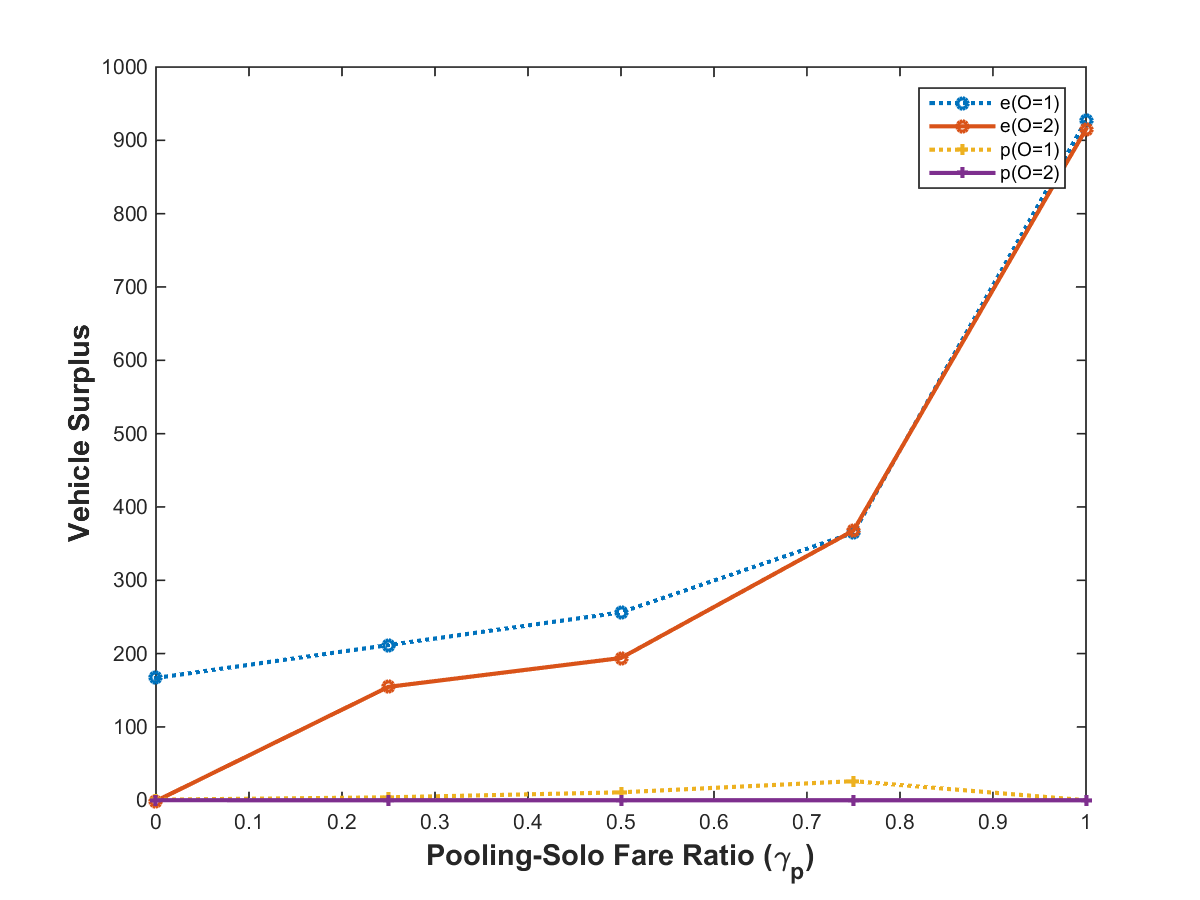}\label{subfig:SF_coeff_surplus50}}
	\caption{Modal disutilities for Scenario~1 Case~2 on Sioux Falls}
	\label{fig:SF_2OD_cap50_u}
\end{figure}

\begin{figure}[H]
	\centering
    \subfloat[Performance measures]{\includegraphics[scale=.30]{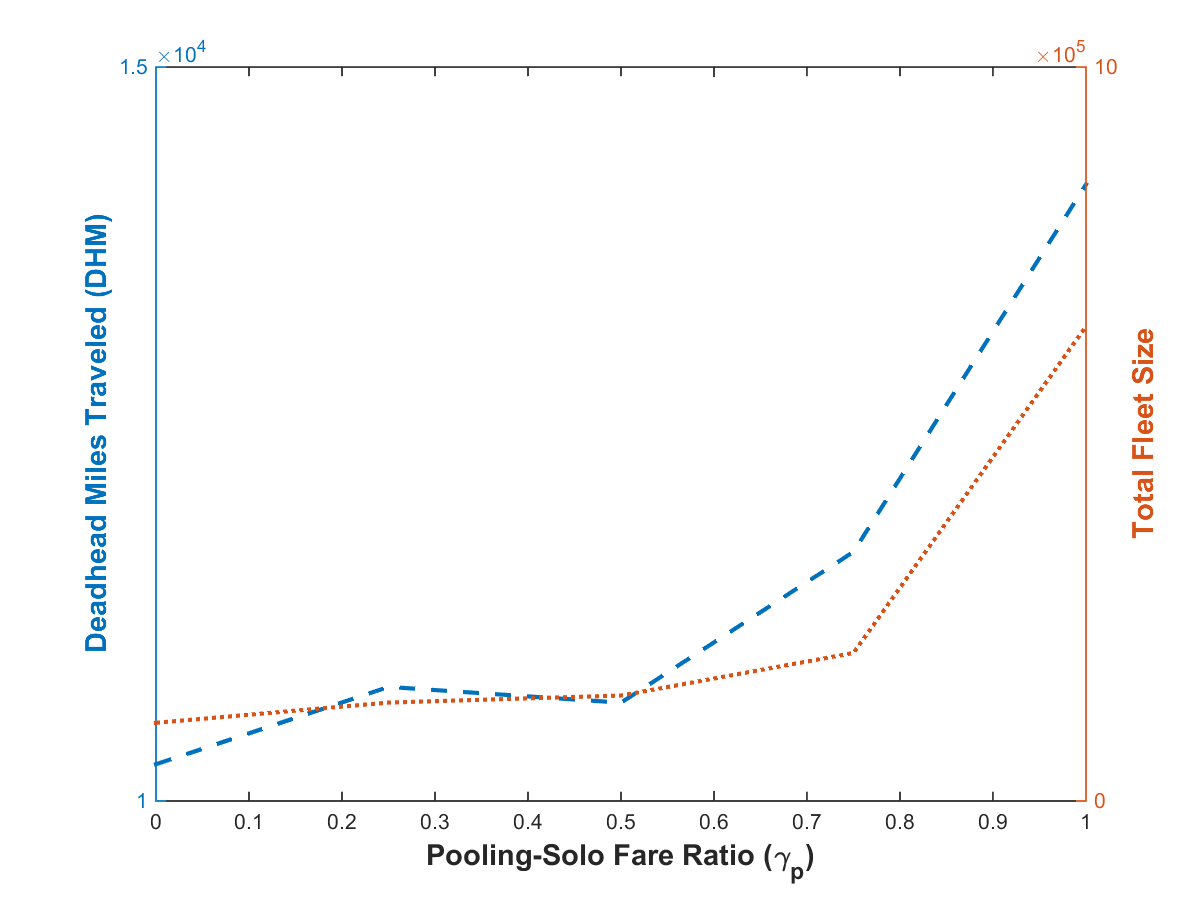}\label{subfig:SF_coeff_perf50}}~
	\subfloat[Minimum travel time by OD]{\includegraphics[scale=.30]{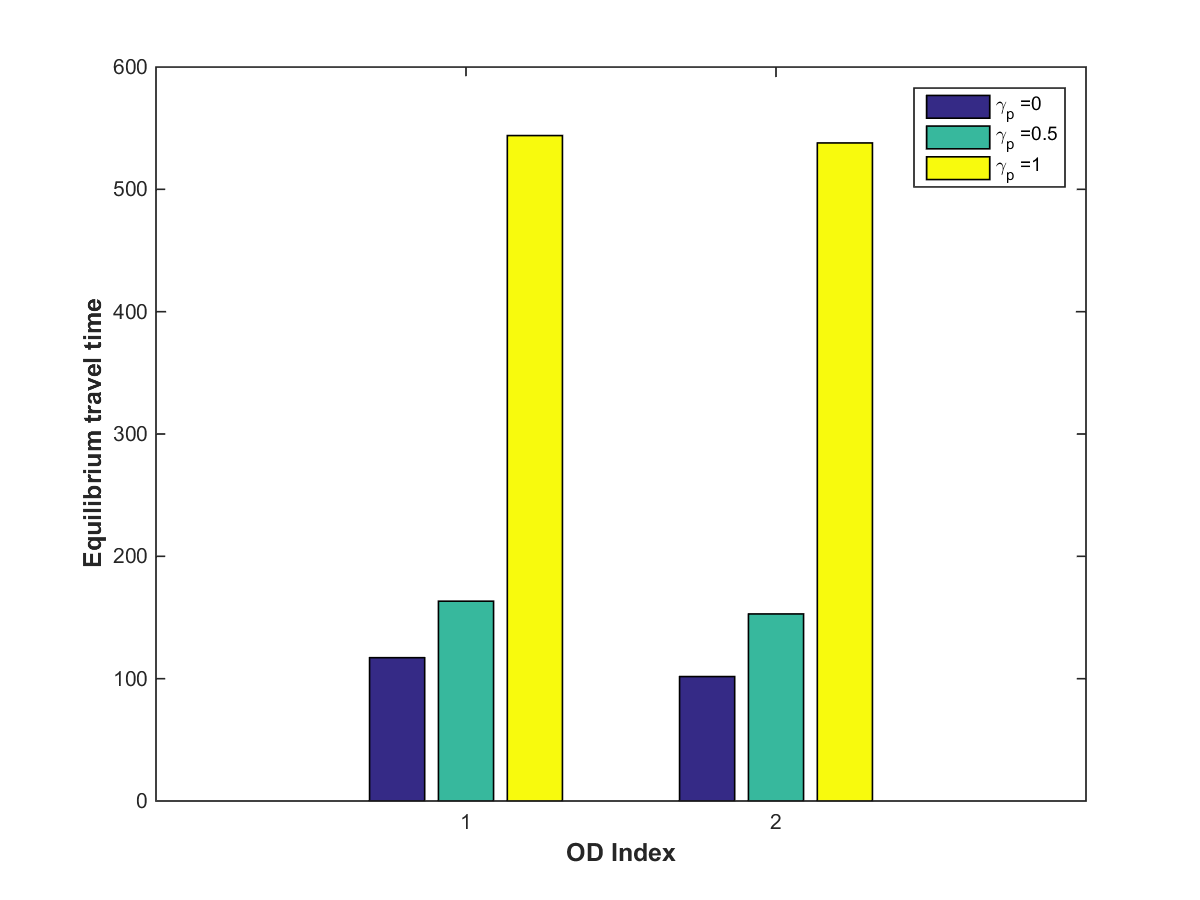}\label{subfig:SF_bar_minc50}}
	\caption{Performance measures for Scenario~1 Case~2 on Sioux Falls}
	\label{fig:SF_2OD_cap50_perf}
\end{figure}

The OD demands (in Fig.~\ref{fig:SF_2OD_cap50_od}), total and itemized disutility (in Fig.~\ref{fig:SF_2OD_cap50_u}), and performance measures (in Fig.~\ref{fig:SF_2OD_cap50_perf}) correspond to those presented in Case~1 sequentially. 
Fig.~\ref{fig:SF_2OD_cap50_od} corresponds to Fig.~\ref{fig:SF_2OD_cap1_od}. There are two major differences between Case 1 and Case 2: (1) When $\gamma_p \leqslant0.5$, the total demand fulfilled by the e-pooling services in Case ~2 (with congestion) is lower than that in Case ~1 (without congestion). This is because the travel time of an e-pooling driver heading to pick up the second passenger increases in a congested network, which leads to the decreasing usage of e-pooling services. (2)  When $\gamma_p \geqslant 0.5$, the curve of OD demands served by e-pooling in Case~2 decreases slower than that in Case~1. This is because the total travel disutility is more sensitive to traffic congestion and less on fare. The more people who select e-solo, the more severe the traffic congestion becomes.

Fig.~\ref{fig:SF_2OD_cap50_u} corresponds to Fig.~\ref{fig:SF_2OD_cap1_u}. 
In Fig.~\ref{subfig:SF_coeff_u50}, the utilities of both e-solo and e-pooling increase as the pooling-to-solo fare ratio increases. So is fare in Fig.~\ref{subfig:SF_coeff_fare50}. Such a trend is in contrast to that in Case~1 (without congestion) when only pooling utilities increase. Compared to Case~1, the supply-demand surplus cost of e-solo service in Case~2 significantly increases (Fig.~\ref{subfig:SF_coeff_surplus50}) when $\gamma_p\geqslant 0.75$. This is because, as the pooling-to-solo fare ratio increases, more vehicles are dispatched to serve the growing e-solo orders and thus incur higher congestion cost and as a result, a relatively lower supply because vehicles have to spend more time on roads.

Fig.~\ref{fig:SF_2OD_cap50_perf} corresponds to Fig.~\ref{fig:SF_2OD_cap1_perf}. 
In Fig.~\ref{subfig:SF_coeff_perf50}, DHM and total fleet size increase faster as the pooling-to-solo fare ratio increases, compared to Case~1 (without congestion).
In Fig.~\ref{subfig:SF_bar_minc50}, equilibrium travel time increases as the pooling-to-solo fare ratio increases for each OD pair. In particular, the equilibrium travel time at the pooling-to-solo fare ratio of one is much higher than when the ratio is less one, which is caused by the increasing vehicles on roads.

\subsection{6 ODs}
For 6 OD pairs $(1,18)$, $(1,20)$, $(2,18)$, $(2,20)$, $(3,18)$ and $(3,20)$, we vary the weight of search friction $\beta_{(se)}$ in order to compare 2 cases: Case~1 with a small search friction coefficient $\beta_{(se)}=1$ and Case~2 with a large search friction coefficient $\beta_{(se)}=10$. The demands of all 6 OD pairs are the same as 300.

\paragraph{Case 1} 

\begin{figure}[H]
	\centering
    \includegraphics[scale=.41]{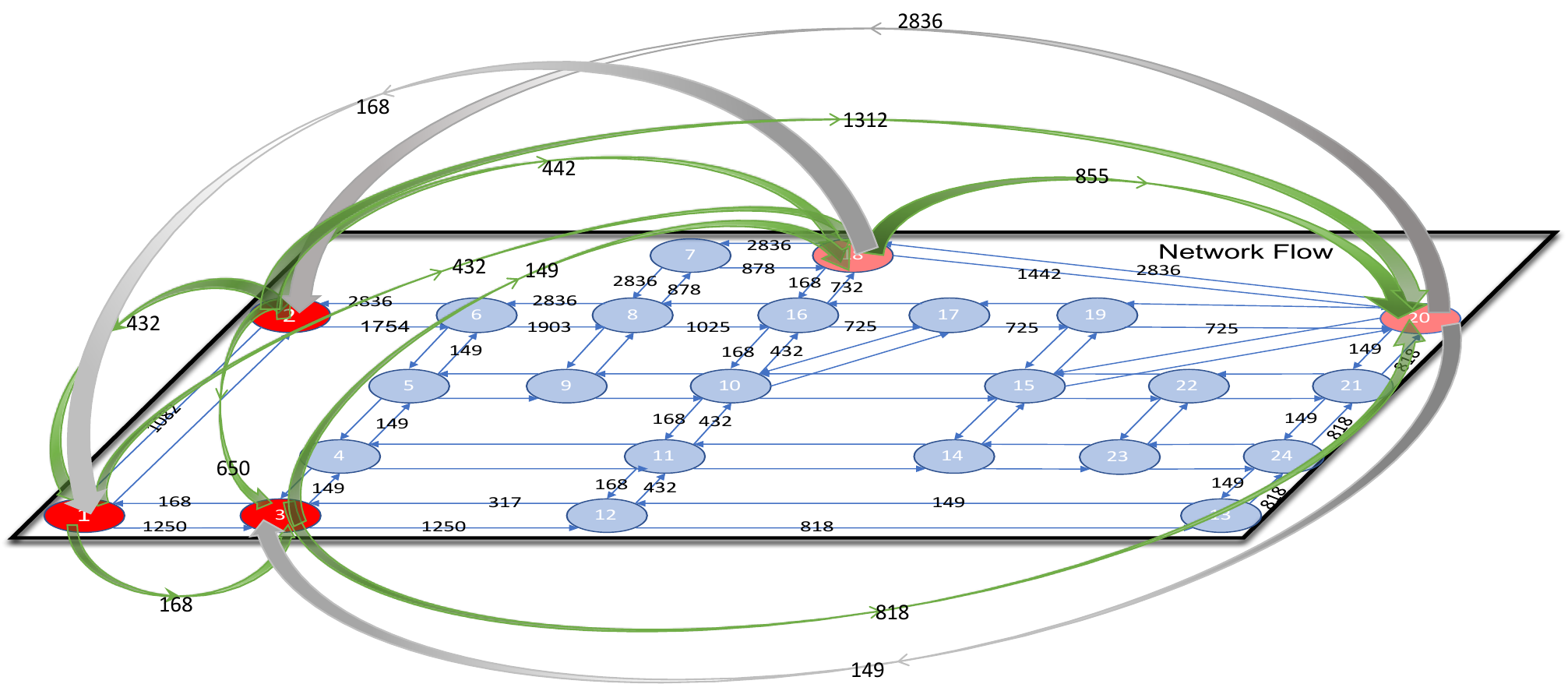}
	\caption{Traffic flow for Scenario~2 Case~1 on Sioux Falls}
	\label{fig:traffic_flow_sioux_fall_6OD}
\end{figure}

We plot the vehicle flow when the weight of search friction is $\beta_{(se)}=1$. In Fig.~\ref{fig:traffic_flow_sioux_fall_6OD}, the traffic flow on the road map (blue links) represents the vehicle flow on the Sioux Falls network. Origins $\{1,2,3\}$ are red solid dots and destinations $\{18,20\}$ are light red ones. Intermediate nodes on the Sioux Falls network are denoted by light blue dots. The 3D arrows (green and grey arrows) represent the vehicle OD flow in the OD graph. The green one is the vehicle OD flow dispatched along OD sequences. The grey one is the rebalancing flow from destinations to origins. Note that the vehicle OD flow in the OD graph (3D arrows) can be projected onto the Sioux Falls network. For example, in the OD graph, 432 vehicles are dispatched from origin 2 to origin 1 along the link $(2 \rightarrow 1)$. 650 vehicles are dispatched from origin 2 to origin 3 along the link $(2 \rightarrow 1 \rightarrow 3)$. Therefore, the vehicle flow on the link $(2 \rightarrow 1)$ is $432+650=1082$, denoted by the blue arrow from node $2$ to $1$ on the Sioux Falls network.

We inspect details of the vehicle OD flow and analyze the OD sequences.

\textbf{Vehicle Dispatch} (green arrows).

\emph{E-pooling}:
\begin{itemize}
    \item $2 \rightarrow 1 \rightarrow 18 \rightarrow 20$ : 432 vehicles are dispatched from node $2$ to node $1$ in order to fulfill demands of OD pairs: (1,18), (1,20) ,(2,18) and (2,20). In other words, vehicles pick up one passenger at node $2$, move to node $1$ along the link $(2 \rightarrow 1)$ on the network and pick up the other passenger at node $1$. After visiting origins $2$ and $1$, vehicles move to node $18$ along the link $(1 \rightarrow 3 \rightarrow 12 \rightarrow  11 \rightarrow 10 \rightarrow 16 \rightarrow 18)$ on the Sioux Falls network, and drop off passengers whose destination is node $18$. Vehicles then move to node $20$ along the link $(18 \rightarrow 20)$ and drop off passengers whose destination is node $20$.
    
    \item $2 \rightarrow 3 \rightarrow 20$ : 650 vehicles are dispatched from node $2$ to node $3$ in order to fulfill demands of OD pairs: (2,20) and (3,20). In other words, vehicles pick up one passenger at node $2$, move to node $3$ along the link $(2 \rightarrow 1 \rightarrow 3)$ on the network and pick up the other passenger at node $3$. After visiting origins $2$ and $3$, vehicles move to node $20$ along the link $(3 \rightarrow 12 \rightarrow  13 \rightarrow 24 \rightarrow 21 \rightarrow 20)$ on the Sioux Falls network, and drop off all passengers at node $20$.
    
    \item $1 \rightarrow 3 \rightarrow 20$ : 168 vehicles are dispatched from node $1$ to node $3$ in order to fulfill demands of OD pairs: (1,20) and (3,20).  Vehicles pick up one passenger at node $1$, move to node $3$ along the link $(1 \rightarrow 3)$ on the network and pick up the other passenger at node $3$. After visiting origins $1$ and $3$, vehicles move to node $20$ along the link $(3 \rightarrow 12 \rightarrow  13 \rightarrow 24 \rightarrow 21 \rightarrow 20)$ on the Sioux Falls network, and drop off all passengers at node $20$.
    
    \item $2 \rightarrow 18 \rightarrow 20$ : There are 442 vehicles dispatched from node $2$ in order to fulfill demands of OD pairs: (2,18) and (2,20). At node $2$, vehicles pick up passengers whose destinations are node $18$ and $20$. After visiting origin $2$, vehicles first move to node $18$ along links $(2 \rightarrow 6 \rightarrow  8 \rightarrow 16 \rightarrow 18)$ and $(2 \rightarrow 6 \rightarrow  8 \rightarrow 7 \rightarrow 18)$ on the Sioux Falls network, and drop off passengers at node $18$. Vehicles then move to node $20$ along the link $18 \rightarrow 20$ and drop off the remaining passengers.
    
    \item  $2 \rightarrow 20 \rightarrow 18$ : Similar to vehicles in the OD sequence $2 \rightarrow 18 \rightarrow 20$, 1312 vehicles are dispatched from node $2$ in order to fulfill demands of OD pairs: (2,18) and (2,20). These vehicles choose to first drop off passengers at destination $20$ and then drop off the remaining passengers at destination $18$. 
\end{itemize}

\emph{E-solo}:
\begin{itemize}
    \item $3 \rightarrow 18$ : There are 149 vehicles dispatched from node $3$ to $18$ to fulfill the travel demand of OD pair (3,18). These vehicles only visit one OD pair in the OD graph and pick up one rider during the trip. 
\end{itemize}

We summarize our findings about vehicle dispatch: at origin $1$, the majority of travel demand (432 out of 600) is fulfilled by vehicles dispatched from origin $2$ and the remaining passengers are picked up by vehicles dispatched from origin $1$.  At origin $2$, all travel demand is fulfilled by vehicles dispatched from origin $2$. At origin $3$, the majority of travel demand is fulfilled by vehicles dispatched from origin $2$ and the remaining demand is fulfilled by the e-solo service where vehicles are dispatched from origin $3$ to destination $18$.


\textbf{Rebalancing Flow} (grey arrows):
\begin{itemize}
    \item $20 \rightarrow 2$: 2836 vehicles at destination 20 plan to pick up passengers at origin $2$ (i.e., travel demands of OD pair $(2,18)$ and $(2,20)$). These vehicles are dispatched from node $20$ to node $2$ along the link $(20 \rightarrow 18 \rightarrow  7 \rightarrow 8 \rightarrow 6  \rightarrow 2)$ on the Sioux Falls network. It is shown that the majority of vehicles at destinations are dispatched to node $2$, which incurs heavy traffic at this origin. 
    \item $20 \rightarrow 3$: 149 vehicles at destination 20 plan to pick up passengers at origin $3$ (i.e., travel demands of OD pair $(3,18)$ and $(3,20)$) and they are dispatched from node $20$ to node $3$ along the link $(20 \rightarrow 21 \rightarrow  24 \rightarrow 13 \rightarrow 12 \rightarrow 3)$.
    \item $18 \rightarrow 1$: 168 vehicles at destination 18 are dispatched to origin $1$ along the link $(18 \rightarrow 16 \rightarrow  10 \rightarrow 11 \rightarrow 12 \rightarrow 3 \rightarrow 1)$ in order to pick up passengers at node 1.
\end{itemize}



\begin{figure}[H]
	\centering
    \includegraphics[scale=.4]{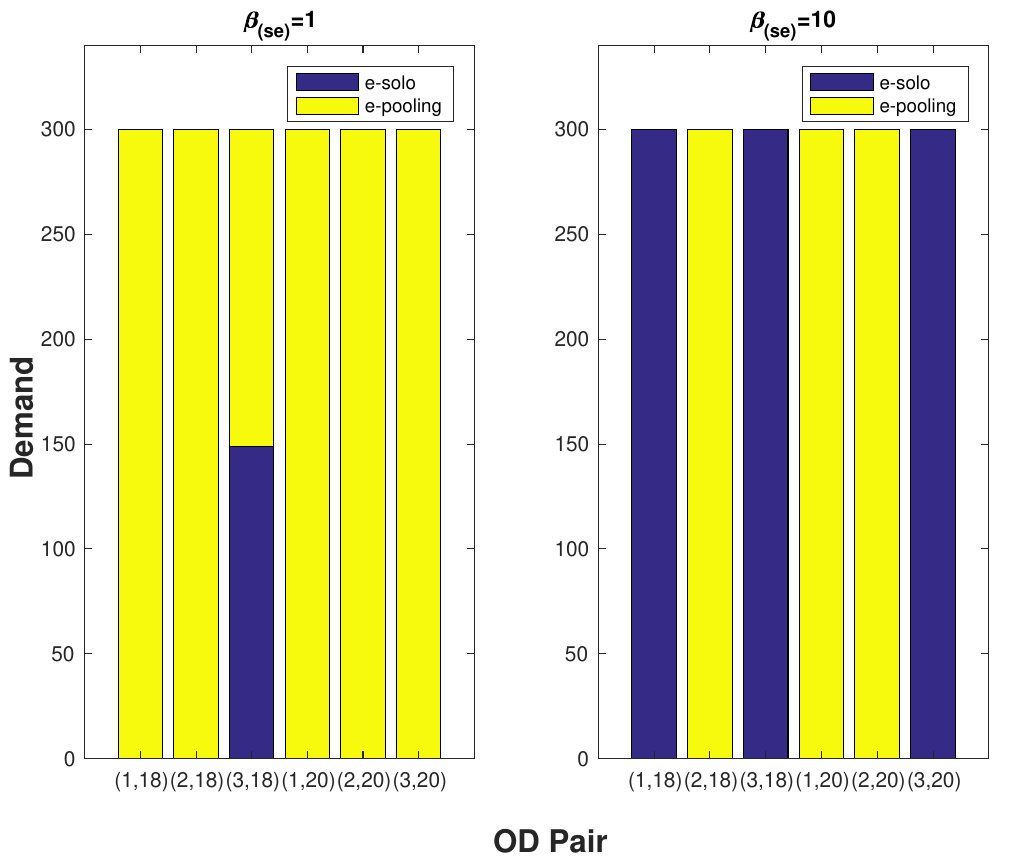}
	\caption{OD demand by mode for Scenario~2 on Sioux Falls}
	\label{fig:mode_6OD}
\end{figure}

\paragraph{Case 2}
We now investigate another case when cost coefficient $\beta_{(se)}$ of search friction is changed from 1 to 10. The vehicle OD flow is demonstrated in \ref{append:sioux}. By making a comparison of Case 1 and 2, we find that: When incurring higher search friction, the number of vehicles in e-pooling service decreases while the number of vehicles in e-solo service increases. To better understand this, we plot OD demands by different modes when varying search friction for drivers in Fig.~\ref{fig:mode_6OD}. Left bars demonstrate the scenario when the cost coefficient of search friction is 1 and right bars represent the scenario when the cost coefficient of search friction is 10. It is shown that when search friction increases, the usage of e-pooling service decreases.



\subsection{More ODs}
\label{sec:more_od}
\begin{figure}[H]
	\centering
    \includegraphics[scale=.25]{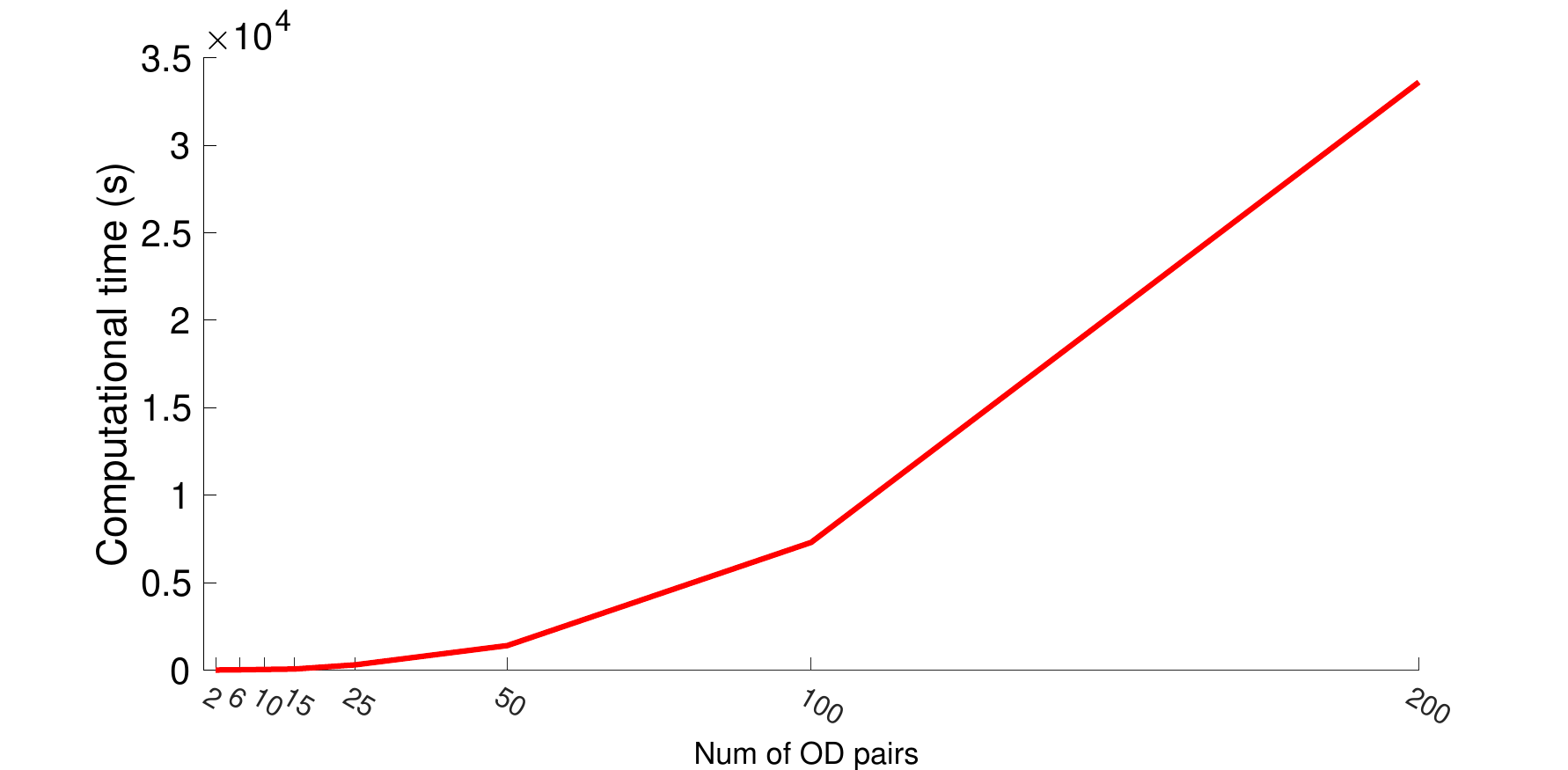}
	\caption{Computational time as the number of OD pairs increases}
	\label{fig:performance}
\end{figure}

Figs.~\ref{fig:performance} demonstrates the computational time against the problem size represented by the OD pair size. 


%

\subsection{Implication for policies and planning}

Relying on the proposed equilibrium framework that depicts the interaction between e-hailing operation, customer modal choice, and network congestion, 
we hope to shed light on policy making and planning for more efficient pooling service.

Summarizing findings from all the numerical experiments, we find that high e-pooling price and high matching friction could lead to higher e-solo demands, which makes roads more congested and in turn hurt e-pooling demands further. To promote pooling service, policy makers and planners should consider taking a series of countermeasures, including subsidy to pooling customers or to e-hailing platforms, congestion pricing to e-solo customers, and regulation. 
In particular, in a competitive e-hailing market with multiple companies, one could consider expanding its fleet sizing for a larger market share. However, an increase in fleet size could result in rising deadhead miles and system travel cost in a transportation system, which in turn leads to worsening congestion and emission. Capping e-hailing companies' fleet size could be one instrument for regulators \citep{NYC2018_cap,matthew2018congestion}.

On the other hand, this paper could offer insights into the e-hailing operation strategies, such as improving matching mechanism and optimizing  pricing and fleet sizing while accounting for congested networks and customer choice.

\section{Conclusions and future research}
\label{sec:conclu}

This paper proposes a unified network equilibrium framework that integrates 3 modules, including travelers' modal choice between e-pooling and e-solo services, e-platforms' decision on vehicle dispatching and driver-passenger matching, and network congestion.  
The first two components are coupled via passengers' waiting cost computed from a market clearance mechanism, 
and the latter two are coupled via equilibrium travel time. 
To model e-platforms' decision on vehicle dispatching and driver-passenger matching, 
the challenge lies in the permutation of potentially pooled OD pairs and the order of pick-up and drop-off among these two pairs. 
The existing literature usually predetermines a combination of OD pairs and fixes the sequence of chaining OD pairs, which limits the flexibility of the pooling service. 
To address the challenge, this paper develops a layered OD graph over which vehicle and passenger OD flows are modeled and computed. 
The minimum-cost circulation network flow problems are formulated to solve optimal vehicle OD flows and passenger OD flows, respectively. 
The total model is tested on a small-sized network and the Sioux Falls network based on the performance measures including deadhead miles, system travel time, and total fleet sizing. 
Sensitivity analysis is performed on parameters associated with fleet sizing, service pricing, and search friction. We demonstrate the performance of our model on a much larger OD graph.
In particular, on the small network, we extend our model to two competing TNCs in the market and evaluate the impact of fleeting sizing on each company's market share. Our key finding is, high e-pooling price and high matching friction could lead to more e-solo customers, which makes roads more congested and in turn hurt e-pooling demands due to the increasing time of seeking multiple riders. Government subsidies or congestion pricing could help promote pooling and mitigate congestion.


This paper will be generalized in the following directions: 
(1) relax the assumption of fixed demands for e-hailing service by incorporating private cars into the framework, so that the unmet orders switch to private driving. And relax the assumption of fixed demands for each OD pair by introducing a demand curve for each OD as a function of its average total travel cost.
(2) integrate emerging shared travel modes, including ad hoc ridesharing of personal vehicles. (3) pooling more than two rides and providing a on-demand shuttle like service will be considered using more complex OD graphs. (4) incorporate drivers' choices including working hours. 
TNC service providers, in other words, drivers, are a crucial type of game players whose entry choice into the TNC market can influence the fleet sizing, customers' waiting time and mode choice. 
Thus incorporation of the driver side will complete the game-theoretic picture of the TNC ecosystem. 

\section*{Acknowledgements}
This work is sponsored by NSF under CAREER award number CMMI-1943998.


\bibliographystyle{elsarticle-harv}

\bibliography{lit,lit_Di}


\newpage
\appendix
\section{Appendices}

\subsection{Performance measures}

We propose the following countermeasures to describe the performance of an e-hailing system: 
\vspace{-2mm}
\begin{subequations}
	\begin{align}
	& \text{Deadhead miles: } 
	\text{\emph{DHM}} \equiv 
	\sum_{v\in{\cal N}^{-}}
	\sum_{u\in{\cal N}^{+}}
	l_{vu} z^{u}_{v}
	+ \sum_{u\in{\cal N}^{+}}
	\sum_{\underline{u}\in{\cal N}^{-}}
	l_{u\underline{u}} z^{\underline{u}}_{u},
	\label{eq:dh}\\
	& \text{System travel cost: } 
	\text{\emph{STC}} \equiv 
	\sum_{v,u\in{\cal N}(m)}
	t_{vu} 
	\left(\sum_{w\in{\cal W}} \sum_{m\in{\cal M}}  y^{(w)[m]}_{vu}\right), 
	\label{eq:th}\\
    & \text{Total vehicle-hour: } 
	\text{\emph{TVH}} \equiv 
	\sum_{v\in{\cal N}^{-}}
	\sum_{u\in{\cal N}^{+}} t_{vu} z^{u}_{v}
	+ 	\sum_{v\in{\cal N}^{-}(m)}
	\sum_{u\in{\cal N}^{+}(m)} \sum_{m\in{\cal M}} t_{vu} z^{[m]}_{vu}, \label{eq:vh}	
	\end{align}
\end{subequations}

\emph{DHM} measures the miles traveled by vacant e-hailing vehicles. 
\emph{STC} measures the total hours consumed by vehicles including service vehicle-hour, which is the total hours both occupied and vacant vehicles spend on road. 
\emph{TVH} measures the total service vehicle hours, which cannot exceed the fleet size. 

\subsection{Traffic entities on roads}
\label{append:entity}

In an e-hailing system, we will first introduce how passengers and vehicles move along a road network. 
Assume there is a set of e-hailing service providers that offers both e-solo and e-pooling modes in the market. 
The network flows include the passenger flow selecting e-solo or e-pooling modes from the provider set and its induced vehicle flow.

 \subsubsection{Passenger flow}
 
Prior to a trip, one traveler selects travel mode with the minimum travel disutility: e-solo 
or e-pooling. 
These travel modes are denoted as ${\cal M}=\left\lbrace \underline{e}_c,\underline{p}_c \right\rbrace$, where $\underline{e}_c$ denotes being an e-solo passenger served by the $c$-th provider and $\underline{p}_c$ denotes being an e-pooling passenger served by the $c$-th provider.

An e-solo passenger is picked up at her origin and directly moved along with the e-solo vehicle to be dropped off at her destination. 
An e-pooling passenger is picked up at her origin. 
If she is the first order in the vehicle, she is moved to a second origin with the e-pooling vehicle to pick up a second passenger order. After the second passenger is picked up, she is either dropped off at her destination and has to visit the destination of the second passenger before being dropped off at her own destination.   
If she is the second order in the vehicle meaning that there is already a passenger inside the vehicle, she is moved to either her destination or the first passenger's destination before being dropped off. 

\begin{rem}
Throughout the paper, we drop the e-hailing provider subscript $c$ for notation simplicity. In other words, our mathematical framework is primarily focused on one e-hailing provider, but it can be easily generalized to multiple providers without loss of generality. The generalized model for multiple providers can be found in  \ref{append:parameters_small}.
We will also demonstrate it on a numerical example in Sec.~\ref{sec:exp}.
\end{rem}

 \subsubsection{Vehicle flow}

Passenger flows induce vehicle flows and thus traffic congestion. 
After an vacant e-hailing vehicle is matched an order, it selects the paths with minimum travel time to rebalance itself for pick-up. 
Occupied vehicle flows consist of e-solo and e-pooling flows. 
For a vacant e-solo vehicle, once it picks up an order and thus becomes occupied, it aims to move to the destination of the order along the paths with minimum travel time. 
For a vacant e-pooling vehicle, once it knows the dispatching and matching plan, it needs to determine in what sequence to visit two origin nodes, so are two destination nodes. The routing plan aims to find the paths with minimum travel time connecting these two OD pairs. 

In summary, 
the vehicle flow on roads (or en route) include 
the occupied e-solo flow (with a passenger occupancy of one), 
the 1-passenger occupied e-pooling flow (with a passenger occupancy of one), 
the 2-passenger occupied flow (with a passenger occupancy of two), 
and vacant rebalancing flow (with a passenger occupancy of zero), respectively. E-hailing vehicle and passenger flows are summarized in \ref{append:table_veh_passen}.

\subsection{Table of vehicle and passenger flows}
\label{append:table_veh_passen}

\begin{table}[H]\centering
\fontsize{9}{0}\selectfont
\caption{E-hailing vehicle and passenger flows}\label{tab:note_e}
\begin{tabular}{p{2cm}||p{1.5cm}|p{3cm}|p{5cm}|p{1cm}|p{1cm}}
\hline
Mode & Traffic flow & State & Description & \multicolumn{2}{l}{Variables}   
\\ \cline{5-6} 
& & & & Driver & Passenger
\\ \hline\hline
E-solo & OD flow & Occupied ($occ=1$) & connecting $w$'s origin to its destination, $u=\bar{w},v=\underline{w}$ & $z^{[e]}_{\bar{w}\underline{w}}$ &  $y^{(w)[e]}_{\bar{w}\underline{w}}$\\ \hline
E-pooling & OO flow & Shared ($occ=2$) & connecting $w$'s origin to $k$'s origin, $u=\bar{w},v=\bar{k}$ & $z^{[p]}_{\bar{w}\bar{k}}$ & $y^{(w)[p]}_{\bar{w}\bar{k}}$ \\ \cline{2-6} 
& OD flow & Shared ($occ=2$) & connecting $w$'s origin to its destination, $u=\bar{w},v=\underline{w}$ & $z^{[p]}_{\bar{w}\underline{w}}$ & $y^{(w)[p]}_{\bar{w}\underline{w}}$ \\ \cline{3-6}
&& Shared ($occ=2$) & connecting $w$'s origin to $k$'s destination, $u=\bar{w},v=\underline{k}$ & $z^{[p]}_{\bar{w}\underline{k}}$ & $y^{(w)[p]}_{\bar{w}\underline{k}}$ \\ \cline{2-6}
& DD flow & Occupied ($occ=1$) & connecting $k$'s destination to $w$'s destination, $u=\underline{k},v=\underline{w}$ & $z^{[p]}_{\underline{k}\underline{w}}$ & $y^{(w)[p]}_{\bar{k}\underline{w}}$ \\ \hline 
Rebalancing & DO flow & Vacant & connecting $w$'s destination to $k$'s origin, $u=\underline{w},v=\bar{k}$ & $z^{\bar{k}}_{\underline{w}}$ & - \\ \hline 
\end{tabular}
\end{table}

\subsection{Relation between an OD graph and a road network}

Both an OD graph and a road network consist of nodes, links, flows, and ODs, but carry different meanings. To clarify, we compare their differences in Table~\ref{tab:term}. 
\begin{table}[H]
	\centering\caption{Comparison of components in an OD graph and a road network}
	\label{tab:term}
	\begin{tabular}{l||l| l}
		\hline
		Term & Road network & OD graph \\ \hline\hline
		Agent & Traveler & e-hailing vehicle or passenger \\ 
		Flow & Travel demand flow & Vehicle or passenger OD flow \\ 
		OD & Travel orders & Augmented OD pairs \\ \hline
		Node & Order O/D, road junction & Order O/D \\ 
		Origin & Travel demand origin & Pick-up node \\ 
		Destination & Travel demand destination & Drop-off node \\ \hline 
		Edge & A road link & An OD/OO/DD/DO path \\
		Edge flow & Link flow & OD flow \\ \hline
		Path & An OD path connecting an order OD & A service path connecting a node pair \\ 
		Path flow & OD flow & Pick-up or drop-off flow \\ \hline
	\end{tabular}
\end{table}

Computation wise, an OD graph only depends on the number of OD pairs, in other words, the cardiality of ${\cal W}^{aug}$. 
Thus, it is agnostic of a real transportation network topology. 
That being said, as long as two road networks share the same number of OD pairs, the topology of their OD graphs remains the same. 

 \paragraph{Difference between a vehicle and a passenger OD graph}

Vehicle OD graphs are closed, meaning that vehicle flows circulate between origins and destinations without leaving the graphs. Thus, the coupled e-solo and e-pooling vehicle OD graph is closed, assuming that no e-hailing vehicle would drop out of the system. 
Solving vehicle flows is formulated as a minimum cost multi-commodity circulation problem as a generalization of network flow problems. 
In contrast, passenger OD graphs are open-ended and those across solo and pooling modes are only connected at virtual origins when a total travel demand is split between e-hailing modes. Thus, e-solo and e-pooling passenger OD graphs are only connected at origin nodes. 
Solving a passenger flow in each layer is formulated as a minimum cost network flow problem. 

\subsection{Preliminaries on minimum-cost circulation network flow problem (MCCNFP)}\label{append:mccnfp}

Network flow problems, a branch of combinatorial optimization, solve optimal edge flows on a graph that respect flow conservation and other designated constraints. 
In particular, the minimum-cost circulation network flow problem (MCCNFP) aims to find a set of edge flows circulating on a graph that can minimize the total flow cost \citep{ahuja1993network}. 
Given a network represetend by a directed graph  $\mathcal{G}=(\mathcal{N},\mathcal{L})$, 
where each edge $(v,u)\in \mathcal{L}$ is associated with flow $z_{vu}$ and cost $C_{vu}$. 
Denote $\mathbf{z}$ and $\mathbf{C}$ as the flow and cost vectors over the graph, respectively. 
The flow $\mathbf{z}$ is bounded below by $\mathbf{b}^L$ and above $\mathbf{b}^U$. The total cost is  $\mathbf{C}^T\cdot \mathbf{z}$. 
In summary, the MCCNFP is formulated as:

\begin{align}
   & min \ \ \mathbf{C}^T\cdot \mathbf{z} \nonumber 
   \\
   & s.t.\ \ \  \sum _{v:(u,v)\in \mathcal{L}}z_{vu}-\sum _{k: (u,k)\in \mathcal{L}}z_{uk}=0, \forall u \in \mathcal{N} \ \ (\text{circulation flow conservation}), \nonumber \\
   & \ \ \ \ 
   \mathbf{b}^L\leq \mathbf{z}\leq \mathbf{b}^U \ \  (\text{capacity constraint}). \nonumber
\end{align}

In Sec.~\ref{sec:1_OD}, we will specify the graph,  edge cost, and capacity bounds for the problem formulation of Module 1.

\subsection{3-node network with 2-OD pair}
\label{append:exmp4.1}

\begin{figure}[H]
  	\centering
  	\includegraphics[scale=.45]{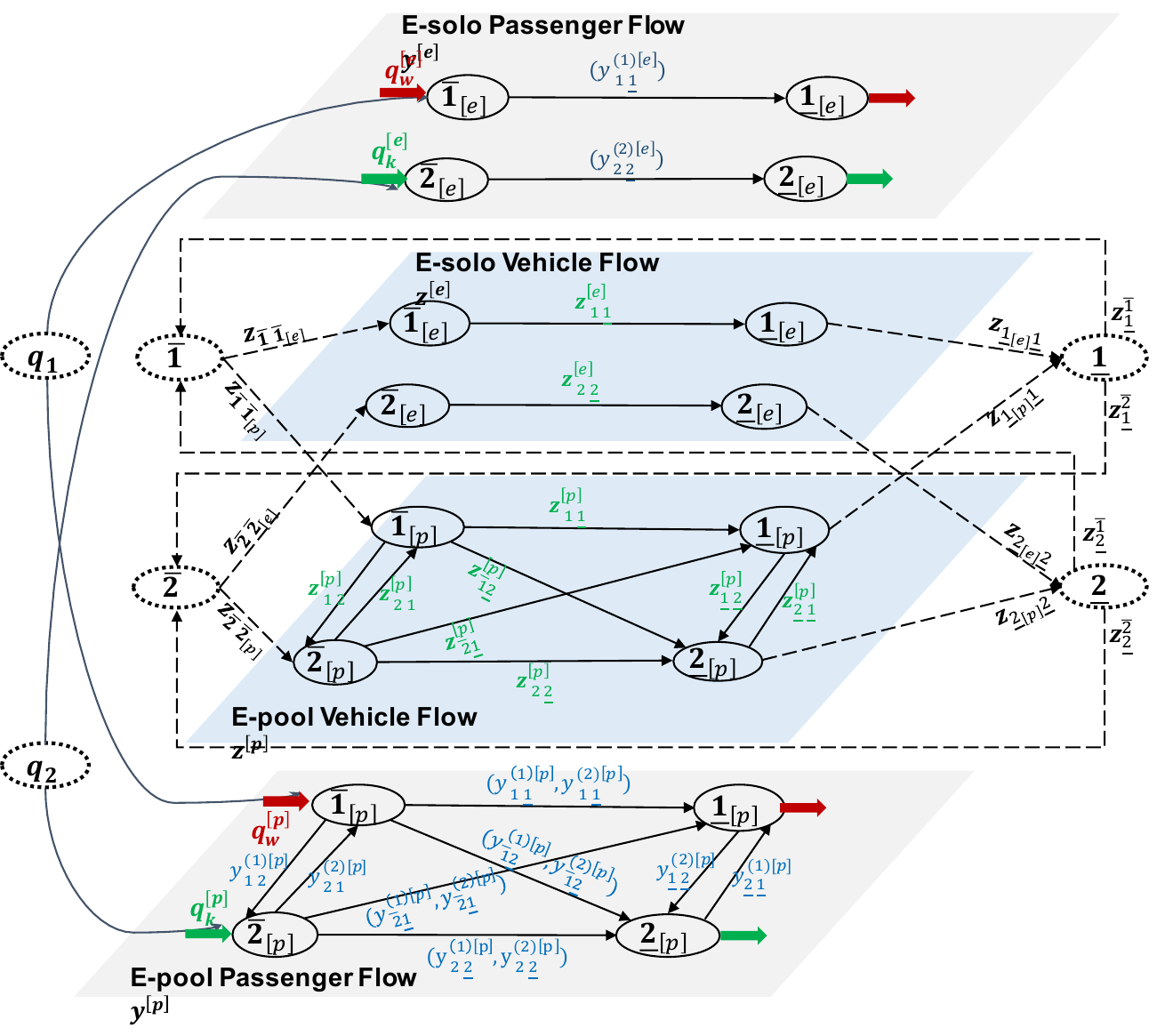}
  	\caption{Layered OD graph with 2-OD pair}
  	\label{fig:3_node_OD}					
\end{figure}

We demonstrate Module 1 on a 3-node network with 2-OD pair (Fig.~\ref{fig:3_node_OD_exp}). The layered OD graph corresponding to the 3-node network with 2-OD pair is shown in Fig.~\ref{fig:3_node_OD}.

Mathematically, the vehicle dispatching module on the OD graph is: 
\begin{subequations}\label{eq:opt_z_ex}
\fontsize{9}{0}\selectfont
	\begin{align}
	& \min_{z^{\bar{1}}_{\underline{1}},z^{\bar{2}}_{\underline{1}},z^{\bar{1}}_{\underline{2}},z^{\bar{2}}_{\underline{2}}} 
	C^{\bar{1}}_{\underline{1}} z^{\bar{1}}_{\underline{1}} + C^{\bar{2}}_{\underline{1}} z^{\bar{2}}_{\underline{1}} + C^{\bar{1}}_{\underline{2}} z^{\bar{1}}_{\underline{2}} + C^{\bar{2}}_{\underline{2}} z^{\bar{2}}_{\underline{2}} + \min_{z^{[e]}_{\bar{1}\underline{1}},z^{[e]}_{\bar{2}\underline{2}},z^{[p]}_{\bar{1}\underline{1}},z^{[p]}_{\bar{2}\underline{2}},z^{[p]}_{\bar{1}\bar{2}},z^{[p]}_{\bar{2}\bar{1}},z^{[p]}_{\bar{1}\underline{2}},z^{[p]}_{\bar{2}\underline{1}},z^{[p]}_{\underline{1}\underline{2}},z^{[p]}_{\underline{2}\underline{1}}} 
	C^{[e]}_{\bar{1}\underline{1}} z^{[e]}_{\bar{1}\underline{1}} + C^{[e]}_{\bar{2}\underline{2}} z^{[e]}_{\bar{2}\underline{2}} \nonumber\\
	& + C^{[p]}_{\bar{1}\underline{1}} z^{[p]}_{\bar{1}\underline{1}} + C^{[p]}_{\bar{2}\underline{2}} z^{[p]}_{\bar{2}\underline{2}}  + C^{[p]}_{\bar{1}\bar{2}} z^{[p]}_{\bar{1}\bar{2}} +C^{[p]}_{\bar{2}\bar{1}} z^{[p]}_{\bar{2}\bar{1}} + C^{[p]}_{\bar{1}\underline{2}} z^{[p]}_{\bar{1}\underline{2}} + C^{[p]}_{\bar{2}\underline{1}} z^{[p]}_{\bar{2}\underline{1}} + C^{[p]}_{\underline{1}\underline{2}} z^{[p]}_{\underline{1}\underline{2}} + C^{[p]}_{\underline{2}\underline{1}} z^{[p]}_{\underline{2}\underline{1}}, \nonumber\\
	& s.t. \nonumber\\
	&  (\textcolor{red}{\bigstar})\   z_{\bar{1}\bar{1}_{[p]}} = z^{[p]}_{\bar{1}\bar{2}},\ z_{\Bar{2}\Bar{2}_{[p]}} = z^{[p]}_{\Bar{2}\Bar{1}}, \nonumber \\
	& (\textcolor{blue}{\bigcirc}) \  z^{[p]}_{\bar{1}\bar{2}}=z^{[p]}_{\bar{2}\underline{1}}+z^{[p]}_{\bar{2}\underline{2}},\  z^{[p]}_{\bar{2}\bar{1}}=z^{[p]}_{\bar{1}\underline{1}}+z^{[p]}_{\bar{1}\underline{2}}, \nonumber \\
	& (\textcolor{yellow}{\bigstar}) \   z^{[p]}_{\bar{1}\underline{1}}+ z^{[p]}_{\bar{2}\underline{1}} = z^{[p]}_{\underline{1}\underline{2}},\ z^{[p]}_{\bar{1}\underline{2}} + z^{[p]}_{\bar{2}\underline{2}}= z^{[p]}_{\underline{2}\underline{1}}, \nonumber\\
	& (\hrectangle)\   z^{[p]}_{\underline{1}\underline{2}}=z_{\underline{2}_{[p]}\underline{2}},\  z^{[p]}_{\underline{2}\underline{1}}=z_{\underline{1}_{[p]}\underline{1}}, \nonumber\\
	& (\triangle) \  z_{\bar{1}\bar{1}_{[p]}} + z^{[p]}_{\bar{2}\bar{1}} \geqslant q^{[p]}_{1},\ z_{\bar{2}\bar{2}_{[p]}} + z^{[p]}_{\bar{1}\bar{2}} \geqslant q^{[p]}_{2}, 
	\nonumber \\
	& (\mdwhtdiamond)\  z_{\underline{1}_{[p]}\underline{1}} + z^{[p]}_{\underline{1}\underline{2}} \geqslant q^{[p]}_{1},\ z_{\underline{2}_{[p]}\underline{2}} + z^{[p]}_{\underline{2}\underline{1}} \geqslant q^{[p]}_{2}, 
	\nonumber \\
	& (\tco{\triangle}) \ z_{\bar{1}\bar{1}_{[e]}}  = z^{[e]}_{\bar{1}\underline{1}}, \ z_{\bar{2}\bar{2}_{[e]}} = z^{[e]}_{\bar{2}\underline{2}}, z^{[e]}_{\bar{1}\underline{1}} = z_{\underline{1}_{[e]}\underline{1}}, \  z^{[e]}_{\bar{2}\underline{2}} = z_{\underline{2}_{[e]}\underline{2}}, z_{\bar{1}\bar{1}_{[e]}}  \geqslant q^{[e]}_{1}, \ z_{\bar{2}\bar{2}_{[e]}}  \geqslant q^{[e]}_{2}, \nonumber \\
	& (\tcr{+})\   z^{\bar{1}}_{\underline{1}} + z^{\bar{1}}_{\underline{2}}=z_{\bar{1}\bar{1}_{[p]}} + z_{\bar{1}\bar{1}_{[e]}}, \  z^{\bar{2}}_{\underline{1}} + z^{\bar{2}}_{\underline{2}}=z_{\bar{2}\bar{2}_{[p]}} + z_{\bar{2}\bar{2}_{[e]}} , z_{\underline{1}_{[e]}\underline{1}} + z_{\underline{1}_{[p]}\underline{1}} = z^{\bar{1}}_{\underline{1}} + z^{\bar{2}}_{\underline{1}}, \ z_{\underline{2}_{[e]}\underline{2}} + z_{\underline{2}_{[p]}\underline{2}} = z^{\bar{1}}_{\underline{2}} + z^{\bar{2}}_{\underline{2}}, \nonumber \\
	& \ \ \ \ \ \  z_{\bar{1}\bar{1}_{[p]}} + z_{\bar{1}\bar{1}_{[e]}}+z_{\bar{2}\bar{2}_{[p]}} + z_{\bar{2}\bar{2}_{[e]}}=z_{\underline{1}_{[e]}\underline{1}} + z_{\underline{1}_{[p]}\underline{1}}+z_{\underline{2}_{[e]}\underline{2}} + z_{\underline{2}_{[p]}\underline{2}}.\nonumber 
	\end{align}
\end{subequations}

\noindent The vehicle-passenger matching module on the OD graph is:

\begin{subequations}\label{eq:opt_y_exp}
\fontsize{9}{0}\selectfont
	\begin{align}
	& \min_{y^{(1)[p]}_{\bar{1}\bar{2}},...,y^{(2)[p]}_{\underline{1}\underline{2}}} 
	\beta_{(in-veh)} \{ t_{\bar{1}\bar{2}}y^{(1)[p]}_{\bar{1}\bar{2}}
	+t_{\bar{2}\bar{1}}y^{(2)[p]}_{\bar{2}\bar{1}}
	+t_{\underline{2}\underline{1}}y^{(1)[p]}_{\underline{2}\underline{1}}
	+t_{\underline{1}\underline{2}}y^{(2)[p]}_{\underline{1}\underline{2}}
	+t_{\bar{1}\underline{1}}(y^{(1)[p]}_{\bar{1}\underline{1}}+y^{(2)[p]}_{\bar{1}\underline{1}})
	+t_{\bar{2}\underline{2}}(y^{(1)[p]}_{\bar{2}\underline{2}}+y^{(2)[p]}_{\bar{2}\underline{2}})
	+t_{\bar{1}\underline{2}}(y^{(1)[p]}_{\bar{1}\underline{2}}+y^{(2)[p]}_{\bar{1}\underline{2}}) \nonumber
	\\& +t_{\bar{2}\underline{1}}(y^{(1)[p]}_{\bar{2}\underline{1}}+y^{(2)[p]}_{\bar{2}\underline{1}}) \}+c_{\bar{1}}^{[p](wt)}(y^{(1)[p]}_{\bar{1}\bar{2}}+y^{(1)[p]}_{\bar{1}\underline{1}}+y^{(1)[p]}_{\bar{1}\underline{2}})+c_{\bar{2}}^{[p](wt)}(y^{(2)[p]}_{\bar{2}\bar{1}}+y^{(2)[p]}_{\bar{2}\underline{1}}+y^{(2)[p]}_{\bar{2}\underline{2}}) \nonumber, \\
	& (\tcr{\bigstar})\ y^{(1)[p]}_{\bar{1}\bar{2}} + y^{(1)[p]}_{\bar{1}\underline{1}}+y^{(1)[p]}_{\bar{1}\underline{2}}
	= q^{[p]}_{1}, \ y^{(2)[p]}_{\bar{2}\bar{1}} + y^{(2)[p]}_{\bar{2}\underline{1}}+y^{(2)[p]}_{\bar{2}\underline{2}}
	= q^{[p]}_{2}, \nonumber \\
	& (\textcolor{orange}{\bigstar})\ y^{(1)[p]}_{\bar{2}\underline{1}} + y^{(1)[p]}_{\bar{1}\underline{1}}+y^{(1)[p]}_{\underline{2}\underline{1}}
	= q^{[p]}_{1}, \ y^{(2)[p]}_{\bar{1}\underline{2}} + y^{(2)[p]}_{\underline{1}\underline{2}}+y^{(2)[p]}_{\bar{2}\underline{2}}
	= q^{[p]}_{2}, \nonumber \\
	& (\tcr{\bigcirc})\ y^{(1)[p]}_{\bar{1}\bar{2}} = y^{(1)[p]}_{\bar{2}\underline{1}}+y^{(1)[p]}_{\bar{2}\underline{2}},\  y^{(2)[p]}_{\bar{2}\bar{1}} = y^{(2)[p]}_{\bar{1}\underline{1}}+y^{(2)[p]}_{\bar{1}\underline{2}},\nonumber \\
	& (\textcolor{orange}{\bigcirc})\ y^{(1)[p]}_{\underline{2}\underline{1}} = y^{(1)[p]}_{\bar{2}\underline{2}}+y^{(1)[p]}_{\bar{1}\underline{2}},\  y^{(2)[p]}_{\underline{1}\underline{2}} = y^{(2)[p]}_{\bar{2}\underline{1}}+y^{(2)[p]}_{\bar{1}\underline{1}},\nonumber
	\\
	& 0 \leqslant y_{\bar{2}\bar{1}}^{(2)[p]}  \leqslant z_{\bar{2}\bar{1}}^{[p]},\ 0 \leqslant y_{\bar{1}\bar{2}}^{(1)[p]} \leqslant z_{\bar{1}\bar{2}}^{[p]},\ 0 \leqslant  y_{\underline{2}\underline{1}}^{(1)[p]} \leqslant z_{\underline{2}\underline{1}}^{[p]},\ 
	0 \leqslant  y_{\underline{1}\underline{2}}^{(2)[p]} \leqslant  z_{\underline{1}\underline{2}}^{[p]},\nonumber \\
	& 0 \leqslant y_{\bar{1}\underline{1}}^{(1)[p]},y_{\bar{1}\underline{1}}^{\bar{2}[p]}  \leqslant z_{\bar{1}\underline{1}}^{[p]}, \ 0 \leqslant y_{\bar{2}\underline{2}}^{(1)[p]},y_{\bar{2}\underline{2}}^{(2)[p]}  \leqslant z_{\bar{2}\underline{2}}^{[p]}, \ 0 \leqslant y_{\bar{1}\underline{2}}^{(1)[p]},y_{\bar{1}\underline{2}}^{(2)[p]}  \leqslant z_{\bar{1}\underline{2}}^{[p]},\ 0 \leqslant y_{\bar{2}\underline{1}}^{(1)[p]},y_{\bar{2}\underline{1}}^{(2)[p]}  \leqslant z_{\bar{2}\underline{1}}^{[p]}. \nonumber
	\end{align}
\end{subequations}

\subsection{Derivation of complementarity conditions for Equ.~\ref{eq:o_vw}}
\label{append:KKT}

The formulation of complementarity conditions (Equ.~\ref{eq:o_vw}) are shown as follows:

\noindent The denominator in $o_{\underline{v}\bar{w}}$ can be zero when there is no vacant vehicle flowing into $\bar{w}$, which causes the waiting cost to be indefinite.
To prevent the denominator from being zero, 
we propose to solve $o_{\underline{v}\bar{w}}$ using a quadratic program:

\begin{equation}\label{eq:theta_vw_QP}
o_{\underline{v}\bar{w}} 
\equiv 
\frac{
z^{\bar{w}}_{\underline{v}} 
}
{
\sum_{\underline{v}'\in {\cal N}^{-}} z^{\bar{w}}_{\underline{v}'}
}
=
\arg\min_{o\in \left[0,1\right]} 
\left\lbrace 
\frac{\sum_{\underline{v}'\in {\cal N}^{-}} z^{\bar{w}}_{\underline{v}'}}{2} 
\cdot
o^2 
- z^{\bar{w}}_{\underline{v}} \cdot
o 
\right\rbrace.
\end{equation}

To solve the above program defined in Equ.~\ref{eq:theta_vw_QP}, we formulate a Lagrangian function as: $L(\cdot)=\frac{\sum_{\underline{v}'\in {\cal N}^{-}} z^{\bar{w}}_{\underline{v}'}}{2} 
\cdot
o^2 
- z^{\bar{w}}_{\underline{v}} \cdot
o 
- \eta^{-}_{\underline{v}\bar{w}}\cdot (1-o)$. 
It can be solved by the following KKT conditions:

\begin{subequations}\label{eq:o_complement_kkt}
	\begin{align}
	& \frac{\partial L}{\partial o}
	=
	\sum_{\underline{v}'\in {\cal N}^{-}} z^{\bar{w}}_{\underline{v}'} \cdot
	o
    - z^{\bar{w}}_{\underline{v}}
    + \eta^{-}_{\underline{v}\bar{w}}=0,\\
	& o \geqslant 0,\  
	\\
	& 1- o\geqslant 0,\  \eta^{-}_{\underline{v}\bar{w}} \geqslant 0, \eta^{-}_{\underline{v}\bar{w}} \cdot (1-o)=0.
	\end{align}
\end{subequations} 

Summarizing these KKT conditions, we get Equ.~\ref{eq:o_vw}.

\subsection{Solution existence}
\label{sec:appendix_exist}

\noindent Before proving Proposition \ref{prop:bounded} and \ref{prop:VI_penalty}, we introduce theorems and lemmas to better prepare for the proof.

\begin{thm} \citep{Nagurney2008VI}
\label{thm:Nagurney2008VI}
In a finite-dimensional variational inequality problem $V(\mathbf{H},\mathcal{F}_{\zeta})$, if $\mathcal{F}_{\zeta}$ is a compact convex set and $\mathbf{H}(F)$ is continuous on $\mathcal{F}_{\zeta}$, then $V(\mathbf{H},\mathcal{F}_{\zeta})$ has at least one solution.
\end{thm}

\begin{thm}\citep{Nagurney2008VI}
\label{thm:Nagurney2008VI_2}
$V(\mathbf{H},\mathcal{F})$ has a solution if and only if there exists a bound $\mathcal{K}>0$ and a solution of $V(\mathbf{H},\mathcal{F}_{\zeta})$, denoted by $\mathcal{F}^{*}$, satisfies: $||\mathcal{F}^{*}||<\mathcal{K}$.
\end{thm}

\begin{lemma}\citep{cottle2009linear}
\label{lemma:cottle}
Consider the following mixed complementarity conditions:
\begin{align}
    & 0 \leqslant \mu_1 \perp G(\mu_1)+A^T \mu_2 +B^T \mu_3 \geqslant 0 \nonumber \\
	& \mu_2\ \text{free} \perp C \mu_1 -E =0 \nonumber \\
	& 0 \leqslant  \mu_3 \perp D \mu_1 -F \geqslant 0 \nonumber
\end{align}
where, $G$ is continuous. Let $\{\mu^k_1, \mu^k_2, \mu^k_3\}$ be a sequence of solutions corresponding  to the convergent sequence $\{ (E^k, F^k) \}$ with $\lim_{k \rightarrow \infty} E^k=E^{\infty}$ and $\lim_{k \rightarrow \infty} F^k=F^{\infty}$. If $\lim_{k \rightarrow \infty} \mu_1^k=\mu_1^{\infty}$, then there exists $(\mu_2^{\infty},\mu_3^{\infty})$ such that $(\mu_1^{\infty},\mu_2^{\infty},\mu_3^{\infty})$ is a solution corresponding to $(E^{\infty},F^{\infty})$.
\end{lemma}


\noindent \emph{Proof of Proposition \ref{prop:bounded}}

\begin{proof}
$\sum_{k,k\neq w} z^{[m]}_{\bar{k}\bar{w}}+\sum_{k,k\neq w} z^{[m]}_{\bar{w}\bar{k}} \leqslant \zeta \cdot  q_w^{[m]}$ implies that the dispatched vehicle flow in the OD graph is bounded by travel demand $q_w, \forall w \in {\cal W}$. Therefore, the vehicle flow on the road network is also bounded. We have $x_{ij} \leqslant \zeta \cdot  \sum_{w \in {\cal W}}q_w, \forall (i,j) \in {\cal L}_{net}$ and $\tau_i^d \leqslant \sum_{(i,j) \in {\cal L}_{net}}t_{ij}(\zeta \cdot  \sum_{w \in {\cal W}}q_w), \forall i \in {\cal N}_{net}, d \in {\cal N}$. The network flow and node potential are bounded. 
\end{proof}

\noindent \emph{Proof of Proposition \ref{prop:VI_penalty}}

\begin{proof}

\noindent Note that the coupled system consists of different modules. We first obtain the whole mixed complementarity problem (MCP) formulation of [\textbf{All}.\mbox{NCP}\mbox{-ESys}] as follows:

\noindent In the vehicle layer, we have
\begin{align}
     \sum_{w\in {\cal W}} z_{\bar{w}\bar{w}_{[p]}} &= \sum_{w\in {\cal W}} \sum_{k:k\neq w} z_{\bar{w}\bar{k}}^{[p]}= \sum_{w\in {\cal W}} \sum_{k:k\neq w} z_{\bar{k}\bar{w}}^{[p]} =\sum_{w\in {\cal W}} \sum_{k} z_{\bar{w}\underline{k}}^{[p]} = \sum_{w\in {\cal W}} \sum_{k} z_{\bar{k}\underline{w}}^{[p]}\nonumber \\
    &=\sum_{w\in {\cal W}} \sum_{k \neq w} z_{\underline{w}\underline{k}}^{[p]} =\sum_{w\in {\cal W}} \sum_{k \neq w} z_{\underline{k}\underline{w}}^{[p]}=\sum_{w\in {\cal W}} z_{\underline{w}_{[p]}\underline{w}}  \nonumber
\end{align}
Similarly, $\sum_{w\in {\cal W}} z_{\bar{w}\bar{w}_{[e]}}=\sum_{w\in {\cal W}} z_{\underline{w}_{[e]}\underline{w}}$ holds. Therefore, we have
\begin{align}
     \sum_{m \in \mathcal{M}}\sum_{w\in {\cal W}} z_{\bar{w}\bar{w}_{[m]}}=\sum_{w\in {\cal W}} z_{\bar{w}\bar{w}_{[p]}} +\sum_{w\in {\cal W}} z_{\bar{w}\bar{w}_{[e]}}= \sum_{w\in {\cal W}} z_{\underline{w}_{[p]}\underline{w}}+\sum_{w\in {\cal W}} z_{\underline{w}_{[e]}\underline{w}}=\sum_{m \in \mathcal{M}} \sum_{w\in {\cal W}} z_{\underline{w}_{[m]}\underline{w}} \nonumber
\end{align}
It means the total flow conservation on virtual nodes holds if the flow conservation on origins and destinations holds. Accordingly, in $[\textbf{M1.1}.\mbox{NCP}\mbox{-VehDispatch}]$, vehicle flow constraints can be generalized as the flow conservation on each node over the layered OD graph. Mathematically, 
\begin{align}
    \sum_{v'} z_{v'u}=\sum_{v} z_{uv}, \forall u \in \mathcal{N}
\end{align}
We look into the demand constraint:
\begin{align}
     q^{[p]}_{w} \leqslant z_{\bar{w}\bar{w}_{[p]}} 
	+ \sum_{\bar{k}\in {\cal N}^+(p),k\neq w} z^{[p]}_{\bar{k}\bar{w}}&=\sum_{k:k\neq w} z_{\bar{w}\bar{k}}^{[p]}+\sum_{\underline{k}\in {\cal N}^{-}(p), k\neq w}  z^{[p]}_{\bar{w}\underline{k}}
    + z^{[p]}_{\bar{w}\underline{w}}\  
    \nonumber \\
    &=  \sum_{k:k\neq w} (z_{\bar{k}\underline{w}}^{[p]}+z_{\bar{k}\underline{k}}^{[p],w})+\sum_{\underline{k}\in {\cal N}^{-}(p), k\neq w}  z^{[p]}_{\bar{w}\underline{k}} +z^{[p]}_{\bar{w}\underline{w}}\nonumber \\
	&= \sum_{k:k\neq w} z_{\bar{k}\underline{w}}^{[p]}+z^{[p]}_{\bar{w}\underline{w}} +\sum_{ k\neq w}(z^{[p]}_{\bar{w}\underline{k}}+z_{\bar{k}\underline{k}}^{[p],w})\nonumber \\
	&=\sum_{\underline{k}\in {\cal N}^{-}(p),k\neq w}
    z^{[p]}_{\underline{w}\underline{k}} +\sum_{ k\neq w}z^{[p]}_{\underline{k}\underline{w}}
    =\sum_{\underline{k}\in {\cal N}^{-}(p),k\neq w}
    z^{[p]}_{\underline{w}\underline{k}} +z_{\underline{w}_{[p]}\underline{w}}
    \nonumber 
\end{align}
It means that in the layered OD graph, the demand constraint at destinations holds if the demand constraint at origins and flow conservation at nodes hold. In $[\textbf{M1.1}.\mbox{NCP}\mbox{-VehDispatch}]$, demand constraints are generalized as:
\begin{align}
    \sum_{k,k\neq w} z^{[m]}_{\bar{k}\bar{w}}+\sum_{k,k\neq w} z^{[m]}_{\bar{w}\bar{k}}-q_w^{[m]}
	 \geqslant 0, \forall w \in \mathcal{W}, m \in \mathcal{M}
\end{align}
Therefore, we reformulate $[\textbf{M1.1}.\mbox{NCP}\mbox{-VehDispatch}]$ as: 
\begin{align}
    & 0 \leqslant z_{uv} \perp C_{uv} 
	+\pi_{u} -\pi_{v} 
	+ \lambda_{fleet}t_{uv}-\phi_{v}\geqslant 0, \forall(u,v)\in \mathcal{L} \nonumber \\
	& 0 \leqslant  \lambda_{fleet} \perp
	S-\sum_{\forall (u,v) \in \mathcal{L} } 
	z_{uv} t_{uv}
	\geqslant 0 \nonumber \\
	& \pi_{u}\ \text{free} \perp \sum_{v'} z_{v'u}-\sum_{v} z_{uv} =0 , \forall u \in \mathcal{N} \nonumber \\
	& 0\leqslant \phi_{u} \perp
	\sum_{v'} z_{v'u}-\sum_{w}\delta^{o}_{u\Bar{w}_{[m]}}q_w^{[m]}
	 \geqslant 0, \forall u \in {\cal N}^+(m). \nonumber
\end{align}
In the passenger layer, we have
\begin{align}
    q_w^{[p]}&=y^{(w)[p]}_{\bar{w}\underline{w}}
	+ \sum_{k\in {\cal W}, k\neq w} 
	\left[y^{(w)[p]}_{\bar{w}\bar{k}} + y^{(w)[p]}_{\bar{w}\underline{k}}\right] = y^{(w)[p]}_{\bar{w}\underline{w}}
	+ \sum_{k\in {\cal W}, k\neq w} \left[y^{(w)[p]}_{\bar{k}\underline{w}} 
	+ y^{(w)[p]}_{\bar{k}\underline{k}}+y^{(w)[p]}_{\bar{w}\underline{k}} \right] \nonumber \\
    &=y^{(w)[p]}_{\bar{w}\underline{w}}
	+ \sum_{k\in {\cal W}, k\neq w} 
	\left[y^{(w)[p]}_{\bar{k}\underline{w}} + y^{(w)[p]}_{\underline{k}\underline{w}}\right] \nonumber 
\end{align}
Therefore, the demand constraint at destinations holds if the demand constraint at origins and flow conservation at nodes hold. We also have
\begin{align}
    q_w^{[p]}
    &\leqslant  y^{(w)[p]}_{\bar{w}\underline{w}}
	+ \sum_{k\in {\cal W}, k\neq w} 
	\left[y^{(w)[p]}_{\bar{w}\bar{k}} + y^{(w)[p]}_{\bar{w}\underline{k}}\right]
	\leqslant  z^{[p]}_{\bar{w}\underline{w}}
	+ \sum_{k\in {\cal W}, k\neq w} 
	\left[z^{[p]}_{\bar{w}\bar{k}} + z^{[p]}_{\bar{w}\underline{k}}\right] \nonumber \\
	&\leqslant  \sum_{k\in {\cal W}, k\neq w}z^{[p]}_{\bar{w}\bar{k}}+ (z^{[p]}_{\bar{w}\underline{w}}+\sum_{k\in {\cal W}, k\neq w}z^{[p]}_{\bar{w}\underline{k}})
	=z_{\bar{w}\bar{w}_{[p]}}+\sum_{k\neq w} z^{[p]}_{\bar{k}\bar{w}}\nonumber
\end{align}
This implies that demand constraints in the vehicle layer hold if supply demand constraints in the passenger layer hold. We now generalize $[\textbf{M1.2}.\mbox{NCP}\mbox{-VehPassMatch}]$ as:
\begin{align}
    & 0 \leqslant 
	y^{(w)}_{uv}
	\perp 
	C^{(w)}_{uv} + \pi^{(w)}_{u}-\pi^{(w)}_{v}
	+ \lambda^{(w)}_{uv}-\phi_u^{+}
	\geqslant 0, \forall w \in \mathcal{W}, (u,v) \in {\cal L}^{+}(m) \cup {\cal L}^{-}(m),\nonumber \\
	& \pi^{(w)}_{u}\ \text{free} \perp  \sum_{v'} y^{(w)}_{v'u}-\sum_{v} y^{(w)}_{uv}=0, \forall w \in \mathcal{W}, u \in {\cal N}^{+}(m) \cup {\cal N}^{-}(m) \nonumber \\
	& 0\leqslant \lambda^{(w)}_{uv} \perp
    z_{uv} - y^{(w)}_{uv}\geqslant 0, 
	\forall w\in {\cal W},
	(u,v) \in {\cal L}^{+}(m) \cup {\cal L}^{-}(m) \nonumber \\
	& \phi_u^{+}\ \text{free}\perp \sum_{v} y^{(w)}_{uv}-\sum_{w}\delta^{o}_{u\Bar{w}_{[m]}}q_w^{[m]} =0, \forall u \in {\cal N}^+(m).  \nonumber
\end{align}
Accordingly,  $[\textbf{All}.\mbox{NCP}\mbox{-ESys}]$  is formulated as $[\textbf{All}.\mbox{MCP}]$:
\begin{align}
\fontsize{9}{0}\selectfont
    & [\textbf{All}.\mbox{MCP}] \nonumber\\
    & 0 \leqslant z_{uv} \perp C_{uv} 
	+\pi_{u} -\pi_{v} -\lambda_{uv}
	+ \lambda_{fleet}t_{uv}-\phi_{v} \geqslant 0, \forall(u,v)\in \mathcal{L} \\
	& 0 \leqslant 
	y^{(w)}_{uv}
	\perp C^{(w)}_{uv} + \pi^{(w)}_{u}-\pi^{(w)}_{v}
	+ \lambda^{(w)}_{uv}-\phi_{u}^{+}
	\geqslant 0, \forall w \in \mathcal{W}, (u,v)\in {\cal L}^{+}(m) \cup {\cal L}^{-}(m) \\
	& 0 \leqslant q^{[m]}_{w}  \perp 
	U^{[m]}_{w} - \mu_{w} 
	\geqslant 0, \forall w\in {\cal W}, m \in {\cal M} \\
	& 0 \leqslant  \lambda_{fleet} \perp
	S-\sum_{\forall (u,v) \in \mathcal{L} } 
	z_{uv} t_{uv}
	  \geqslant 0  \label{subeq:penalty_z} \\
	&0 \leqslant x^{d}_{ij} \perp \tau^{d}_{j} + t_{ij}(\mathbf{x}) - \tau^{d}_{i} \geqslant 0,
	\forall (i,j)\in {\cal L}_{(net)},
	d\in {\cal N}  \\
	& 0 \leqslant \tau^{d}_{i} \perp \sum_{j:(i,j)\in {\cal L}_{(net)}} x^{d}_{ij} - \sum_{k:(k,i)\in {\cal L}_{(net)}} x^{d}_{ki}  - q_{(i,d)}\geqslant 0, \forall i\in {\cal N}_{net}, d\in {\cal N}  \\
	& 0 \leqslant o_{uv}
	\perp \sum_{u} z_{uv} \cdot o_{uv}- z_{uv} + \eta_{uv} \geqslant 0, 
	\forall (u,v) \in \mathcal{L}, u \in {\cal N}^-  \\
	& 0 \leqslant o^{(w)}_{uv}
	\perp \sum_{u} y^{(w)}_{uv} \cdot o^{(w)}_{uv}- y^{(w)}_{uv} + \eta^{(w)}_{uv} \geqslant 0, 
	\forall w \in {\cal N}, (u,v) \in \mathcal{L}, u \in {\cal N}^{+}(m)  \\
	& \pi_{u}\ \text{free} \perp \sum_{v'} z_{v'u}-\sum_{v} z_{uv} =0 , \forall u \in \mathcal{N} \\
	& \pi^{(w)}_{u}\ \text{free} \perp \sum_{v'} y^{(w)}_{v'u}-\sum_{v} y^{(w)}_{uv}=0, \forall w \in \mathcal{W}, u \in {\cal N}^{+}(m) \cup {\cal N}^{-}(m) \\
	&  \mu_{w} \ \text{free} \perp
    \sum_{m\in {\cal M}} q^{[m]}_{w} - q_{w}=0,  \forall w\in {\cal W} \\
    & \phi_u^{+}\ \text{free}\perp \sum_{v} y^{(w)}_{uv}-\sum_{w}\delta^{o}_{u\Bar{w}_{[m]}}q_w^{[m]} =0, \forall u \in {\cal N}^+(m)  \label{subeq:penalty_y} \\
	& 0\leqslant \phi_{u} \perp
	\sum_{v'} z_{v'u}-\sum_{w}\delta^{o}_{u\Bar{w}_{[m]}}q_w^{[m]}
	 \geqslant 0, \forall u \in {\cal N}^+(m) \\
	& 0\leqslant \lambda^{(w)}_{uv} \perp
    z_{uv} - y^{(w)}_{uv}\geqslant 0, 
	\forall w\in {\cal W},
	(u,v)\in {\cal L} \\
	& 0 \leqslant  \eta_{uv} \perp 1-o_{uv} \geqslant 0, 
	\forall (u,v) \in \mathcal{L}, u \in  {\cal N}^{-} \\
	& 0 \leqslant  \eta^{(w)}_{uv} \perp 1-o^{(w)}_{uv} \geqslant 0, 
	\forall w \in {\cal W}, (u,v) \in \mathcal{L}, u \in {\cal N}^{+}(m).
\end{align}
To handle the lack of symmetry of multipliers $\lambda_{fleet}$ (\ref{subeq:penalty_z}) and $\phi_w^{+}$ (\ref{subeq:penalty_y}), we add penalties $\epsilon_z$ and $\epsilon_y$ to constraints $S-\sum_{\forall (u,v) \in \mathcal{L} }z_{uv} t_{uv} \geqslant 0$ and $\sum_{v} y^{(w)}_{uv}=\sum_{w}\delta^{o}_{u\Bar{w}_{[m]}}q_w^{[m]}$, respectively. We then have a variational inequality $V(\mathbf{H}^{\epsilon},\mathcal{F})$ where $\epsilon=[\epsilon_z,\epsilon_y]$. We denote $s_+=\max(0,s)$ for any scalar $s$. Mathematically, 
\begin{align}
    & [V(\mathbf{H}^{\epsilon},\mathcal{F})] \nonumber\\
    & \mathbf{H}^{\epsilon}(F) =\begin{Bmatrix}
    & C_{uv}+\epsilon_z t_{uv}\left[ \sum_{(u,v)}z_{uv}t_{uv} -S\right]_{+}, \forall (u,v) \in \mathcal{L} \nonumber \\
    &C^{(w)}_{uv} +\epsilon_y \left[ \sum_{v} y^{(w)}_{uv}-\sum_{w}\delta^{o}_{u\Bar{w}_{[m]}}q_w^{[m]} \right]_+ ,\forall w \in \mathcal{W}, (u,v) \in {\cal L}^{+}(m) \cup {\cal L}^{-}(m) \nonumber \\
    & U_w^{[m]}, \forall w \in \mathcal{W}, m \in \mathcal{M} \nonumber \\
    & t_{ij}(x)-(\tau_i^d-\tau_j^d), \forall (i,j) \in \mathcal{L}_{(net)}, d \in \mathcal{N} \nonumber \\
    & \sum_{j:(i,j)\in {\cal L}_{(net)}} x^{d}_{ij} - \sum_{k:(k,i)\in {\cal L}_{(net)}} x^{d}_{ki}  - q_{(i,d)}, \forall i \in \mathcal{N}_{net}, d \in \mathcal{N} \nonumber \\
    & \sum_{u} z_{uv} \cdot o_{uv}- z_{uv}, \forall (u,v) \in \mathcal{L},  u \in  {\cal N}^{-} \nonumber \\
    & \sum_{u} y^{(w)}_{uv} \cdot o^{(w)}_{uv}- y^{(w)}_{uv} , 
	\forall w \in {\cal W}, (u,v) \in \mathcal{L}, u \in {\cal N}^{+}(m) \nonumber
    \end{Bmatrix}
\end{align}
$\mathcal{F}$ is the feasible set of $F=(\mathbf{z,y,q,x,\tau,o})$ and $(\mathbf{z,y,q,o})\geqslant 0$ satisfy:
\begin{align}
    &\sum_{v'} z_{v'u}=\sum_{v} z_{uv}, \forall u \in \mathcal{N}  \nonumber\\
    & \sum_{v'} y^{(w)}_{v'u}-\sum_{v} y^{(w)}_{uv}=0, \forall w \in \mathcal{W}, u \in {\cal N}^{+}(m) \cup {\cal N}^{-}(m) \nonumber \\
    & \sum_{m\in {\cal M}} q^{[m]}_{w} = q_{w}, \forall w \in \mathcal{W} \nonumber\\
    & 	\sum_{v'} z_{v'u}  \geqslant \sum_{w}\delta^{o}_{u\Bar{w}_{[m]}}q_w^{[m]}
	 , \forall u \in {\cal N}^+(m)  \nonumber\\
    &y^{w}_{uv} \leqslant z_{uv}, \forall w\in {\cal W},
	(u,v) \in {\cal L}^{+}(m) \cup {\cal L}^{-}(m) \nonumber  \\ 
    &o_{uv} \leqslant 1, \forall (u,v) \in \mathcal{L}, u \in {\cal N}^{-} \nonumber \\
     &o^{(w)}_{uv} \leqslant 1, \forall w \in {\cal W}, (u,v) \in \mathcal{L}, u \in {\cal N}^{+}(m). \nonumber
\end{align}
We now demonstrate the solution existence of $V(\mathbf{H}^{\epsilon},\mathcal{F})$. Note that the feasible set ${\cal F}$ is unbounded. We first consider a modified variational inequality $V(\mathbf{H}^{\epsilon},\mathcal{F}_{\zeta})$ with assumptions in Proposition \ref{prop:bounded}. We have 
\begin{align}
    y^{w}_{uv} \leqslant z_{uv} \leqslant \zeta \sum_{w \in {\cal W}}\cdot q_w, \forall w\in {\cal W},
	(u,v) \in {\cal L}^{+}(m) \cup {\cal L}^{-}(m)
\end{align}
The feasible set $\mathcal{F}_{\zeta}$ of $F=(\mathbf{z,y,q,x,\tau,o})$ is closed and bounded (i.e., compact) polyhedron. According to Thm \ref{thm:Nagurney2008VI}, $V(\mathbf{H}^{\epsilon},\mathcal{F}_{\zeta})$ has a solution for every $\epsilon>0$. Therefore, $V(\mathbf{H}^{\epsilon},\mathcal{F})$ at least has one solution $\forall \epsilon>0$ (Thm \ref{thm:Nagurney2008VI_2}).

\noindent We then show the solution of $[\textbf{All}.\mbox{MCP}]$ is recovered by $[V(\mathbf{H}^{\epsilon},\mathcal{F})]$ when $\epsilon_z,\epsilon_y \rightarrow \infty$. An MCP formulation of  $V(\mathbf{H}^{\epsilon},\mathcal{F})$ is obtained as follows: We denote  {\small $\lambda^{\epsilon}_{fleet}=\epsilon_z \left[ \sum_{(u,v)}z_{uv}t_{uv} -S\right]_{+}$} and {\small $\phi_{w}^{\epsilon+}=\epsilon_y\left[\sum_{v} y^{(w)}_{uv}-\sum_{w}\delta^{o}_{u\Bar{w}_{[m]}}q_w^{[m]}\right]_{+}$}. Mathematically,
\begin{align}
\fontsize{9}{0}\selectfont
    & [\textbf{All}.\mbox{MCP}\mbox{-$V(\mathbf{H}^{\epsilon},\mathcal{F})$}] \nonumber\\
    & 0 \leqslant z_{uv} \perp C_{uv} 
	+\pi_{u} -\pi_{v} -\lambda_{uv}
	+ \lambda_{fleet}^{\epsilon}t_{uv}-\phi_{v} \geqslant 0, \forall(u,v)\in \mathcal{L} \nonumber \\
	& 0 \leqslant 
	y^{(w)}_{uv}
	\perp C^{(w)}_{uv} + \pi^{(w)}_{u}-\pi^{(w)}_{v}
	+ \lambda^{(w)}_{uv}-\phi_{u}^{\epsilon+}
	\geqslant 0, \forall w \in \mathcal{W}, (u,v)\in {\cal L}^{+}(m) \cup {\cal L}^{-}(m) \nonumber \\
	& 0 \leqslant q^{[m]}_{w}  \perp 
	U^{[m]}_{w} - \mu_{w} 
	\geqslant 0, \forall w\in {\cal W}, m \in {\cal M} \nonumber \\
	&0 \leqslant x^{d}_{ij} \perp \tau^{d}_{j} + t_{ij}(\mathbf{x}) - \tau^{d}_{i} \geqslant 0,
	\forall (i,j)\in {\cal L}_{(net)},
	d\in {\cal N}  \nonumber \\
	& 0 \leqslant \tau^{d}_{i} \perp \sum_{j:(i,j)\in {\cal L}_{(net)}} x^{d}_{ij} - \sum_{k:(k,i)\in {\cal L}_{(net)}} x^{d}_{ki}  - q_{(i,d)}\geqslant 0, \forall i\in {\cal N}_{net}, d\in {\cal N} \nonumber \\
	& 0 \leqslant o_{uv}
	\perp \sum_{u} z_{uv} \cdot o_{uv}- z_{uv} + \eta_{uv} \geqslant 0, 
	\forall (u,v) \in \mathcal{L}, u \in {\cal N}^-  \nonumber \\
	& 0 \leqslant o^{(w)}_{uv}
	\perp \sum_{u} y^{(w)}_{uv} \cdot o^{(w)}_{uv}- y^{(w)}_{uv} + \eta^{(w)}_{uv} \geqslant 0, 
	\forall w \in {\cal N}, (u,v) \in \mathcal{L}, u \in {\cal N}^{+}(m)  \nonumber \\
	& \pi_{u}\ \text{free} \perp \sum_{v'} z_{v'u}-\sum_{v} z_{uv} =0 , \forall u \in \mathcal{N} \nonumber \\
	& \pi^{(w)}_{u}\ \text{free} \perp \sum_{v'} y^{(w)}_{v'u}-\sum_{v} y^{(w)}_{uv}=0, \forall w \in \mathcal{W}, u \in {\cal N}^{+}(m) \cup {\cal N}^{-}(m) \nonumber \\
	&  \mu_{w} \ \text{free} \perp
    \sum_{m\in {\cal M}} q^{[m]}_{w} - q_{w}=0,  \forall w\in {\cal W} \nonumber \\
	& 0\leqslant \phi_{u} \perp
	\sum_{v'} z_{v'u}-\sum_{w}\delta^{o}_{u\Bar{w}_{[m]}}q_w^{[m]}
	 \geqslant 0, \forall u \in {\cal N}^+(m) \nonumber \\
	& 0\leqslant \lambda^{(w)}_{uv} \perp
    z_{uv} - y^{(w)}_{uv}\geqslant 0, 
	\forall w\in {\cal W},
	(u,v)\in {\cal L} \nonumber \\
	& 0 \leqslant  \eta_{uv} \perp 1-o_{uv} \geqslant 0, 
	\forall (u,v) \in \mathcal{L}, u \in  {\cal N}^{-} \nonumber \\
	& 0 \leqslant  \eta^{(w)}_{uv} \perp 1-o^{(w)}_{uv} \geqslant 0, 
	\forall w \in {\cal W}, (u,v) \in \mathcal{L}, u \in {\cal N}^{+}(m). \nonumber
\end{align}
We follow \citep{ban2019general} by applying a limit argument to the penalization $\epsilon^b=[\epsilon^b_z,\epsilon^b_y]$. $F^b=(\mathbf{z}^b,\mathbf{y}^b,\mathbf{q}^b,\mathbf{x}^b,\mathbf{\tau}^b,\mathbf{o}^b)$ represents a solution of $[V(\mathbf{H}^{\epsilon^b},\mathcal{F})]$ for each $b$. We have $\forall F \in \mathcal{F}, \ \mathbf{H}^{\epsilon^b}(F^b)^{T}(F-F^b)\geqslant 0$. We denote $\bar{F}^b=(\bar{\mathbf{z}}^b,\bar{\mathbf{y}}^b,\mathbf{q}^b,\mathbf{x}^b,\mathbf{\tau}^b,\mathbf{o}^b), \bar{F}^b \in \mathcal{F}$. $\bar{\mathbf{z}}^b$ and $\bar{\mathbf{y}}^b$ satisfy
\begin{align}
      &S-\sum_{\forall (u,v) \in \mathcal{L} } 
	\bar{z}^b_{uv} t^b_{uv}
	  \geqslant \eta \nonumber \\
	  &\sum_{v} \Bar{y}^{(w),b}_{uv}-\sum_{w}\delta^{o}_{u\Bar{w}_{[m]}}q_w^{[m],b} =0, \forall w \in \mathcal{W}, u \in {\cal N}^+(m) \cup {\cal N}^-(m) \nonumber 
\end{align}
Therefore, 
\begin{align}
    0 \leqslant & \mathbf{H}^{\epsilon^b}(\mathcal{F}^b)^{T}(\bar{\mathcal{F}}^b-\mathcal{F}^b) \nonumber \\
    = & \text{bounded term}+\sum_{(u,v)} \epsilon^b_z t_{uv}\left[ \sum_{(u,v)}z^b_{uv}t^b_{uv} -S\right]_{+} (\bar{z}^{b}_{uv}-z^{b}_{uv}) + \sum_{w}\sum_{(u,v)}  \epsilon^b_y  (\sum_{v} \Bar{y}^{(w),b}_{uv}-\sum_{w}\delta^{o}_{u\Bar{w}_{[m]}}q_w^{[m],b} )\  (\bar{y}^{(w),b}_{uv}-y^{(w),b}_{uv}) \nonumber \\
    =& \text{bounded term}+\epsilon^b_z \left[ \sum_{(u,v)}z^b_{uv}t^b_{uv} -S\right]_{+}\sum_{(u,v)} t^b_{uv} (\bar{z}^{b}_{uv}-z^{b}_{uv}) - \epsilon^b_y \sum_{w} (\sum_{v} y^{(w),b}_{uv}-\sum_{w}\delta^{o}_{u\Bar{w}_{[m]}}q_w^{[m],b})^2 \nonumber 
\end{align}
We have $ \sum_{(u,v)} t^b_{uv} (\bar{z}^{b}_{uv}-z^{b}_{uv})=\sum_{(u,v)} t^b_{uv} \bar{z}^{b}_{uv}-\sum_{(u,v)} t^b_{uv} z^{b}_{uv} \leqslant S- \sum_{(u,v)} t^b_{uv} z^{b}_{uv} -\eta$. Accordingly,
\begin{align}
   0\leqslant& \text{bounded term}-\epsilon^b_z \left[ \sum_{(u,v)}z^b_{uv}t_{uv} -S\right]_{+}^2-\eta \cdot  \epsilon_z^b \left[ \sum_{(u,v)}z_{uv}t_{uv} -S\right]_{+} - \epsilon^b_y \sum_{w} (\sum_{v} y^{(w),b}_{uv}-\sum_{w}\delta^{o}_{u\Bar{w}_{[m]}}q_w^{[m]})^2 \nonumber 
\end{align}
When $\lim_{b \rightarrow \infty}\epsilon^b_z=\infty$ and $\lim_{b \rightarrow \infty}\epsilon^b_y =\infty$, we have $\lambda_{fleet}^{b}=\epsilon_z^b \left[ \sum_{(u,v)}z^b_{uv}t^b_{uv} -S\right]_{+}$ is bounded and 
\begin{align}
      &\lim_{b\rightarrow \infty}\left[\sum_{\forall (u,v) \in \mathcal{L} } 
	z^b_{uv} t^b_{uv}-S \right]_+
	  = 0 \nonumber \\
	  &\lim_{b\rightarrow \infty} (\sum_{v} y^{(w),b}_{uv}-\sum_{w}\delta^{o}_{u\Bar{w}_{[m]}}q_w^{[m],b})=0, \forall w \in \mathcal{W},u \in {\cal N}^+(m) \cup {\cal N}^-(m) \nonumber
\end{align}
Let $E_u^{(w),b}$ denote $\sum_{v} y^{(w),b}_{uv}-\sum_{w}\delta^{o}_{u\Bar{w}_{[m]}}q_w^{[m],b}$ and $a^b$ denote $\left[\sum_{\forall (u,v) \in \mathcal{L} }z^b_{uv} t^b_{uv}-S \right]_+$. We have the following MCP:
\begin{align}
\fontsize{9}{0}\selectfont
    & [\textbf{All}.\mbox{MCP}\mbox{-$V(\mathbf{H}^{\epsilon^b},\mathcal{F})$}] \nonumber\\
    & 0 \leqslant z^b_{uv} \perp C^b_{uv} 
	+\pi^b_{u} -\pi^b_{v} -\lambda^b_{uv}
	+ \lambda_{fleet}^bt^b_{uv}-\phi^b_{v} \geqslant 0, \forall(u,v)\in \mathcal{L} \nonumber \\
	& 0 \leqslant 
	y^{(w),b}_{uv}
	\perp C^{(w),b}_{uv} + \pi^{(w),b}_{u}-\pi^{(w),b}_{v}
	+ \lambda^{(w),b}_{uv}-\phi_{u}^{\epsilon^b +}
	\geqslant 0, \forall w \in \mathcal{W}, (u,v)\in {\cal L}^{+}(m) \cup {\cal L}^{-}(m) \nonumber \\
	& 0 \leqslant q^{[m],b}_{w}  \perp 
	U^{[m],b}_{w} - \mu^b_{w} 
	\geqslant 0, \forall w\in {\cal W}, m \in {\cal M} \nonumber \\
	& 0 \leqslant  \lambda_{fleet}^b \perp
	a^b+S-\sum_{\forall (u,v) \in \mathcal{L} } 
	z^b_{uv} t^b_{uv} \geqslant 0 \nonumber \\
	&0 \leqslant x^{d,b}_{ij} \perp \tau^{d,b}_{j} + t^b_{ij}(\mathbf{x}) - \tau^{d,b}_{i} \geqslant 0,
	\forall (i,j)\in {\cal L}_{(net)},
	d\in {\cal N}  \nonumber \\
	& 0 \leqslant \tau^{d,b}_{i} \perp \sum_{j:(i,j)\in {\cal L}_{(net)}} x^{d,b}_{ij} - \sum_{k:(k,i)\in {\cal L}_{(net)}} x^{d,b}_{ki}  - q^b_{(i,d)}\geqslant 0, \forall i\in {\cal N}_{net}, d\in {\cal N} \nonumber \\
	& 0 \leqslant o^b_{uv}
	\perp \sum_{u} z^b_{uv} \cdot o^b_{uv}- z^b_{uv} + \eta^b_{uv} \geqslant 0, 
	\forall (u,v) \in \mathcal{L}, u \in {\cal N}^-  \nonumber \\
	& 0 \leqslant o^{(w),b}_{uv}
	\perp \sum_{u} y^{(w),b}_{uv} \cdot o^{(w),b}_{uv}- y^{(w),b}_{uv} + \eta^{(w),b}_{uv} \geqslant 0, 
	\forall w \in {\cal N}, (u,v) \in \mathcal{L}, u \in {\cal N}^{+}(m)  \nonumber \\
	& \pi^b_{u}\ \text{free} \perp \sum_{v'} z^b_{v'u}-\sum_{v} z^b_{uv} =0 , \forall u \in \mathcal{N} \nonumber \\
	& \pi^{(w),b}_{u}\ \text{free} \perp \sum_{v'} y^{(w),b}_{v'u}-\sum_{v} y^{(w),b}_{uv}=0, \forall w \in \mathcal{W}, u \in {\cal N}^{+}(m) \cup {\cal N}^{-}(m) \nonumber \\
	&  \mu^b_{w} \ \text{free} \perp
    \sum_{m\in {\cal M}} q^{[m],b}_{w} - q_{w}=0,  \forall w\in {\cal W} \nonumber \\
    & \phi_u^{\epsilon^b +}\ \text{free}\perp \sum_{v} y^{(w),b}_{uv}-\sum_{w}\delta^{o}_{u\Bar{w}_{[m]}}q_w^{[m],b} =E_u^{(w),b}, \forall u \in {\cal N}^+(m)  \nonumber \\
	& 0\leqslant \phi^b_{u} \perp
	\sum_{v'} z^b_{v'u}-\sum_{w}\delta^{o}_{u\Bar{w}_{[m]}}q_w^{[m],b}
	 \geqslant 0, \forall u \in {\cal N}^+(m) \nonumber \\
	& 0\leqslant \lambda^{(w),b}_{uv} \perp
    z^b_{uv} - y^{(w),b}_{uv}\geqslant 0, 
	\forall w\in {\cal W},
	(u,v)\in {\cal L} \nonumber \\
	& 0 \leqslant  \eta^b_{uv} \perp 1-o^b_{uv} \geqslant 0,
	\forall (u,v) \in \mathcal{L}, u \in  {\cal N}^{-} \nonumber \\
	& 0 \leqslant  \eta^{(w),b}_{uv} \perp 1-o^{(w),b}_{uv} \geqslant 0, 
	\forall w \in {\cal W}, (u,v) \in \mathcal{L}, u \in {\cal N}^{+}(m). \nonumber
\end{align}
Therefore, $[\textbf{All}.\mbox{MCP}\mbox{-$V(\mathbf{H}^{\epsilon^b},\mathcal{F})$}]$ when $\epsilon^b \rightarrow \infty$ satisfies assumptions of Lemma \ref{lemma:cottle}. The solution of $[\textbf{All}.\mbox{NCP}\mbox{-ESys}]$ exists. Proposition \ref{prop:VI_penalty} holds.

\end{proof}

\subsection{Numerical results on the small network}
\label{append:parameters_small}

\textbf{Goal:} There are two TNC companies (Company~1 and Company~2) in the market and each provides both e-solo and e-pooling services. 
We would like to investigate how fleet sizing of companies impacts their market shares. The competition between e-hailing providers is not modeled directly. 
Instead, e-hailing providers select their revenue structures and fleet sizing first and leave it for travelers to decide which provider to take. 
Once the demand for each provider is determined, the role each platform plays is simply to match orders to vacant vehicles. 
Before delving into the example, we first present the generalized model to multiple e-platforms.
Denote a set of e-hailing service providers that offers both e-solo and e-pooling modes as ${\cal C}=\left\lbrace c \right\rbrace_{c\in\mathbb{N}^+}$. 
Then the vehicle flows, passenger flows, and customer mode choice are augmented with an additional subscript $c$ (from $m$ to $m_c$) for all the modules. 

\begin{subequations}
\fontsize{9}{0}\selectfont
\begin{align}
	& [\textbf{M1.1}.\mbox{NCP}\mbox{-VehDispatch-c}] \nonumber\\
	& 0 \leqslant z^{\bar{w},c}_{\underline{k}} \perp C^{\bar{w},c}_{\underline{k}} 
	-\pi_{\bar{w},c} +\pi_{\underline{k},c} 
	+ \lambda_{fleet,c} 
	\geqslant 0, \forall w \in {\cal W}, c \in {\cal C} \nonumber \\
	& 0 \leqslant z^{[p_c]}_{\bar{w} \bar{k}}
	\perp 
	 C^{[p_c]}_{\bar{w} \bar{k}} 
	+\pi^{[p_c]+}_{\bar{w}} -\pi^{[p_c]-}_{\bar{k}}
	- \phi^{[p_c]}_{\bar{k}} 
	- \lambda^{[p_c]}_{\bar{w}\bar{k}}
	+ \lambda_{fleet,c}
	\geqslant 0, \forall w,k\in {\cal W},k\neq w, c \in {\cal C} \nonumber\\
	& 0 \leqslant z^{[p_c]}_{\bar{k}\underline{w}} 
	\perp C^{[p_c]}_{\bar{k}\underline{w}} 
	+\pi^{[p_c]-}_{\bar{k}}
	-\pi^{[p_c]+}_{\underline{w}}
    -\lambda^{[p_c]}_{\bar{k}\underline{w}}
	+\lambda_{fleet,c} \geqslant 0, \forall w,k\in {\cal W}, k\neq w,c \in {\cal C} \nonumber \\
	& 0 \leqslant z^{[p_c]}_{\bar{w}\underline{w}} 
	\perp C^{[p_c]}_{\bar{w}\underline{w}} 
	+\pi^{[p_c]-}_{\bar{w}}
	-\pi^{[p_c]+}_{\underline{w}}
	-\lambda^{[p_c]}_{\bar{w}\underline{w}}
	+\lambda_{fleet,c} \geqslant 0, \forall w\in {\cal W}, c \in {\cal C} \nonumber\\
	& 0 \leqslant z^{[p_c]}_{\underline{w}\underline{k}} 
	\perp C^{[p_c]}_{\underline{w}\underline{k}}  
	+\pi^{[p_c]+}_{\underline{w}}
	-\lambda^{[p_c]}_{\underline{w}\underline{k}}
	-\pi^{[p_c]-}_{\underline{k}}
	- \phi^{[p_c]}_{\underline{w}} 
	+\lambda_{fleet,c} \geqslant 0, \forall w,k\in {\cal W}, k\neq w, c \in {\cal C} \nonumber\\
    & 0\leqslant z_{\Bar{w}\Bar{w}_{[p_c]}} \perp 
    C_{\Bar{w}\Bar{w}_{[p_c]}} 
    - \pi^{[p_c]+}_{\Bar{w}} 
    - \phi^{[p_c]}_{\Bar{w}} 
    - \pi_{\Bar{w},c}
    -\lambda_c
    \geqslant 0, \forall w\in {\cal W}, c \in {\cal C}
    \nonumber\\
	& 0\leqslant z_{\underline{w}_{[p_c]}\underline{w}} \perp 
	C_{\underline{w}_{[p_c]}\underline{w}} 
	+ \pi^{[p_c]-}_{\underline{w}} 
	- \phi^{[p_c]}_{\underline{w}} 
	+ \pi_{\underline{w},c} 
	+\lambda_c
	\geqslant 0, \forall w\in {\cal W}, c \in {\cal C}
	\nonumber	\\
	& 0 \leqslant z^{[e_c]}_{\bar{w}\underline{w}} 
	\perp C^{[e_c]}_{\bar{w}\underline{w}} 
	+ \pi^{[e_c]}_{\bar{w}} - \pi^{[e_c]}_{\underline{w}}
	-\lambda^{[e_c]}_{\bar{w}\underline{w}}
	+ \lambda_{fleet,c}
	\geqslant 0, \forall w \in {\cal W}, c \in {\cal C} \nonumber\\
	& 0\leqslant z_{\Bar{w}\Bar{w}_{[e_c]}} \perp 
    C_{\Bar{w}\Bar{w}_{[e_c]}} - \pi^{[e_c]}_{\Bar{w}} + \pi_{\Bar{w},c}
    - \phi^{[e_c]}_{w} 
    -\lambda_c
    \geqslant 0, \forall w\in {\cal W}, c \in {\cal C}
	\nonumber \\
	& 0\leqslant z_{\underline{w}_{[e_c]}\underline{w}} \perp 
	C_{\underline{w}_{[e_c]}\underline{w}} + \pi^{[e_c]}_{\underline{w}} - \pi_{\underline{w},c} 
	+\lambda_c 
	\geqslant 0, \forall w\in {\cal W}, c \in {\cal C}
	\nonumber \\
	& 0=z_{\Bar{w}\Bar{w}_{[p_c]}} - \sum_{k\in {\cal W},k\neq w} z^{[p_c]}_{\Bar{w}\Bar{k}} \perp \pi^{[p_c]+}_{\Bar{w}} \mbox{ free}, 
	\forall w\in {\cal W}, c \in {\cal C} 
	\nonumber \\
	& 0=\sum_{k\in {\cal W},k\neq w} z^{[p_c]}_{\Bar{k}\bar{w}} - \sum_{k\in {\cal W}, k \neq w} z^{[p_c]}_{\Bar{w}\underline{k}}-z^{[p_c]}_{\Bar{w}\underline{w}}
	\perp \pi^{[p_c]-}_{\Bar{w}} \mbox{ free}, 
	\forall w\in {\cal W}, c \in {\cal C} 
	\nonumber \\
    & 0 = z^{[p_c]}_{\bar{w}\underline{w}} 
	+ \sum_{k \in {\cal W}, k\neq w} z^{[p_c]}_{\bar{k}\underline{w}} - 
    \sum_{k \in {\cal W}, k\neq w}
    z^{[p_c]}_{\underline{w}\underline{k}}
    \perp \pi^{[p_c]+}_{\underline{w}} \mbox{ free},
	\forall  w\in {\cal W}, c \in {\cal C} 
	\nonumber \\
	& 
	0 = \sum_{k\neq w} 
    z^{[p_c]}_{\underline{k}\underline{w}}
	- z_{\underline{w}_{[p_c]}\underline{w}}
	\perp \pi^{[p_c]-}_{\underline{w}} \mbox{ free}, 
	\forall w\in {\cal W},c \in {\cal C}
	\nonumber \\
	& 0\leqslant z_{\bar{w}\bar{w}_{[p_c]}} 
	+ \sum_{k\in {\cal W},k\neq w} z^{[p_c]}_{\Bar{k}\bar{w}}
	-q^{[p_c]}_{w} 
	\perp \phi^{[p_c]}_{\bar{w}} \geqslant 0, 
	\forall w\in {\cal W}, c \in {\cal C}
	\nonumber \\
	& 0\leqslant z_{\underline{w}_{[p_c]}\underline{w}} 
	+ \sum_{k\in {\cal W},k\neq w} z^{[p_c]}_{\underline{w}\underline{k}} - q^{[p_c]}_{w}
	\perp \phi^{[p_c]}_{\underline{w}} \geqslant 0, 
	\forall w\in {\cal W}, c \in {\cal C}
	\nonumber \\
	& 0=z_{\bar{w} \bar{w}_{[e_c]}} - z^{[e_c]}_{\bar{w}\underline{w}}
	\perp \pi^{[e_c]}_{\bar{w}} \mbox{ free}, 
	\forall w\in {\cal W}, c \in {\cal C}
	\nonumber \\
	& 0=z^{[e_c]}_{\bar{w}\underline{w}} - z_{\underline{w}_{[e_c]}\underline{w}}
	\perp \pi^{[e_c]}_{\underline{w}} \mbox{ free}, 
	\forall w\in {\cal W}, c \in {\cal C}
	\nonumber \\
	& 0\leqslant z_{\bar{w}\bar{w}_{[e_c]}}-q^{[e_c]}_{w} 
	\perp \phi^{[e_c]}_{w} \geqslant 0, 
	\forall w\in {\cal W}, c \in {\cal C}
	\nonumber \\
	& 0=\sum_{k\in {\cal W}} z^{\Bar{w},c}_{\underline{k}}
	-\sum_{m\in {\cal M}} z_{\Bar{w}\Bar{w}_{[m_c]}}
	\perp \pi_{\Bar{w},c} \mbox{ free},  
	\forall w\in {\cal W}, c \in {\cal C}
	\nonumber \\
	& 0=\sum_{m\in {\cal M}} z_{\underline{w}_{[m_c]}\underline{w}}
	-\sum_{k \in {\cal W}} z^{\Bar{k},c}_{\underline{w}} 
	\perp \pi_{\underline{w},c} \mbox{ free},  
	\forall w \in {\cal W}, c \in {\cal C}
	\nonumber \\
    & 0= \sum_{m_c\in {\cal M}} \sum_{w\in {\cal W}} z_{\bar{w}\bar{w}_{[m_c]}} 
	- \sum_{m_c\in {\cal M}} \sum_{w\in {\cal W}} z_{\underline{w}_{[m_c]}\underline{w}}\perp \lambda_c  \mbox{ free}, \forall c \in {\cal C}
	\nonumber \\
	& 0\leqslant 
	S_c
	-\sum_{k,w \in {\cal W}} 
	z^{\Bar{w},c}_{\underline{k}} t_{\underline{k}\Bar{w}}
	-\sum_{m_c\in {\cal M}}\sum_{(v,u)\in {\cal L}(m_c)} z^{[m_c]}_{vu}t_{vu}
	\perp \lambda_{fleet,c} \geqslant 0, \forall c \in {\cal C}
    \nonumber
\end{align}
\end{subequations}
The parameter values are shown in Table~(\ref{tab:para_val}). 
\begin{table}[H]\centering
	\caption{Parameter values for Small Network}\label{tab:para_val}
	\begin{tabular}{l|l|l}
		\hline
		Name & Notations & Values  \\\hline\hline
		Total travel demands & $q_{1},q_{2}$ & $2,1$ \\ \hline 
		Fleet sizing for Company~1 & $Cap \times 3.5$ & $Cap\in\left[0,2\right]$\\
		Fleet sizing for Company~2 &  & $7$ \\ \hline
		Fare discount (or multiplier) & $\gamma_p$ & $.8$ \\
		Fixed fare & $F_{1},F_{2}$ & .2, 0.5  \\ 
		Time-based fare rate & $\alpha^{[e]}_1,\alpha^{[p]}_1$ & .3, .3 \\ 
		Distance-based fare rate & $\alpha^{[e]}_2,\alpha^{[p]}_2$ & 1,1 \\ \hline 
		Value of time while traveling & $\beta^{[e]}_1,\beta^{[p]}_1$ & .2, .2 \\ 
		Value of time while waiting for pickup & $\beta^{[e]}_{(wt)},\beta^{[p]}_{(wt)}$ & .5, .5 \\ 
		Value of time while matching to drivers & $\beta_{(se-veh)},\beta_{(se-pas)}$ & 1, 1 \\ 
		Supply-demand surplus cost & $\beta^{[m]}_{(surp-pas)}$ & 0, 0 \\
		Conversion rate from in-vehicle travel time to cost & $\beta_{(in-veh)}$ & .2 \\ 
		Inconvenience cost for e-pooling & $\beta_{1(inc)},\beta_{2(inc)}$ & .2, .2 \\ \hline 
	\end{tabular}
\end{table}

The link performance functions are the Bureau of Public Roads (BPR) function, i.e., 

\begin{equation}
	t_{ij}(x_{ij})=t^0_{ij}\left[1+A\left(\frac{x_{ij}}{CAPACITY_{ij}}\right)^{B}\right],
\end{equation}

where $x_{ij}$ is the vehicular flow, $t^0_{ij}$ is the free-flow travel time, $A,B$ are coefficients, and $CAPACITY_{ij}$ is the link capacity (i.e., maximum vehicular flow).
The coefficients in the BPR link cost functions are listed in Table~(\ref{tab:cost_BR}).

\begin{table}[H]\centering
	\caption{Cost coefficients}\label{tab:cost_BR}
	\begin{tabular}{c|c|c c c c}
		\hline
Link & Distance & $t^0_{ij}$  &   $A$    &   $B$  &   CAPACITY  \\\hline\hline
(1,2)  &  1 &  1 &  0.15  &  4  & .8 \\
(1,3)  &  1.5 &  1.5 & 0.15  &  4  &  .8 \\
(2,1)  &  .5 &  .5 & 0.15  &  4  &  .8 \\
(2,3)  &  1 &  1 & 0.15  &  4  &  .8 \\
(3,2)  &  1 &  1 & 0.15  &  4  &  .8 \\ \hline
\end{tabular}
\end{table}

\begin{rem}
\begin{enumerate}
\item With this parameter setting, all passengers select e-solo and there is no e-pooling order. 
\item Company 1 offers service at a lower fare rate than Company 2. Thus, as long as Company 1 has capacity, it serves orders first. 
Only when Company 1's fleet sizing is limited (i.e., less than the required fleets), passengers have no option but to select Company 2 for service. 
\end{enumerate}
\end{rem}

\begin{figure}[H]
	\centering
	\subfloat[Demand served by Company 1]{\includegraphics[scale=.47]{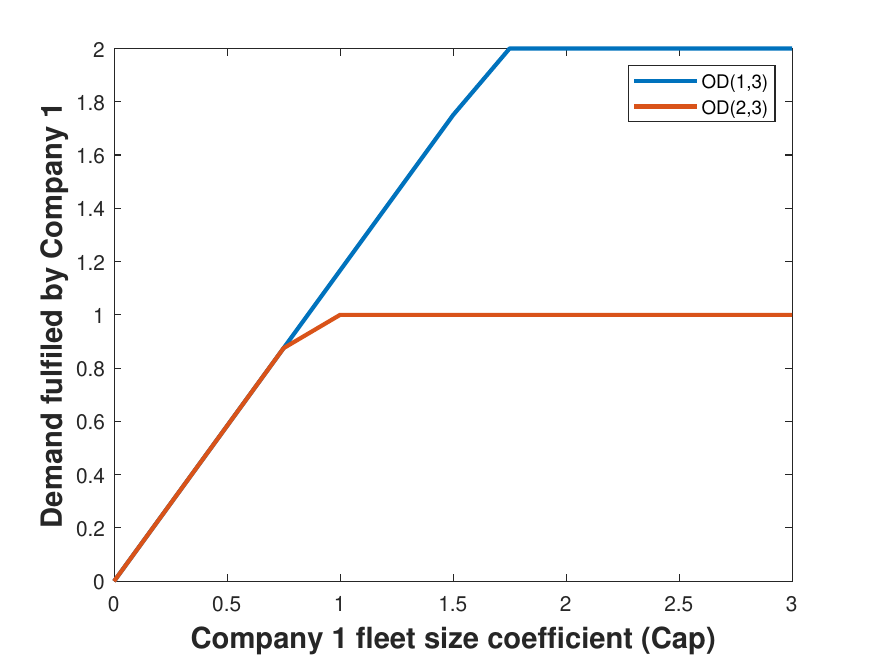}\label{subfig:2comp_dm}}~
	\hspace{3mm}
	\subfloat[Demand proportion by company]{\includegraphics[scale=.47]{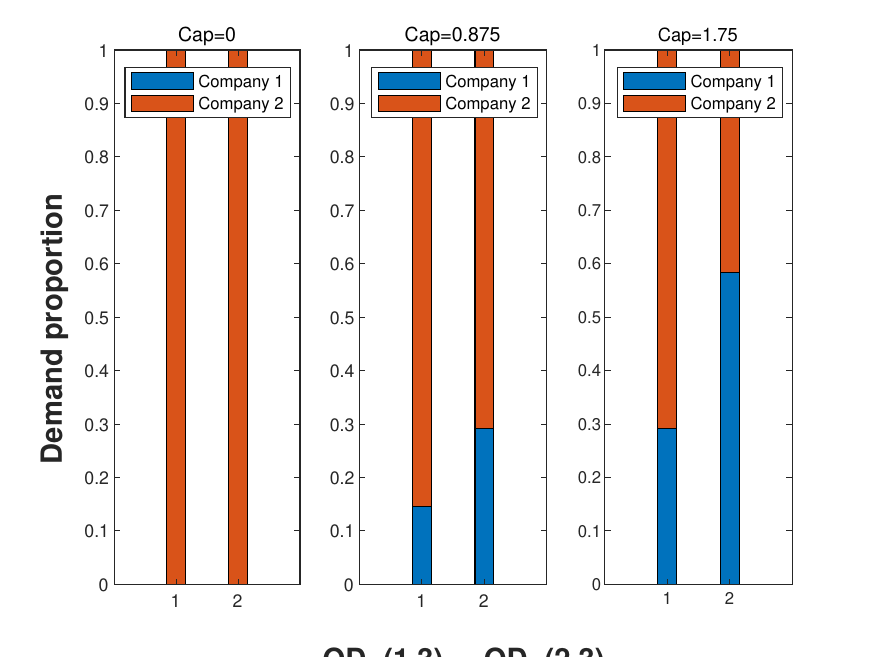}\label{subfig:2comp_bar_dm}}
	
	\subfloat[Cost]{\includegraphics[scale=.47]{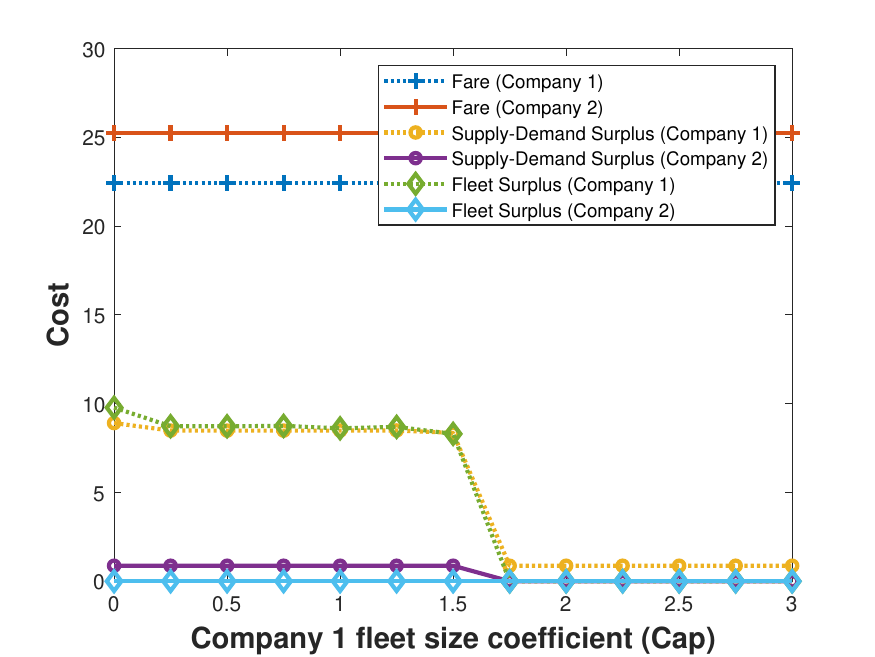}\label{subfig:2comp_cost}}~
	\hspace{3mm}
	\subfloat[Performance by company]{\includegraphics[scale=.47]{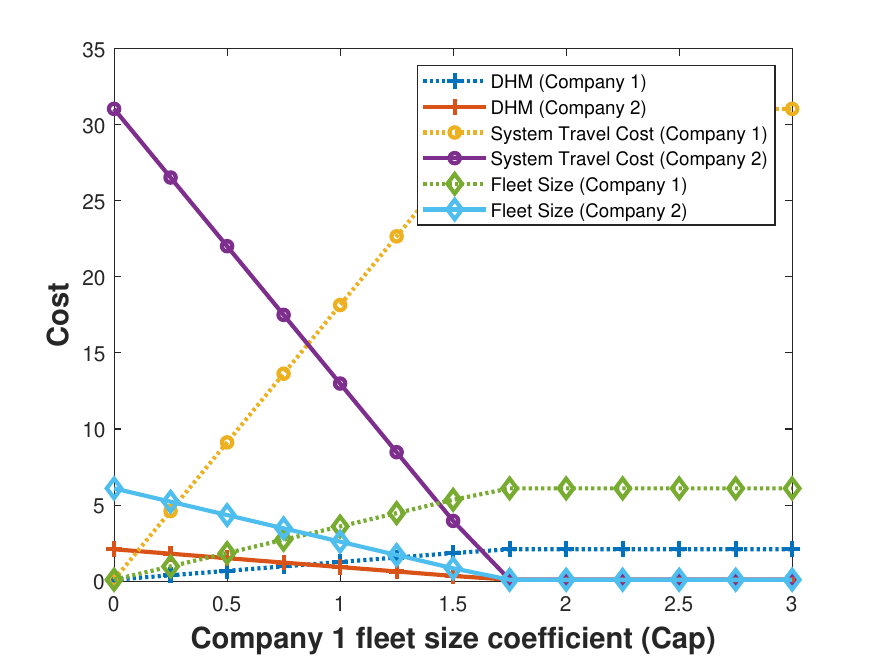}\label{subfig:2comp_perf2}}
	\caption{Performance measures on small network}
	\label{fig:2comp}
\end{figure}

Running GAMS 24.4.6 on a DELL desktop with Intel i7 CPU \@ 3.40GHz and 16GB RAM. The algorithm stops with a residual less than $10^{-7}$.  
The demands, cost, and performances are demonstrated in Fig.~\ref{fig:2comp}.

In Fig.\ref{subfig:2comp_dm}, the x-axis is the fleet coefficient of Company~1 growing from 0 to 3 and the y-axis is the demand level served by Company~1.  
Travel demands fulfilled by Company~1 increase as its fleet size increases.
Now let us focus on two extremes. 
When Company~1 does not contain any fleet, no demand is fulfilled by Company~1.
When its fleet size increases to the maximum, all orders are served by Company~1. 

In Fig.\ref{subfig:2comp_bar_dm}, each stacked bar indicates the market share of 2 companies for OD (1,3) and OD (2,3), respectively. 
Each subfigure plots the both OD shares given a specified fleet size of Company~1, of which the value is indicated in the title. 
The market share of Company~1 is depicted by a blue bar and that of Company~2 by a yellow bar. 
As the fleet size of Company~1 increases (from left to right subfigures), the market share of Company~1 for OD (1,3) increases accordingly.

In Fig.\ref{subfig:2comp_cost}, the x-axis is the fleet size of Company~1, and the y-axis lists 
fare (the dotted blue line with cross makers for Company~1 and the solid red line with cross markers for Company~2), 
supply-demand surplus (the orange dotted line with circle makers for Company~1 and the purple line with circle markers for Company~2), 
and fleet size surplus (the green dotted line with diamond makers for Company~1 and the blue line with diamond markers for Company~2). 
The fares for both companies remain constant across the fleet size coefficient. This is because total travel time and distance that factor into the fare computation stay the same. 
The fare of Company~1 is lower than that of Company~2, because its overall fare coefficients are smaller. 
With a zero fleet size of Company~1, passengers are forced to choose Company~2 paying higher fares. 
The supply-demand surplus represents the fee paid by passengers to drivers when the demand is higher than the supply. 
A zero value means that the vehicle supply can fulfill the demand. 
As the fleet size of Company~1 increases, Company~1's supply-demand surplus is always greater than zero, meaning that all its fleet size is used to serve the demand and there is no extra supply. 
The fleet surplus represents the incentive from a company to its drivers to expand the supply size. 
Because Company~2 has a sufficient fleet size, its fleet surplus is always zero. 
For Company~1, when its fleet size coefficient is less than 2 (i.e., insufficient fleet sizes), its fleet surplus is positive (indicating that Company~1 struggles with recruiting drivers and needs to pay incentives to drivers) 
and is reduced to zero when greater than 2. 

In Fig.\ref{subfig:2comp_perf2}, the x-axis is the fleet size coefficient, and the y-axis lists 
DHM (with the blue dotted line with cross makers for Company~1 and the solid red line with cross markers for Company~2), 
STC (with the orange dotted line with circular makers for Company~1 and the purple line with circular markers for Company~2), 
and TVH (with the green dotted line with diamond makers for Company~1 and the blue line with diamond markers for Company~2). 
Consistent with the common sense, as Company~1's fleet size rises, DHM, STC, and TVH all increase, while those of Company~2's taper down.



\subsection{Numerical results of 6 ODs in  Sioux Falls network}
\label{append:sioux}

\begin{table}[H]\centering
	\caption{Parameter values for Sioux Falls network}\label{tab:para_SiouxFalls}
	\begin{tabular}{l|l|l}
		\hline
		Name & Notations & Values  \\\hline\hline
		Total travel demands & 2 ODs & $500,200$ \\ 
		& 6 ODs & $300,300,300,300,300,300$ \\ \hline 
		Fare discount (or multiplier) & $\gamma_p$ & $[0,1]$ \\
		Fixed fare & $F^{(e)}$ $F^{(p)}$ & .5, .3 \\
		Time-based fare rate & $\alpha^{[e]}_1,\alpha^{[p]}_1$ & .3, .3 \\ 
		Distance-based fare rate & $\alpha^{[e]}_2,\alpha^{[p]}_2$ & 1,1 \\ \hline 
		Value of time while traveling & $\beta_{(in-veh)}$ & 1.0 \\ 
		Value of time while waiting for pickup & $\beta^{[e]}_{(wt)},\beta^{[p]}_{(wt)}$ & .5, .5 \\ 
		Value of time while matching to drivers & $\beta_{(se-veh)},\beta_{(se-pas)}$ & 1, 1 \\ 
		Supply-demand surplus cost & $\beta^{[m]}_{(surp-pas)}$ & 0, 0 \\
		Conversion rate from in-vehicle travel time to cost & $\beta_{(in-veh)}$ & .1 \\ 
		Inconvenience cost for e-pooling & $\beta_{1(inc)},\beta_{2(inc)}$ & .5, .5 \\ \hline 
	\end{tabular}
\end{table}


\begin{figure}[H]
	\centering
    \includegraphics[scale=.41]{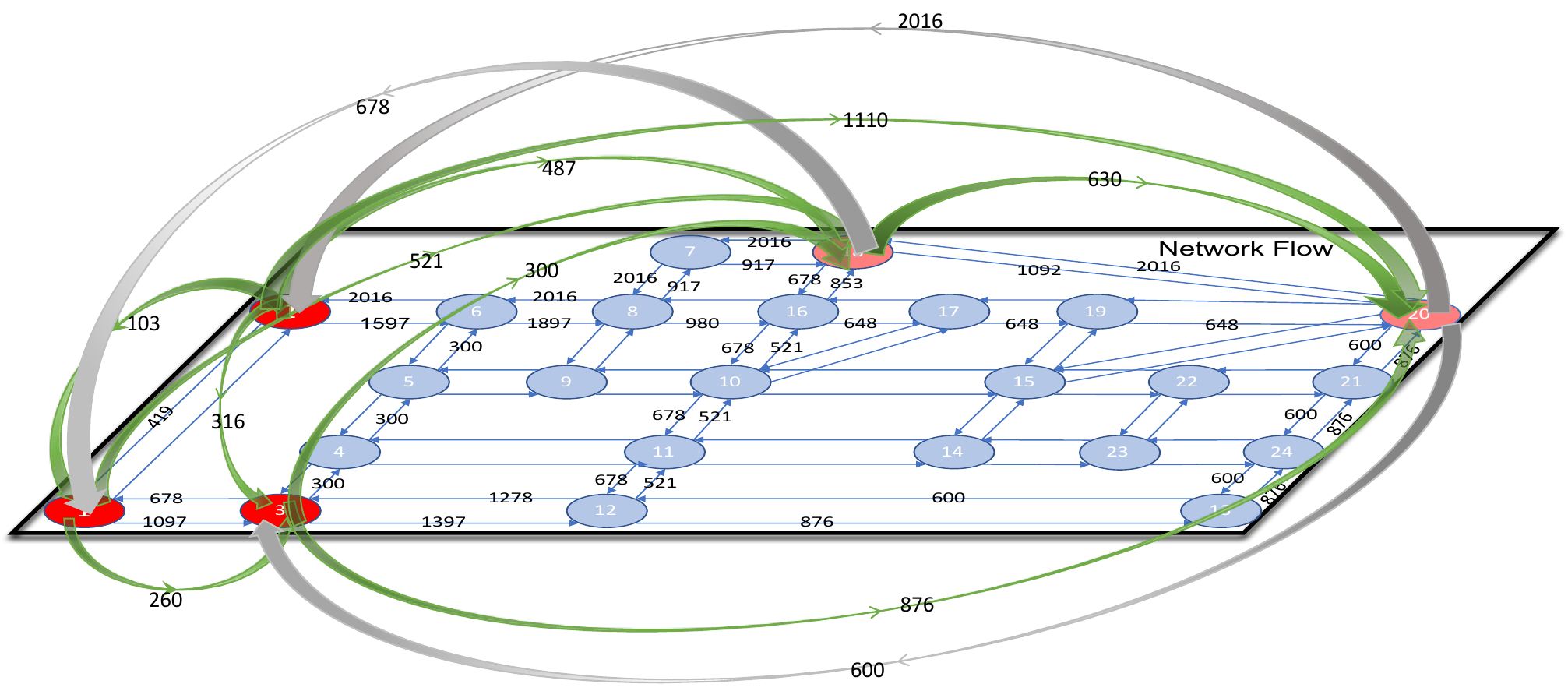}
	\caption{Traffic flow for Scenario~2 Case~2 on Sioux Falls}
	\label{fig:traffic_flow_sioux_fall_6OD_2}
\end{figure}

We inspect details of the vehicle OD flow and analyze the OD sequences.

\textbf{Vehicle Dispatch} (green arrows).

\emph{E-pooling}:

\begin{itemize}
    \item $2 \rightarrow 1 \rightarrow 18 \rightarrow 20$ : 103 vehicles are dispatched from node $2$ to node $1$ in order to fulfill demands of OD pairs: (1,18), (1,20), (2,18) and (2,20). Compared to the traffic flow in Fig.~\ref{fig:traffic_flow_sioux_fall_6OD}, the number of vehicles in e-pooling service decreases. 
    
    \item $2 \rightarrow 3 \rightarrow 20$ : 316 vehicles are dispatched from node $2$ to fulfill demands of OD pairs: (2,20) and (3,20). Compared to the traffic flow in Fig.~\ref{fig:traffic_flow_sioux_fall_6OD}, the number of vehicles in e-pooling service decreases.
    
    \item $1 \rightarrow 3 \rightarrow 20$ : 260 vehicles are dispatched from node $1$ to node $3$ in order to fulfill demands of OD pairs: (1,20) and (3,20). Compared to the traffic flow in Fig.~\ref{fig:traffic_flow_sioux_fall_6OD}, the number of vehicles in e-pooling service increases.
    
    \item $2 \rightarrow 18 \rightarrow 20$ : 487 vehicles are dispatched from node $2$ in order to fulfill demands of OD pairs: (2,18) and (2,20). Compared to the traffic flow in Fig.~\ref{fig:traffic_flow_sioux_fall_6OD}, the number of vehicles in e-pooling service increases.
    
    \item  $2 \rightarrow 20 \rightarrow 18$ : 1110 vehicles are dispatched from node $2$ in order to fulfill demands of OD pairs: (2,18) and (2,20). Compared to the traffic flow in Fig.~\ref{fig:traffic_flow_sioux_fall_6OD}, the number of dispatched vehicles decreases. 
\end{itemize}

\emph{E-solo}:
\begin{itemize}
    \item $1 \rightarrow 18$ : 300 vehicles are dispatched to fulfill the travel demand of OD pair (1,18).
    \item $3 \rightarrow 18$ : 300 vehicles  are dispatched to fulfill the travel demand of OD pair (3,18).
    \item $3 \rightarrow 20$ : 300 vehicles are dispatched to fulfill the travel demand of OD pair (3,20).
\end{itemize}

\textbf{Rebalancing Flow} (grey arrows):
\begin{itemize}
    \item $20 \rightarrow 2$: 2016 vehicles at destination 20 plan to pick up passengers at origin $2$.
    \item $20 \rightarrow 3$:  600 vehicles at destination 20 plan to pick up passengers at origin $3$.
    \item $18 \rightarrow 1$: 678 vehicles at destination 18 plan to pick up passengers at origin $1$. 
\end{itemize}

\end{document}